\documentclass[a4paper, amsfonts, amssymb, amsmath, reprint, showkeys, nofootinbib, twoside]{revtex4-1}
\usepackage[english]{babel}
\usepackage[utf8]{inputenc}
\usepackage{rwd-drafting}
\usepackage{amsthm}
\usepackage{mathtools}
\usepackage{physics}
\usepackage{xcolor}
\usepackage{graphicx}
\usepackage[left=23mm,right=13mm,top=35mm,columnsep=15pt]{geometry} 
\usepackage{adjustbox}
\usepackage{placeins}
\usepackage[T1]{fontenc}
\usepackage{lipsum}
\usepackage{csquotes}
\usepackage{tikz}
\usepackage{float}
\usepackage{xspace}
\usepackage[colorlinks,linkcolor=blue,anchorcolor=blue,citecolor=blue,urlcolor=blue]{hyperref} 
\usepackage{cleveref}
\usetikzlibrary{quantikz2}
\usepackage{svg}
\usepackage{float}

\newcommand{\prepare}{\mbox{P{\scriptsize REPARE}}\xspace}
\newcommand{\prep}{\mbox{P{\scriptsize REP}}\xspace}
\newcommand{\select}{\mbox{S\scriptsize ELECT}\xspace}
\newcommand{\sel}{\mbox{S\scriptsize EL}\xspace}
\newcommand{\lcu}{\mbox{LCU}}

\newcommand{\mctrl}[1]{
  \gate[style={inner sep=-1, minimum height=1pt, minimum width=1pt}]{}\vqw{#1}
}

\usepackage{xspace}

\newcommand{\VV}{\ensuremath{\overline{V}}\xspace}
\newcommand{\PP}{\ensuremath{\overline{P}}\xspace}
\newcommand{\CC}{\ensuremath{\mathcal{C}}\xspace}

\newcommand{\ry}{\ensuremath{R_y}\xspace}
\newcommand{\rz}{\ensuremath{R_z}\xspace}
\newcommand{\cx}{\ensuremath{\textsc{cx}}\xspace}
\newcommand{\ccx}{\ensuremath{\textsc{ccx}}\xspace}

\newcommand{\x}{\ensuremath{X}\xspace}

\newcommand{\z}{\ensuremath{Z}\xspace}

\newcommand{\identity}{\ensuremath{I}\xspace}

\newcommand{\Z}{\z}
\newcommand{\Ry}{\ry}
\newcommand{\Rz}{\rz}
\newcommand{\CX}{\cx}

\begin{document}
\title{Hamiltonian dynamics simulation using linear combination of unitaries \\ on an ion trap quantum computer}

\author{Michelle Wynne Sze}
\email[]{michelle.sze@quantinuum.com}
\author{Yao Tang}
\email[]{yao.tang@quantinuum.com}
\author{Silas Dilkes}
\email[]{silas.dilkes@quantinuum.com}
\author{David Muñoz Ramo}
\email[]{david.munozramo@quantinuum.com}
\author{Ross Duncan}
\email[]{ross.duncan@quantinuum.com}
\author{Nathan Fitzpatrick}
\email[]{nathan.fitzpatrick@quantinuum.com}
\affiliation{Quantinuum, 13-15 Hills Road, CB2 1NL, Cambridge, United Kingdom}

\date{\today}

\begin{abstract}
The linear combination of unitaries (LCU) method has proven to scale better than existing product formulas in simulating long time Hamiltonian dynamics. However, given the number of multi-control gate operations in the standard prepare-select-unprepare architecture of LCU, it is still resource-intensive to implement on the current quantum computers. In this work, we demonstrate LCU implementations on an ion trap quantum computer for calculating squared overlaps $|\langle \psi(t=0)|\psi(t>0)\rangle|^2$ of time-evolved states. This is achieved by an optimized LCU method, based on pre-selecting relevant unitaries, coupled with a compilation strategy which makes use of quantum multiplexor gates, leading to a significant reduction in the depth and number of two-qubit gates in circuits. For $L$ Pauli strings in a Taylor series expanded $n$-qubit-mapped time evolution operator, we find a two-qubit gate count of $2^{\lceil log_2(L)\rceil}(2n+1)-n-2$. We test this approach by simulating a Rabi-Hubbard Hamiltonian.
\end{abstract}

\maketitle
\section{Introduction}

Hamiltonian dynamics simulations serve as an excellent testbed for quantum computing applications due to the potential exponential speedup on offer. It is indispensable in studying electronic structures and calculating the energy spectra of large molecules\cite{Kassal2010, Wecker2013, Reiher2016}, understanding thermalization\cite{Kaufman2016}, phase transitions\cite{Figueroa2018} and out-of-equilibrium properties\cite{Defenu2024} of quantum many-body systems. However, the larger the many-body system, the more intractable it becomes for a classical computer to handle due to an exponentially expanding Hilbert space \cite{Wecker2013} and entanglement of the time-evolved states \cite{Defenu2024}. An ideal quantum computer is believed to outperform a classical one by overcoming these difficulties \cite{Childs2018, Kassal2010}.

Many quantum algorithms have been developed for the purpose of simulating Hamiltonian dynamics by implementing the unitary evolution operator $U=\exp(-iHt)$, where $H$ is a system Hamiltonian and $t$ is the evolution time.\footnote{We assume time-independent $H$ here. In general, it can be time-dependent and quantum algorithms are extended for such $H$ \cite{sharma2024hamiltoniansimulationinteractionpicture, low2019hamiltoniansimulationinteractionpicture, PRXQuantum.2.030342,Zhao_2024, Ikeda_2023, PRXQuantum.5.040316,cao2023quantumsimulationtimedependenthamiltonians, An_2022, Berry_2020_tdH}.} Approaches include the product formulas, or Trotterization \cite{Trotter1959, Suzuki1976} method (and its different flavors \cite{Suzuki1990, Granet2024, Childs2018}), linear combination of unitaries (LCU) \cite{ChildsLCU} or sum formula, quantum singular value transform (QSVT) \cite{qsvtguilyen,grandunification}, and combinations of these algorithms \cite{Low2019, Zeng2022, YuanSu, Zhuk2023, Vazquez2023}. Trotterization is widely studied and utilized, partially due to its straightforward implementation as a series of small angle Pauli exponentials with no extra ancilla qubits. Compact quantum circuits can be constructed with a variety of methods, such as alternative decompositions \cite{Haah2018}, arranging terms by exploiting commutativity \cite{Childs2019}, or pre-selecting Hamiltonian terms to reduce gate counts and depth from gate cancellations \cite{Wecker2015}. Hamiltonians which exhibit local interactions resulting in numerous sets of commuting terms, such as the Ising model, make good candidates for Trotter application \cite{YuanSu, Childs_2021, Tran_2020}. While the precision depends on the time resolution, the linear scaling with time for a fixed time step size is the main contributor to deep circuit. Techniques which rely on randomization have been exploited to ameliorate this issue \cite{Cho2022, Childs_Ostrander_Su2018, Campbell2018, Chen2021}, but the improved performance is also dependent on specific types of Hamiltonian. Recent experimental studies have found that Trotter errors appear to be constant for certain time scales but this, again, depends on the type of initial states and Hamiltonian \cite{Chertkov2024, Granet2024b, chen2024trottererrortimescaling}. A promising algorithm that can perform Hamiltonian dynamics due to better asymptotic scaling is the LCU via truncated Taylor series \cite{Meister2022tailoringterm, Berry_etal_2015, Berry1}. The time evolution operator is expressed as a linear combination of Pauli matrices (or any unitaries, in general), and the sum is encoded via two main subroutines -- one that prepares coefficients by writing them to basis states on an ancillary prepare register, and a second that carries out the Pauli gate operations conditioned on the prepare register. This algorithm calls for an extra number of qubits which is at least logarithmic in the number of terms in LCU. Moreover, since the success of LCU relies highly on its shot-based measurements, an amplitude amplification is often used. Including the amplitude amplification in the circuit implementation costs at least threefold\footnote{See Appendix \ref{appendix:oaa}.}, but requires fewer measurements. The LCU method is versatile as it offers scope for extensibility and improvement; for example, extending to a time dependent Hamiltonian in the interaction picture \cite{low2019hamiltoniansimulationinteractionpicture,Loaiza_2023} or reducing the $\mathcal{\ell}_1$-norm to increase the success probability \cite{Loaiza_2023}. Molecular or chemical Hamiltonians are commonly used in theoretical studies of LCU methods. Furthermore, due to its generality, LCU has been used as a component in  many quantum
algorithms, including 
phase estimation~\cite{BabbushPhase, LeeTensorHyper, Babbush1Q, ShokrianZini2023quantumsimulationof},
power series expansion~\cite{Kalev2021quantumalgorithm},
quantum walks~\cite{Berry2},
Trotterization~\cite{YuanSu},
amplitude amplification~\cite{Amp1},
differential equation solvers~\cite{dif1,dif2,Childs2021highprecision},
ground state preparation~\cite{CiracGS,keen2021quantum,PhysRevA.106.032420},
measurement reduction in variational algorithms~\cite{AlexisLCU},
linear response~\cite{RallPhysical},
Greens functions~\cite{LinLinGFLCU},
Gibbs sampling~\cite{SommaGibbsSampling}, and
semidefinite programming solvers~\cite{vanApeldoorn2020quantumsdpsolvers}.
It can also be used as an input model for the Quantum Singular Value
Transform (QSVT)~\cite{qsvtguilyen, grandunification}. QSVT, considered as a generalized quantum signal processing (QSP) algorithm, shows great promise in terms of asymptotic error, complexity and resource scaling \cite{Childs2018, toyoizumi2023hamiltoniansimulationusingquantum, Berry2024}.  QSVT utilizes the LCU method to implement the linear combination of $\cos{(H't)}$ and $\sin{(H't)}$, where $H'$ is the diagonalized $H$. The trigonometric functions are expanded in appropriate polynomials that lead to efficient circuit implementations. Indeed, the efficiency of QSVT is best demonstrated on a large quantum computer with high fidelity operations. However, on noisy near-term devices, large quantum circuits from many QSP iterations may prove prohibitively expensive. Here, we consider collapsing a truncated Taylor series expansion of the evolution operator into a single LCU as an alternative with smaller quantum circuits.

While considerable efforts have been put into designing and improving quantum algorithms for Hamiltonian dynamics simulations, experimental progress is sparse in demonstrating these computing techniques. In these few experiments the Trotter method dominates as it requires no more qubits than the quantum system, with Trotter experiments \cite{Brown2006, Barends2015, Lanyon2011, Raeisi2012} using either a few Trotter steps or short time scale dynamics. Unsurprisingly, the LCU method has been less utilized \cite{yu2022simulating, Zeng2022} owing to its restrictively expensive resource overhead and large two-qubit gate counts; thus, LCU functioned only as a complement to the Trotter method in these experiments.  Few strategies \cite{Boyd2023, Meister2022tailoringterm, Chakraborty2023, loaiza2023blockinvariantsymmetryshiftpreprocessing, Loaiza_2023} have been developed to make LCU more amenable to a resource-limited device. Some of the proposed techniques include exploiting commuting terms in the LCU and executing these commuting observables in parallel on a circuit level \cite{Boyd2023}, preprocessing procedures \cite{Loaiza_2023} that lower the $\ell_1-$norm of LCU which is inversely related to the success probability of the algorithm, or importance sampling that utilizes single-ancilla LCU \cite{Chakraborty2023}. 

In this work, we simulate the dynamics of a simple light-matter interaction by running LCU experiments on Quantinuum's H1-1 trapped ion computer \cite{quantinuum_h11}.  We carefully choose a quantum system such that LCU circuits benefit from a circuit construction technique using multiplexor gates. Since LCU is known to suffer high overhead costs, our goal is to reduce the circuit resource requirements that meet the capacity of the H1-1 device which has a non-negligible error rate. We achieve this in two levels. First, a classical pre-selection procedure is employed where only relevant terms in the observable are chosen by exploiting the symmetry properties of the system. On the circuit level, we use quantum multiplexor gates to implement the largest subroutine in the LCU construction, the
\select oracle, which for our problem reduces the circuit size in comparison to unary iteration\cite{Babbush_2018}. We focus on the number of two-qubit (2Q) gates in the circuit, a good indicator of overall circuit size, and the key to performance using noisy physical gates.  The core of the method is an alternative exact synthesis of the large multi-controlled gates which define the oracle circuit. To date, the multiplexor construction, which has been employed previously to demonstrate experimentally the QSP algorithm \cite{Kikuchi_23}, may not be the most efficient for asymptotically large quantum systems where unary iteration is. It is, however, more economical for the number of qubits in the prepare and system registers of and the total gate count for our chosen test case, hence proving applicable in a pre-large scale quantum computing device.

The rest of the paper is organized as follows. We review the LCU method and describe our modified truncated Taylor series approach in Sec.~\ref{sec:tts_lcu}. We discuss quantum multiplexor gates for the $\select$ oracle compilation and the reduced squared overlap strategy in Sec.~\ref{sec:method}. We present our results using the Rabi-Hubbard Hamiltonian in Sec.~\ref{sec:results}. Finally, in Sec.~\ref{sec:discussion}, we summarize some notable aspects of the pioneering experiment and suggest improvements for possible future work.

\section{Truncated Taylor series and linear combination of unitaries}\label{sec:tts_lcu}

We adapt the truncated Taylor series method for the evolution operator described in Ref.~\cite{Berry_etal_2015}. For a quantum system with a Hamiltonian $H$ evolved over time $t$, the time evolution operator is given by $U = e^{-iHt} =  (e^{-iH\tau})^m $, segmented into $m$ time steps of width $\tau=t/m$. The segment $e^{-iH\tau}$ is then Taylor expanded and truncated up to order $K$:
\begin{align}{\label{eq:U_dt}}
    U_{\tau,K} = \sum_{k=0}^K \frac{(-iH\tau)^k}{k!} = \sum_{k=0}^K  \frac{1}{k!} \left(-i\tau\sum_{j}^{L_H}\beta_j H_j\right)^k,
\end{align}
where $H_j$ are the Pauli strings that make up the qubit representation of the Hamiltonian and $L_H$ is the number of Pauli strings. Powers of linear combination of Pauli strings, represented by the right-hand-side of Eq.~(\ref{eq:U_dt}), result in another linear combination of Pauli terms. The original conception of the method \cite{Berry_etal_2015} utilizes  $K+1$ ancilla registers on top of the system register - one of which has $K+1$ qubits, and the rest has each $\lceil \log_2(L_H)\rceil$ qubits - a series of LCUs within another LCU. We diverge from this approach by expanding classically the powers of $H$ further, thereby constructing only a single LCU for $U_{\tau,K}$. Calculating the moments of large system qubit Hamiltonian is not trivial~\cite{Vallury2020}, but here we assume that this can be accomplished.  Consequently, after exponentiation of the resulting linear combination $U_{\tau,K}$ to power $m$, the result is another linear combination of Pauli terms $\mathcal{P}_j$ with corresponding complex coefficients $\alpha_j$:
\begin{align}{\label{eq:U_app}}
    U \approx \widetilde{U} =  (U_{\tau,K})^m = \sum_{j}\alpha_j \mathcal{P}_j.
\end{align}
Note that because of the truncation, $\widetilde{U}$ may not be unitary.

\begin{figure}[hbt!]
\centering
\begin{quantikz}
  \lstick{$|\bar{0}\rangle$}
  & \gate{\prep(\alpha)}
  & \ctrl{1}
  & \gate{\prep(\alpha)^\dagger}
  & \meterD{\bar{0}}
  \\
  \lstick{$\ket{\psi}$}
  & \qw
  & \gate{\sel(\Upsilon)}
  & \qw
  & \qw \rstick{$\frac{\Upsilon}{|\alpha|_1}\ket{\psi}$}
  \\
\end{quantikz}
\caption{LCU Circuit for an operator $\Upsilon$, when a successful postselection
  occurs.} 
\label{fig:lcu_figure0}
\end{figure}
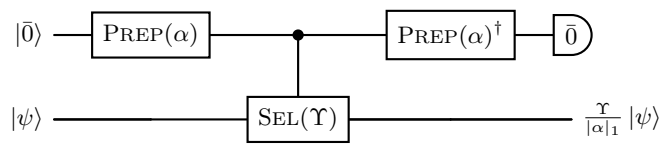

The linear combination of unitaries (LCU) approach allows a
\emph{non}-unitary operator
\begin{align}\label{eq:lcu}
\Upsilon = \sum_{\ell=0}^{L-1}\alpha_\ell \mathcal{P}_\ell
\end{align}
to be applied to a quantum state
$\ket{\psi}$.  This works by embedding $\Upsilon$ as a block in a larger
unitary,
\begin{equation}\label{equation:blockform}
  U_{LCU} =
  \begin{pmatrix}
    \frac{\Upsilon}{|\alpha|_1} & * \\ 
*  & *
\end{pmatrix},
\end{equation}
$|\alpha|_1=\sum_\ell |\alpha_\ell|$, acting on a larger quantum register and by measuring and
post-selecting on ancilla qubits, projecting the state into the subspace
where $\Upsilon$ is applied\footnote{ The block encoded operator $\Upsilon/|\alpha|$ is $\Upsilon$ rescaled
  by its $\ell_1$-norm
  to ensure unitarity.}. This is achieved by using two
oracles, $\prepare$ and $\select$, which encode the coefficients $\alpha_\ell$ and unitaries $\mathcal{P}_\ell$ as
shown in Fig.~\ref{fig:lcu_figure0}. In general, $\alpha_\ell$'s are complex coefficients, but in the LCU implementation, the phases of these coefficients are absorbed in the operators $\mathcal{P}_\ell$ so that from here on, we assume $\alpha_\ell \in \mathbb{R^+}$. The $\prepare(\alpha)$ oracle maps the all-zero state of the ancilla or control register to a state containing the weighted coefficients ${\sqrt{\alpha_\ell/|\alpha|_1}}$, i.e.,
\begin{align}
\prepare(\alpha): |\bar{0}\rangle \mapsto \sum_{\ell=0}^{L-1} \sqrt{\frac{\alpha_\ell}{|\alpha|_1}} |\ell\rangle.
\end{align}
On the other hand, the \select($\Upsilon$) is a multi-controlled unitary $\mathcal{P}_\ell$ acting on a
target or system register and conditioned on the control register, i.e.,
\begin{equation}
\select(\Upsilon):=\sum^{L-1}_{\ell=0} |\ell\rangle \langle \ell | \otimes \mathcal{P}_\ell.
\label{equation:select}
\end{equation}
Operating these two oracles together as illustrated in Fig. \ref{fig:lcu_figure0}, one obtains the state $\Upsilon|\psi\rangle$ successfully upon a measurement of $|\bar{0}\rangle$ on the prepare, or ancilla, register. Therefore, the LCU method has a measurement overhead due to a success probability given by
\begin{equation}
\label{eq:lcu_success_prob}
    p = \frac{1}{|\alpha|
_1^2}\langle \psi| \Upsilon^\dagger \Upsilon | \psi \rangle,  
\end{equation}
which is related to how close to unitary $\Upsilon$ is and the initial state overlap. Both the \prepare and \select oracles can be implemented using
\emph{multiplexor} gates which we discuss in Sec.~\ref{subsec:multiplexor}.

To boost the LCU success probability in Eq.~(\ref{eq:lcu_success_prob}), oblivious amplitude amplification (OAA) is often applied. Details of OAA and sample application outcome are included in Appendices~\ref{appendix:oaa} and~\ref{appendix:jc}.

\section{Method}\label{sec:method}

\subsection{Quantum multiplexor gates}\label{subsec:multiplexor}

A \emph{quantum multiplexor gate} is a generalisation of familiar controlled
gates like the \CX, but which acts on its target register with many
possible unitary operations.  To be precise, let
$\VV :=[V_0,\cdots,V_{2^k-1}]$ be a list of unitary operators acting on a
$n$-qubit register; then the $k$-control multiplexed-$\VV$ gate is
defined as:
\begin{equation}\label{def:multiplexor}
  M^{k}_{n}(\VV) 
  = \sum^{2^k-1}_{i=0}\ket{i}\bra{i}_\CC \otimes V_i,
\end{equation}
where integers $i$ are mapped to binary strings defining computational basis states.



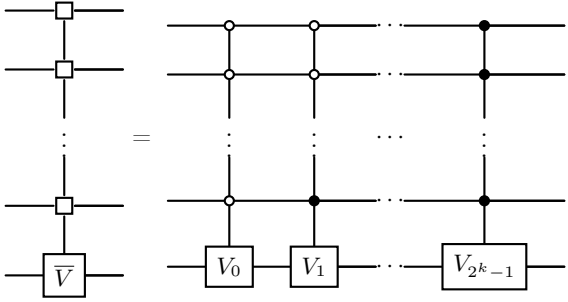
\begin{figure}
    \begin{quantikz}[wire types={q,q,n,q,q}]
        \qw& \mctrl{1}& \qw \\
        \qw& \mctrl{1}& \qw \\
        & \vdots\vqw{1}& \\
        \qw& \mctrl{1}& \qw \\
        \qw& \gate{\VV}& \qw \\
    \end{quantikz} 
    =
    \begin{quantikz}[wire types={q,q,n,q,q}]
        \qw& \octrl{1}& \octrl{1}& \qw \cdots& \ctrl{1}& \qw \\
        \qw& \octrl{1}& \octrl{1}& \qw \cdots & \ctrl{1}& \qw \\
        \qw& \vdots\vqw{1}& \vdots\vqw{1}& \cdots & \vdots\vqw{1} &  \\
        \qw& \octrl{1} &\ctrl{1} & \qw \cdots& \ctrl{1}& \qw \\
        \qw& \gate{V_0}& \gate{V_1} & \qw \cdots & \gate{V_{2^k -1}}& \qw \\
    \end{quantikz}
    \caption{A $k$-controlled multiplexed gate  $M^\CC_n(V)$ implemented as a sequence of multi-controlled gates. The white square boxes designate multiplexed controls.}
    \label{fig:expmulitplex}
\end{figure}

A multiplexor is equivalent to a series of multi-controlled
quantum gates covering all control conditions spanned by the $2^k$
basis states, as in Fig.~\ref{fig:expmulitplex}. For example, a \ccx
gate is a $2$-control multiplexor gate
\begin{equation}
  \ccx = M^2_1(\VV)
  = \sum^{2}_{i=0}\ket{i}\bra{i}\otimes \identity + \ket{3}\bra{3}\otimes \x
\end{equation}
whose single qubit target gates are  $\VV = [\identity,\identity,\identity,\x]$.

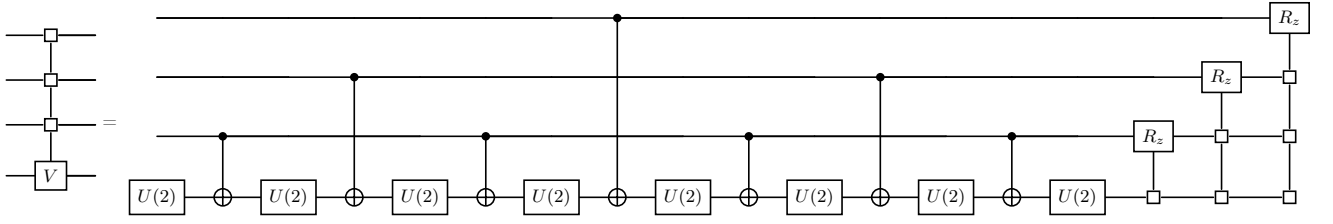
\begin{figure*}[!ht]
\centering
\adjustbox{width=\textwidth}{
  \begin{quantikz}
    \qw    &\mctrl{1}       &\qw\\
    \qw    &\mctrl{1}       &\qw\\
    \qw    &\mctrl{1}       &\qw\\
    \qw    &\gate{V}              &\qw\\ 
  \end{quantikz}
  =
  \begin{quantikz}
    \qw&\qw&\qw&\qw&\qw&\qw&\qw&\ctrl{3}&\qw&\qw&\qw&\qw&\qw&\qw&\qw&\qw&\qw&\gate{\Rz}\\
    \qw&\qw&\qw&\ctrl{2}&\qw&\qw&\qw&\qw&\qw&\qw&\qw&\ctrl{2}&\qw&\qw&\qw&\qw&\gate{\Rz}&\mctrl{-1}\\
    \qw&\ctrl{1}&\qw&\qw&\qw&\ctrl{1}&\qw&\qw&\qw&\ctrl{1}&\qw&\qw&\qw&\ctrl{1}&\qw&\gate{\Rz}&\mctrl{-1}&\mctrl{-1}\\
    \gate{U(2)}&\targ{}&\gate{U(2)}&\targ{}&\gate{U(2)}&\targ{}&\gate{U(2)}&\targ{}&\gate{U(2)}&\targ{}&\gate{U(2)}&\targ{}&\gate{U(2)}&\targ{}&\gate{U(2)}&\mctrl{-1}&\mctrl{-1}&\mctrl{-1}\\
  \end{quantikz}
}
\caption{Decomposition for $M^\CC_n(V)$ where $V$ is
  a list of U(2) unitaries, $n=1$, and $\CC=3$.}
    \label{fig:vdecomp}
\end{figure*}

A $k$-control multiplexor targeting a single qubit can be recursively
decomposed into a quantum circuit with $2^k - 1$ \cx gates, $2^k$
single qubit gates, and a cascade of $k$ multiplexed-\Rz
gates~\cite{Bergholm_2005}, as illustrated in Fig.~\ref{fig:vdecomp}.  The
multiplexed-\Rz gates can themselves be implemented 
recursively~\cite{Shende_2006}, requiring an additional $2^{k+1} -2$
\CX gates.  See Appendix \ref{sec:app_m} for details of these
constructions. 

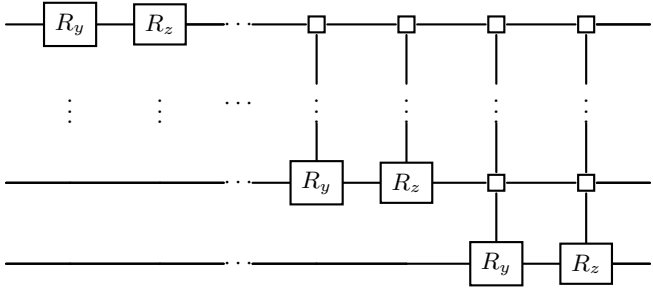
\begin{figure}
    \begin{quantikz}[wire types={q,n,q,q}]
\qw   & \gate{\Ry} & \gate{\Rz}         & \qw \cdots & \mctrl{1}    & \mctrl{1}     & \mctrl{1}     & \mctrl{1}     & \qw \\
& \vdots    & \vdots     & \cdots     & \vdots\vqw{1}& \vdots\vqw{1} & \vdots\vqw{1} & \vdots\vqw{1} & \\
\qw   & \qw       & \qw        & \qw \cdots & \gate{\Ry}    & \gate{\Rz}    & \mctrl{1}     & \mctrl{1}     & \qw \\
\qw   & \qw       & \qw  & \qw \cdots & \qw          & \qw          & \gate{\Ry}     & \gate{\Rz}     & \qw \\
    \end{quantikz}
\caption{Circuit construction that prepares an arbitrary state
  $\ket{\psi}$ from a initial $\ket{\bar0}$ state.}
\label{fig:complexprep}
\end{figure}

An arbitrary complex quantum state preparation can be constructed from
a series of interleaved multiplexed-\Rz and multiplexed-\Ry
gates~\cite{Shende_2006} as shown in Fig.~\ref{fig:complexprep}.
$\prepare(\alpha)$ is an arbitrary $k$-qubit \emph{real}-valued quantum state
preparation.  Since only real coefficients are required, the circuit
can be constructed with only multiplexed-\Ry gates, requiring at most $2^{k}-2$
\CX gates.


%

\subsection{The $\select$ oracle}
\label{sec:select-oracle}

While any operator can, in principle, be written as a linear
combination of unitaries, in this work we will focus on Hamiltonians
expressed as sums of Paulis.  Since we require that the coefficients
in Eq.~(\ref{eq:lcu}) are real, we have
\begin{equation}\label{eq:phased-pauli-terms}
  \mathcal{P}_\ell = e^{\text{i}\theta_\ell} \bigotimes_{j=0}^{n-1} P_{\ell,j},
\end{equation}
where $n$ is the number of qubits in the system, and each $P_{\ell,j}$
is a single-qubit Pauli operator; i.e.\! $P_{\ell,j} \in \{I, X, Y, Z\}$.
Therefore, for an observable $\Upsilon$ given as Pauli terms, $\select(\Upsilon)$ is a
multiplexor gate whose target operators are Pauli words with a
complex phase.  Setting $\PP = [\mathcal{P}_0,\ldots,\mathcal{P}_\ell]$, we have:
\begin{equation}\label{eq:select-h-phased-paulis}
  \begin{split}
    M^k_n(\PP)
    & = \sum^{2^k-1}_{\ell=0} \ket{\ell}\bra{\ell} \otimes \mathcal{P}_\ell  \\
    & = \sum^{2^k-1}_{\ell=0} e^{\text{i}\theta_\ell}
      \ket{\ell}\bra{\ell} \otimes P_{\ell,0}\otimes ...\otimes
      P_{\ell,n-1}      
  \end{split}
\end{equation}
To simplify notation, in the following we will assume that the phase
factor $e^{\text{i} \theta_\ell}$ is included in the Pauli matrix $P_{\ell,0}$.

Since the target terms are all tensor products of Pauli matrices, this
multiplexor can be rewritten as a product of multiplexors with single
qubit targets,
\begin{equation}    \label{equation:commutingpaulis}
\begin{split}
  M^k_n(\PP)
  & = \prod^{n-1}_{j=0}\sum^{2^k-1}_{\ell=0}\ket{\ell}\bra{\ell}\otimes
  P_{\ell, j}
  \\
  & = \prod^{n-1}_{j=0} M^k_1(\PP_j) 
\end{split}
\end{equation}
where $\PP_j = [P_{0,j},\ldots,P_{\ell,j}]$, the $j$th tensor factor of
\PP, acting on the $j$th qubit of the target register.



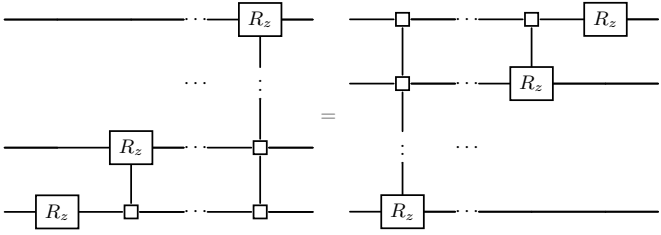
\begin{figure}
    \begin{adjustbox}{width=0.5\textwidth}
    \begin{quantikz}[wire types={q,n,q,q}]
        \qw&\qw&\qw&\qw\cdots&\gate{\Rz}\vqw{1}&\qw\\
           &   &   &   \cdots&\vdots&   \\
        \qw&\qw&\gate{\Rz}&\qw\cdots&\mctrl{-1}&\qw\\
        \qw&\gate{\Rz}&\mctrl{-1}&\qw\cdots&\mctrl{-1}&\qw\\
    \end{quantikz}
    =
    \begin{quantikz}[wire types={q,q,n,q}]
        \qw&\mctrl{1}&\qw\cdots&\mctrl{1}&\gate{\Rz}&\qw\\
        \qw&\mctrl{1}&\qw\cdots&\gate{\Rz}&\qw&\qw\\
           &\vdots&   \cdots&   &   &   \\
        \qw&\gate{\Rz}\vqw{-1}&\qw\cdots&\qw&\qw&\qw\\
    \end{quantikz}
    \end{adjustbox}
    \caption{A diagonal operator can be resynthesized to change the order of multiplexed-\Rz gates.}
    \label{fig:rzrewrite}
\end{figure}

As with any multiplexed-U(2) gates, each $M^k_1(\PP_j)$ can be
implemented with \CX, multiplexed-\Rz and single-qubit gates, as in
Fig.~\ref{fig:complexprep}.  With this decomposition, the final
sequence of multiplexed-\Rz gates defines a diagonal operator.  This
operator can be resynthesised such that only one multiplexed-\Rz acts
on the target register, with remaining multiplexed-\Rz acting on the
control register, shown in Fig. \ref{fig:rzrewrite}.
When synthesising $M^k_n(\PP)$, the control register part of this
diagonal operator can be commuted through the $M^k_1(\PP_j)$
controls and merged. This gives a significant gate count reduction, as
multiplexed-\Rz on the control register only need to be synthesised
once.

Using this decomposition, $\select(\Upsilon)$ requires at most $2^k(2n+1)-n-2$ \CX gates where $k$ is the size of the
control register and $n$ the size of the target register.





\subsection{Reduced squared overlap}\label{sec:sq_overlap}

Here, we outline an efficient method for calculating the squared overlap  $|\langle \psi_0| \Upsilon | \psi_0 \rangle|^2$, if $\Upsilon \approx e^{-iHt}$, as defined in Eqs.~(\ref{eq:U_dt}) and ~(\ref{eq:U_app}), and where $|\psi_0\rangle$ is a given normalized initial state. 

As shown in Sec.~\ref{sec:select-oracle}, the number of \CX gates scales linearly with the the number of terms. We can reduce the gate overhead with a preselection scheme that removes LCU terms in $\Upsilon$ with a zero overlap. We write $\Upsilon = \Upsilon_{\parallel} + \Upsilon_{\perp}$, where $\langle \psi_0 | \Upsilon_{\parallel} | \psi_0 \rangle \neq 0$ and 
$\langle \psi_0 | \Upsilon_{\perp} | \psi_0 \rangle = 0$. The final state $|\psi_f\rangle$ then is 
\begin{align}
    |\psi_f \rangle = \frac{\left(\Upsilon_{\parallel} + \Upsilon_{\perp} \right) |\psi_0 \rangle}{\sqrt{\langle \psi_0 | \Upsilon^\dagger  \Upsilon | \psi_0 \rangle }},
\end{align}
where the denominator preserves the normalization of the returned final state.
The overlap with $|\psi_0\rangle$ is
\begin{align}
\label{eq:overlap}
    \langle \psi_0 | \psi_f \rangle &= \frac{1}{\sqrt{\langle \psi_0 | \Upsilon^\dagger  \Upsilon | \psi_0 \rangle }} \left(\langle \psi_0 | \Upsilon_{\parallel}|  \psi_0 \rangle + \langle \psi_0 | \Upsilon_{\perp}|  \psi_0 \rangle \right) \nonumber\\
    &= \frac{1}{\sqrt{\langle \psi_0 | \Upsilon^\dagger  \Upsilon | \psi_0 \rangle }} \langle \psi_0 | \Upsilon_{\parallel}|  \psi_0 \rangle .
\end{align}
Then, we can implement the LCU only for $\Upsilon_\parallel$, and the returned final state is
\begin{align}
    |\psi_f ^ \parallel \rangle = \frac{\Upsilon_{\parallel}  |\psi_0 \rangle}{\sqrt{\langle \psi_0 | \Upsilon_\parallel^\dagger  \Upsilon_\parallel | \psi_0 \rangle }}.
\end{align}
The reduced squared overlap $|\langle \psi_0 |\psi_f ^ \parallel \rangle |^2$ is then given by
\begin{align}
    |\langle \psi_0 |\psi_f ^ \parallel \rangle |^2 = \frac{1}{\langle \psi_0 | \Upsilon_\parallel^\dagger  \Upsilon_\parallel | \psi_0 \rangle } | \langle \psi_0 | \Upsilon_{\parallel}|  \psi_0 \rangle |^2.
\end{align}
Using this to calculate the squared overlap in Eq.~(\ref{eq:overlap}), we get 
\begin{align}\label{eq:sq_overlap_from_reduced}
    |\langle \psi_0 |\psi_f \rangle |^2 = \frac{\langle \psi_0 | \Upsilon_\parallel^\dagger  \Upsilon_\parallel | \psi_0 \rangle}{\langle \psi_0 | \Upsilon^\dagger  \Upsilon | \psi_0 \rangle} |\langle \psi_0 |\psi_f ^ \parallel \rangle |^2.
\end{align}
In our problem, since $\Upsilon \approx e^{-iHt}$, we assume $\Upsilon^\dagger \Upsilon \approx I$. From the definition of LCU success probability in Eq.~(\ref{eq:lcu_success_prob}), we recognize that 
\begin{align}\label{eq:p_parallel}
    \langle \psi_0 | \Upsilon_\parallel^\dagger  \Upsilon_\parallel | \psi_0 \rangle = p_\parallel |\alpha_\parallel|_1^2,
\end{align}
where $p_\parallel$ is the LCU success probability of $\Upsilon_\parallel$, and $|\alpha_\parallel|_1$ is the $\ell_1$ norm of the  $\Upsilon_\parallel$ coefficients. Equations~(\ref{eq:sq_overlap_from_reduced}) and ~(\ref{eq:p_parallel}) allow one to compute $|\langle \psi_0 |\psi_f \rangle |^2$ by performing LCU for the reduced $\Upsilon_\parallel$. 

\begin{figure}[h!]
\centering
\begin{quantikz}
\lstick{$|0\rangle_p$} & \qwbundle{} &  & \gate[2]{\lcu (\Upsilon_\parallel)}&  &\meterD{0\cdots0}  \\
\lstick{$|0\rangle_\psi$} & \qwbundle{} & \gate{U_{\psi}} &  & \gate{U_{\psi}^\dagger}  & \meterD{0\cdots0}  \end{quantikz}
\caption{Circuit for $| \langle 0 |_\psi U_\psi^\dagger \langle 0|_p \lcu(\Upsilon_\parallel) |0\rangle_p U_\psi |0\rangle_\psi|^2$.}
\label{fig:sq_overlap}
\end{figure}
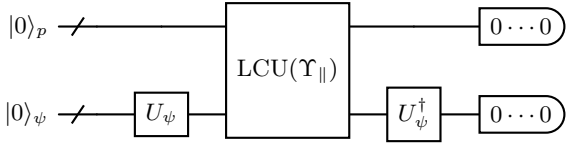

With a circuit, the squared overlap $|\langle \psi_0 |\psi_f ^ \parallel \rangle |^2 = |\langle \psi_0 |\Upsilon_\parallel|\psi_0 \rangle |^2 $ can be calculated using the vacuum test \cite{Tudorovskaya_2024}. Let $U_\psi$ be the unitary that prepares the initial state $|\psi_0\rangle$ such that $U_\psi|\widetilde{0}\rangle = |\psi_0\rangle$ and $\langle \widetilde{0} |U_\psi^\dagger = \langle\psi_0|$.  Then, $|\langle \psi_0 |\Upsilon_\parallel|\psi_0 \rangle |^2 = |\langle \widetilde{0} | {U_\psi}^\dagger \Upsilon_\parallel U_\psi|\widetilde{0}\rangle |^2$. Figure~\ref{fig:sq_overlap} illustrates the circuit used to calculate the squared overlap $|\langle \psi_0 | \Upsilon_\parallel | \psi_0 \rangle|^2$, where both the ancilla and system registers are initialized to $|0,0,...\rangle$. 

Grouping terms between $\Upsilon_\parallel$ and $\Upsilon_\perp$ can be straightforwardly done by calculating
 $\mathcal{P}_j |\psi_0 \rangle$ for each $j$ and determining if the new state is in the basis of $|\psi_0\rangle$. However, this is only viable for smaller system. For larger systems, another approach is to exploit symmetry of the system (the initial state and the observable) to group the terms.  detail this process in Appendix~\ref{appendix:grouping} for the system used in this paper.
\section{Results}\label{sec:results}

For the Hamiltonian in $e^{-iHt}$, we choose the Rabi-Hubbard (RH), a spin-boson system describing light-matter interaction that is a frequently used model Hamiltonian for observing quantum phase transition. The most elementary Rabi-Hubbard system consists of an array of two cavities, each containing at most $N=2$ photons interacting with a two-level system atom with strength $g$. Photon tunneling between two cavities is possible and is controlled by a hopping parameter $J$. This RH-model Hamiltonian is expressed as\footnote{For neatness, we set $\hbar=1$ so it is not explicit in the expression.}
\begin{align}
    H_{RH} =& \sum_{i=1}^2 \left[ \omega_c \hat{b}_i^\dagger \hat{b}_i + \omega_a \hat{\sigma}_{+,i}\hat{\sigma}_{-,i} + g \hat{\sigma}_{x,i}\left( \hat{b}_i + \hat{b}_i^\dagger \right)\right] \nonumber \\
      & - J\sum_{i,j} \left(\hat{b}_i \hat{b}_j^\dagger + \hat{b}_i^\dagger  \hat{b}_j \right),
\end{align}
where $\omega_c$ and $\omega_a$ are the cavity and atomic transition frequencies. The light field mode is expressed in terms of the bosonic creation and annihilation operators $\hat{b}^\dagger$, $\hat{b}$, while the two-level atom is treated as a spin-half system, where $\hat{\sigma}_\pm = \frac{1}{2}\left( \hat{\sigma}_x \pm i \hat{\sigma}_y \right)$ is the raising or lowering operator of the atom\footnote{$\hat{\sigma}_+ = |e\rangle\langle g| = |0\rangle\langle1|$ and $\hat{\sigma}_- = |g\rangle\langle e| = |1\rangle\langle 0| $}, and $\hat{\sigma}_z = |e\rangle\langle e| - |g\rangle\langle g| $ is the atomic inversion operator.  Binary encoding \cite{Sawaya_2020} is employed to map the field operators into qubit Pauli operators, while the atomic mode operators are already expressed in qubit Paulis. The corresponding transformed qubit Hamiltonian has $49$ Pauli terms. Further, we choose a Mott-insulator initial state, $|\psi_0 \rangle = (\cos \theta |N=0, e\rangle - \sin \theta |N=1, g\rangle)^{\otimes 2} $, which takes a qubit form 
\begin{align}
    |\psi_0\rangle = & \cos^2 \theta |011011 \rangle + \sin^2 \theta | 000000 \rangle \nonumber \\
     &-\cos \theta \sin \theta \left( |011000 \rangle + |000011 \rangle\right),
\end{align}
with $\tan (2\theta) = 2g/\delta $, $\delta = \omega_a -\omega_c$. By exploiting symmetry properties of the Hamiltonian, the number of system qubits can be tapered down to five \cite{Bravyi:2017eoo}, with a tapered initial state
\begin{align}
    |\psi_0\rangle = & \cos^2 \theta |01011 \rangle + \sin^2 \theta | 00000 \rangle \nonumber \\
     &-\cos \theta \sin \theta \left( |01000 \rangle + |00011 \rangle\right).
\end{align}

Our target observable is $|\langle \psi_0| e^{-iH_{RH}t} | \psi_0 \rangle|^2$ for a time interval $Jt \in \left(0,1\right]$. Figure~\ref{fig:sq_overlap} illustrates the circuit used to compute the squared overlap. Truncating terms with coefficients $|\alpha_\ell| < 10^{-8}$ ($K=8$) in the Taylor series expanded $e^{-iH_{RH}t}$, we find that the number of terms $L$ in the resulting propagator varies between $400$ and $500$ within the time interval, with the approximated propagator having precision $|1 - \sqrt{\langle \psi_0 | \Upsilon^\dagger \Upsilon| \psi_0\rangle}| \leq 10^{-6}$ (See Fig.~\ref{fig:RH_lcu_precision}).  Then, by applying the strategy in Sec.~\ref{sec:sq_overlap}, the relevant terms $\Upsilon_\parallel$ can be expressed using at most $11$ Pauli strings, meaning $4$ ancilla qubits are required in the LCU implementation. Appendix~\ref{appendix:grouping} illustrates how we do this explicitly. This results in a $98\%$ reduction in circuit depth and 2Q-gates as illustrated in Fig.~\ref{fig:RH_lcu_resources_opt1}. Moreover, we find that the multiplexed $\select$ compilation uses a significantly lower number of 2Q-gates (about $\frac{1}{6}$ of the gates and about $\frac{1}{12}$ of the depth) than unary iteration for this number of terms. Table~\ref{tab:Exp_setup} outlines the parameters used in our experiments.
 
\begin{table}[!htb]
    \centering
\begin{tabular}{l|r}
\hline
  Parameter & Value \\ 
\hline
$J$ & $0.1 \omega_c$ \\
$g$ & $0.1 \omega_c$ \\
$\Delta$ & $0.1 \omega_c$ \\
$\tau$ & $0.05/\omega_c$ \\
Prepare qubits & 4 \\
System qubits & 5 \\
Circuit depth & $(150 - 163)$ \\
2Q gates & $(175 - 180)$ \\
Shots (emulator H1E) & 20480 \\
Shots (hardware H1-1) & 2048 \\
\hline
\end{tabular}
\caption{LCU experimental setup parameters for a two-cavity Rabi-Hubbard system.}
\label{tab:Exp_setup}
\end{table}


\begin{figure}[p]
     \centering
         \includegraphics[width=0.52\textwidth]{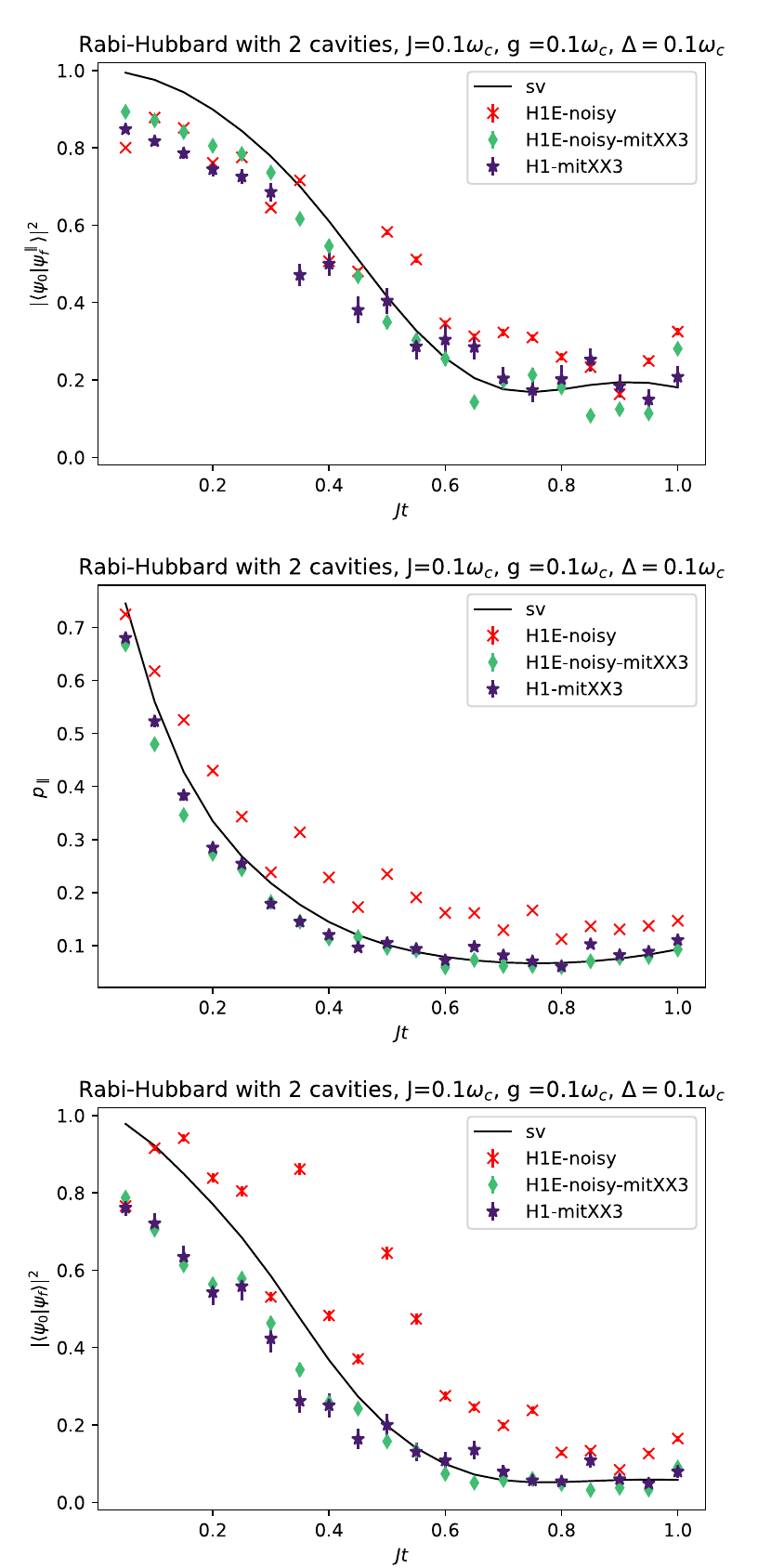}
     \caption{Squared overlap calculations from reduced LCU. Squared overlap (bottom) calculated using Eqs.~(\ref{eq:sq_overlap_from_reduced}) and~(\ref{eq:p_parallel}) and with data from reduced squared overlap (top) and success probability of the reduced LCU (middle).
     The red x data points are results from the noisy emulator (20480 shots/circuit), green diamond from memory error mitigated circuits on noisy emulator (20480 shots/circuit), and violet star from memory error mitigated circuits on the H1-1 quantum computer (2048 shots/circuit). Hardware results deviate from the theoretical prediction in black curve (sv) for early times $Jt$.  }
        \label{fig:sq_overlap_rh}
\end{figure}

\begin{figure}[!htbp]
         \centering
         \includegraphics[width=0.5\textwidth]{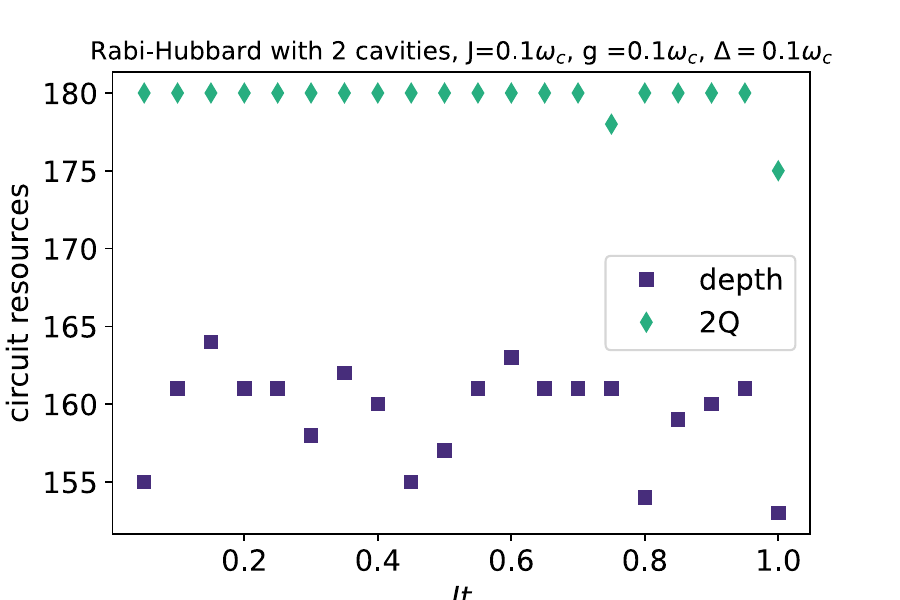}
         \caption{LCU circuit resources for the Rabi-Hubbard propagator.}
         \label{fig:RH_lcu_resources}
\end{figure}

The plots in Fig.~\ref{fig:sq_overlap_rh} show the $|\langle \psi_0| \psi_f^\parallel \rangle|^2$ (top), $p_\parallel$ (middle) as functions of time from our computational experiments, and the main target observable $|\langle \psi_0| \psi_f \rangle|^2$ (bottom) derived from the preceding data using Eq.(~\ref{eq:sq_overlap_from_reduced}). The black curves are the state vector results. $20480$ shots are executed for the initial unmitigated emulator experiment, labelled `H1E-noisy' in the legend. There are two main contributions to the noise, a dominant memory error effecting idling qubits, and a depolarising error on two qubit gates. We combat memory error by inserting pairs of X gates (single X at every third instance of idling qubits), similar to the technique in Ref.~\cite{PhysRevResearch.6.013221}, as shown in the data points labeled `H1E-noisy-mitXX3’. Finally, we conduct the same experiment with memory error mitigation and with $2048$ shots on our H1-1 machine. The results are represented by `H1-mitXX3' labeled data in the figures. Circuit resources utilized in the experiments are shown in Fig.~\ref{fig:RH_lcu_resources}. We find almost constant circuit depth and 2Q counts for the simulation time interval. The optimized empirical 2Q count is consistent with the derived $2^k(2n+1)-n-2$ \CX gates asymptotic scaling, given $k=4$ prepare qubits and $n=5$ system qubits. In Appendix~\ref{appendix:jc}, we compare some quantitative results of LCU (with and without OAA) and naïve Trotter approach using a simpler Hamiltonian. We observe linear scaling of circuit resources with respect to time from Trotter as expected and almost constant from LCU. For larger Rabi-Hubbard systems, we show the estimated circuit resources in Fig.~\ref{fig:LargeRH} in Appendix~\ref{appendix:more_figs}.

The source code used in producing the results in this section can be found in the Repository~\cite{github}.

\section{Discussion}\label{sec:discussion}

Intractable quantum many-body systems such as those found in quantum chemistry are usually the useful target model Hamiltonian to perform dynamics with on a quantum computer. However, since current state-of-the-art quantum devices are not yet ready for them, a Hubbard Hamiltonian, with many lattice sites, is a prototypical alternative. In this work, we deviate slightly from such a Hamiltonian and use the Rabi-Hubbard with two cavities to test the LCU algorithm on a near term device. This is not the first attempt to demonstrate cavity quantum electrodynamics on a quantum computer \cite{Tudorovskaya_2024, Shaw2020quantumalgorithms,yu2022simulating}, but, to our knowledge, is the first demonstration of Hamiltonian dynamics simulations using purely LCU on an ion-trap quantum device. While it is possible to simulate the chosen Rabi-Hubbard system on a classical computer \cite{Figueroa2018, Norambuena_2020}, a distinct feature of this Hamiltonian is the combination of spin and bosonic systems, which is rarely used as a test case. 

Our experiments exhibit interesting results, with some flaws. A notable observation is the constant circuit resource requirements with respect to simulation time. The caveats here are the initial state is known and the system considered is not large such that fast saturation to the maximum size (upper bounded by $4^n$ Pauli terms, where $n$ is the system size) of $\Upsilon_\parallel$ is reached in short simulation time. Moreover, this is achieved by collapsing multiple LCU's in the Taylor series expansion into a single LCU at the expense of doing classical expansion of moments of the Hamiltonian and of $(\exp^{-iH\tau})^m$. For a much larger system, increase of circuit depth with time is possible. However, a linear growth in circuit depth, which is expected from a Trotter approach, is highly unlikely. One could argue that for the chosen cutoff $K$ or given available 2Q gates, the resulting algorithmic error from LCU might still be significant. Indeed, algorithmic error is an important aspect that our demonstration did not quantify. Some interesting physical phenomena that may be observed in a cavity quantum electrodynamics system, such as quantum phase transition, are qualitative in nature depending on the $g/\Delta$ parameter. This is in contrast to most target observables in chemistry, which require high precision. The memory error mitigation may have benefited from a systematic approach, though we found that inserting X gates at every third instance of qubit idling was a good setting for the majority of the $Jt$ values. However, this was not always the case, as was observed by comparing the noisy emulator results for various placements of XX gates against the ground truth solution. 

As illustrated, the key to a realizable implementation of the LCU algorithm on a near term device is twofold. For one, our classical preprocessing procedure to narrow down the number of LCU terms proves to be effective. We note that similar reduction method will work as long as the state to be evolved possesses some symmetry which is the case for chemical or physical systems. It is remarkable that Eq.~(\ref{eq:sq_overlap_from_reduced}) enables the computation of what supposed to be a large LCU from just a small fraction of this LCU because the `renormalization' factor $p_\parallel |\alpha_\parallel|_1^2$ can be obtained practically for free once the LCU for $\Upsilon_\parallel$ is calculated. The only downside is the propagation of error from $p_\parallel$. The target observable $\Upsilon \approx \exp(-iHt)$ in our experiments also works favorably. Had we chosen another $\Upsilon$ which is highly non-unitary, then the denominator in Eq.~(\ref{eq:sq_overlap_from_reduced}) had to be thought of carefully. Secondly, the quantum multiplexor synthesis has proven effective for optimizing the LCU circuit in our use case. Asymptotically, multiplexor will not be more superior than the unary iteration compilation in the fault-tolerant regime when the qubit requirement (both for the system and prepare register ancilla) becomes $\gg\mathcal{O}(10^2)$. See Appendix~\ref{appedix:scaling}. However, the multiplexor compilation strategy suits well for a near term device in the early fault tolerant era. Performing LCU for a more complex use case such as the electronic structure Hamiltonian will be an anticipated scientific  endeavor when a high capacity and high fidelity quantum device becomes available.

\section*{Acknowledgements} \label{sec:acknowledgements}

We thank Iakov Polyak and Pablo Andrés-Martínez for their work on \texttt{pytket-cutensornet}. We also thank Pablo, Eric Brunner and Marcello Benedetti for discussions regarding the circuit construction, and Kentaro Yamamoto for memory error mitigation. We thank David Zsolt Manrique and Yuta Kikuchi for providing feedback on the paper. In addition to \texttt{pytket}, \texttt{inquanto}\cite{inquanto} was also utilized in the computational experiments. We thank Robert Anderson for providing efficient tools for calculating high powers of qubit operators. 

\bibliographystyle{apsrev4-1.bst}
\bibliography{references.bib}

\begin{thebibliography}{87}%
\makeatletter
\providecommand \@ifxundefined [1]{%
 \@ifx{#1\undefined}
}%
\providecommand \@ifnum [1]{%
 \ifnum #1\expandafter \@firstoftwo
 \else \expandafter \@secondoftwo
 \fi
}%
\providecommand \@ifx [1]{%
 \ifx #1\expandafter \@firstoftwo
 \else \expandafter \@secondoftwo
 \fi
}%
\providecommand \natexlab [1]{#1}%
\providecommand \enquote  [1]{``#1''}%
\providecommand \bibnamefont  [1]{#1}%
\providecommand \bibfnamefont [1]{#1}%
\providecommand \citenamefont [1]{#1}%
\providecommand \href@noop [0]{\@secondoftwo}%
\providecommand \href [0]{\begingroup \@sanitize@url \@href}%
\providecommand \@href[1]{\@@startlink{#1}\@@href}%
\providecommand \@@href[1]{\endgroup#1\@@endlink}%
\providecommand \@sanitize@url [0]{\catcode `\\12\catcode `\$12\catcode `\&12\catcode `\#12\catcode `\^12\catcode `\_12\catcode `\%12\relax}%
\providecommand \@@startlink[1]{}%
\providecommand \@@endlink[0]{}%
\providecommand \url  [0]{\begingroup\@sanitize@url \@url }%
\providecommand \@url [1]{\endgroup\@href {#1}{\urlprefix }}%
\providecommand \urlprefix  [0]{URL }%
\providecommand \Eprint [0]{\href }%
\providecommand \doibase [0]{http://dx.doi.org/}%
\providecommand \selectlanguage [0]{\@gobble}%
\providecommand \bibinfo  [0]{\@secondoftwo}%
\providecommand \bibfield  [0]{\@secondoftwo}%
\providecommand \translation [1]{[#1]}%
\providecommand \BibitemOpen [0]{}%
\providecommand \bibitemStop [0]{}%
\providecommand \bibitemNoStop [0]{.\EOS\space}%
\providecommand \EOS [0]{\spacefactor3000\relax}%
\providecommand \BibitemShut  [1]{\csname bibitem#1\endcsname}%
\let\auto@bib@innerbib\@empty
\bibitem [{\citenamefont {Kassal}\ \emph {et~al.}(2011)\citenamefont {Kassal}, \citenamefont {Whitfield}, \citenamefont {Perdomo-Ortiz}, \citenamefont {Yung},\ and\ \citenamefont {Aspuru-Guzik}}]{Kassal2010}%
  \BibitemOpen
  \bibfield  {author} {\bibinfo {author} {\bibfnamefont {I.}~\bibnamefont {Kassal}}, \bibinfo {author} {\bibfnamefont {J.~D.}\ \bibnamefont {Whitfield}}, \bibinfo {author} {\bibfnamefont {A.}~\bibnamefont {Perdomo-Ortiz}}, \bibinfo {author} {\bibfnamefont {M.-H.}\ \bibnamefont {Yung}}, \ and\ \bibinfo {author} {\bibfnamefont {A.}~\bibnamefont {Aspuru-Guzik}},\ }\href {\doibase https://doi.org/10.1146/annurev-physchem-032210-103512} {\bibfield  {journal} {\bibinfo  {journal} {Annual Review of Physical Chemistry}\ }\textbf {\bibinfo {volume} {62}},\ \bibinfo {pages} {185} (\bibinfo {year} {2011})}\BibitemShut {NoStop}%
\bibitem [{\citenamefont {Wecker}\ \emph {et~al.}(2013)\citenamefont {Wecker}, \citenamefont {Bauer}, \citenamefont {Clark}, \citenamefont {Hastings},\ and\ \citenamefont {Troyer}}]{Wecker2013}%
  \BibitemOpen
  \bibfield  {author} {\bibinfo {author} {\bibfnamefont {D.}~\bibnamefont {Wecker}}, \bibinfo {author} {\bibfnamefont {B.}~\bibnamefont {Bauer}}, \bibinfo {author} {\bibfnamefont {B.~K.}\ \bibnamefont {Clark}}, \bibinfo {author} {\bibfnamefont {M.~B.}\ \bibnamefont {Hastings}}, \ and\ \bibinfo {author} {\bibfnamefont {M.}~\bibnamefont {Troyer}},\ }\href {\doibase 10.1103/PhysRevA.90.022305} {\bibfield  {journal} {\bibinfo  {journal} {Phys. Rev. A}\ } (\bibinfo {year} {2013}),\ 10.1103/PhysRevA.90.022305}\BibitemShut {NoStop}%
\bibitem [{\citenamefont {Reiher}\ \emph {et~al.}(2016)\citenamefont {Reiher}, \citenamefont {Wiebe}, \citenamefont {Svore}, \citenamefont {Wecker},\ and\ \citenamefont {Troyer}}]{Reiher2016}%
  \BibitemOpen
  \bibfield  {author} {\bibinfo {author} {\bibfnamefont {M.}~\bibnamefont {Reiher}}, \bibinfo {author} {\bibfnamefont {N.}~\bibnamefont {Wiebe}}, \bibinfo {author} {\bibfnamefont {K.~M.}\ \bibnamefont {Svore}}, \bibinfo {author} {\bibfnamefont {D.}~\bibnamefont {Wecker}}, \ and\ \bibinfo {author} {\bibfnamefont {M.}~\bibnamefont {Troyer}},\ }\href {\doibase 10.1073/pnas.1619152114} {\bibfield  {journal} {\bibinfo  {journal} {Proc. Natl. Acad. Sci. U.S.A.}\ } (\bibinfo {year} {2016}),\ 10.1073/pnas.1619152114}\BibitemShut {NoStop}%
\bibitem [{\citenamefont {Kaufman}\ \emph {et~al.}(2016)\citenamefont {Kaufman}, \citenamefont {Tai}, \citenamefont {Lukin}, \citenamefont {Rispoli}, \citenamefont {Schittko}, \citenamefont {Preiss},\ and\ \citenamefont {Greiner}}]{Kaufman2016}%
  \BibitemOpen
  \bibfield  {author} {\bibinfo {author} {\bibfnamefont {A.~M.}\ \bibnamefont {Kaufman}}, \bibinfo {author} {\bibfnamefont {M.~E.}\ \bibnamefont {Tai}}, \bibinfo {author} {\bibfnamefont {A.}~\bibnamefont {Lukin}}, \bibinfo {author} {\bibfnamefont {M.}~\bibnamefont {Rispoli}}, \bibinfo {author} {\bibfnamefont {R.}~\bibnamefont {Schittko}}, \bibinfo {author} {\bibfnamefont {P.~M.}\ \bibnamefont {Preiss}}, \ and\ \bibinfo {author} {\bibfnamefont {M.}~\bibnamefont {Greiner}},\ }\href {\doibase 10.1126/science.aaf6725} {\bibfield  {journal} {\bibinfo  {journal} {Science}\ } (\bibinfo {year} {2016}),\ 10.1126/science.aaf6725}\BibitemShut {NoStop}%
\bibitem [{\citenamefont {Figueroa}\ \emph {et~al.}(2018)\citenamefont {Figueroa}, \citenamefont {Rogan}, \citenamefont {Valdivia}, \citenamefont {Kiwi}, \citenamefont {Romero},\ and\ \citenamefont {Torres}}]{Figueroa2018}%
  \BibitemOpen
  \bibfield  {author} {\bibinfo {author} {\bibfnamefont {J.}~\bibnamefont {Figueroa}}, \bibinfo {author} {\bibfnamefont {J.}~\bibnamefont {Rogan}}, \bibinfo {author} {\bibfnamefont {J.~A.}\ \bibnamefont {Valdivia}}, \bibinfo {author} {\bibfnamefont {M.}~\bibnamefont {Kiwi}}, \bibinfo {author} {\bibfnamefont {G.}~\bibnamefont {Romero}}, \ and\ \bibinfo {author} {\bibfnamefont {F.}~\bibnamefont {Torres}},\ }\href {\doibase 10.1038/s41598-018-30789-9} {\bibfield  {journal} {\bibinfo  {journal} {Scientific Reports}\ }\textbf {\bibinfo {volume} {8}} (\bibinfo {year} {2018}),\ 10.1038/s41598-018-30789-9}\BibitemShut {NoStop}%
\bibitem [{\citenamefont {Defenu}\ \emph {et~al.}(2024)\citenamefont {Defenu}, \citenamefont {Lerose},\ and\ \citenamefont {Pappalardi}}]{Defenu2024}%
  \BibitemOpen
  \bibfield  {author} {\bibinfo {author} {\bibfnamefont {N.}~\bibnamefont {Defenu}}, \bibinfo {author} {\bibfnamefont {A.}~\bibnamefont {Lerose}}, \ and\ \bibinfo {author} {\bibfnamefont {S.}~\bibnamefont {Pappalardi}},\ }\href {\doibase 10.1016/j.physrep.2024.04.005} {\bibfield  {journal} {\bibinfo  {journal} {Physics Reports}\ }\textbf {\bibinfo {volume} {1074}},\ \bibinfo {pages} {1} (\bibinfo {year} {2024})}\BibitemShut {NoStop}%
\bibitem [{\citenamefont {Childs}\ \emph {et~al.}(2018{\natexlab{a}})\citenamefont {Childs}, \citenamefont {Maslov}, \citenamefont {Nam}, \citenamefont {Ross},\ and\ \citenamefont {Su}}]{Childs2018}%
  \BibitemOpen
  \bibfield  {author} {\bibinfo {author} {\bibfnamefont {A.~M.}\ \bibnamefont {Childs}}, \bibinfo {author} {\bibfnamefont {D.}~\bibnamefont {Maslov}}, \bibinfo {author} {\bibfnamefont {Y.}~\bibnamefont {Nam}}, \bibinfo {author} {\bibfnamefont {N.~J.}\ \bibnamefont {Ross}}, \ and\ \bibinfo {author} {\bibfnamefont {Y.}~\bibnamefont {Su}},\ }\href {\doibase 10.1073/pnas.1801723115} {\bibfield  {journal} {\bibinfo  {journal} {Proceedings of the National Academy of Sciences of the United States of America}\ }\textbf {\bibinfo {volume} {115}},\ \bibinfo {pages} {9456} (\bibinfo {year} {2018}{\natexlab{a}})}\BibitemShut {NoStop}%
\bibitem [{\citenamefont {Sharma}\ and\ \citenamefont {Tran}(2024)}]{sharma2024hamiltoniansimulationinteractionpicture}%
  \BibitemOpen
  \bibfield  {author} {\bibinfo {author} {\bibfnamefont {K.}~\bibnamefont {Sharma}}\ and\ \bibinfo {author} {\bibfnamefont {M.~C.}\ \bibnamefont {Tran}},\ }\href {https://arxiv.org/abs/2404.02966} {\enquote {\bibinfo {title} {Hamiltonian simulation in the interaction picture using the magnus expansion},}\ } (\bibinfo {year} {2024}),\ \Eprint {http://arxiv.org/abs/2404.02966} {arXiv:2404.02966 [quant-ph]} \BibitemShut {NoStop}%
\bibitem [{\citenamefont {Low}\ and\ \citenamefont {Wiebe}(2019)}]{low2019hamiltoniansimulationinteractionpicture}%
  \BibitemOpen
  \bibfield  {author} {\bibinfo {author} {\bibfnamefont {G.~H.}\ \bibnamefont {Low}}\ and\ \bibinfo {author} {\bibfnamefont {N.}~\bibnamefont {Wiebe}},\ }\href {https://arxiv.org/abs/1805.00675} {\enquote {\bibinfo {title} {Hamiltonian simulation in the interaction picture},}\ } (\bibinfo {year} {2019}),\ \Eprint {http://arxiv.org/abs/1805.00675} {arXiv:1805.00675 [quant-ph]} \BibitemShut {NoStop}%
\bibitem [{\citenamefont {Chen}\ \emph {et~al.}(2021{\natexlab{a}})\citenamefont {Chen}, \citenamefont {Kalev},\ and\ \citenamefont {Hen}}]{PRXQuantum.2.030342}%
  \BibitemOpen
  \bibfield  {author} {\bibinfo {author} {\bibfnamefont {Y.-H.}\ \bibnamefont {Chen}}, \bibinfo {author} {\bibfnamefont {A.}~\bibnamefont {Kalev}}, \ and\ \bibinfo {author} {\bibfnamefont {I.}~\bibnamefont {Hen}},\ }\href {\doibase 10.1103/PRXQuantum.2.030342} {\bibfield  {journal} {\bibinfo  {journal} {PRX Quantum}\ }\textbf {\bibinfo {volume} {2}},\ \bibinfo {pages} {030342} (\bibinfo {year} {2021}{\natexlab{a}})}\BibitemShut {NoStop}%
\bibitem [{\citenamefont {Zhao}\ \emph {et~al.}(2024)\citenamefont {Zhao}, \citenamefont {Bukov}, \citenamefont {Heyl},\ and\ \citenamefont {Moessner}}]{Zhao_2024}%
  \BibitemOpen
  \bibfield  {author} {\bibinfo {author} {\bibfnamefont {H.}~\bibnamefont {Zhao}}, \bibinfo {author} {\bibfnamefont {M.}~\bibnamefont {Bukov}}, \bibinfo {author} {\bibfnamefont {M.}~\bibnamefont {Heyl}}, \ and\ \bibinfo {author} {\bibfnamefont {R.}~\bibnamefont {Moessner}},\ }\href {\doibase 10.1103/physrevlett.133.010603} {\bibfield  {journal} {\bibinfo  {journal} {Physical Review Letters}\ }\textbf {\bibinfo {volume} {133}} (\bibinfo {year} {2024}),\ 10.1103/physrevlett.133.010603}\BibitemShut {NoStop}%
\bibitem [{\citenamefont {Ikeda}\ \emph {et~al.}(2023)\citenamefont {Ikeda}, \citenamefont {Abrar}, \citenamefont {Chuang},\ and\ \citenamefont {Sugiura}}]{Ikeda_2023}%
  \BibitemOpen
  \bibfield  {author} {\bibinfo {author} {\bibfnamefont {T.~N.}\ \bibnamefont {Ikeda}}, \bibinfo {author} {\bibfnamefont {A.}~\bibnamefont {Abrar}}, \bibinfo {author} {\bibfnamefont {I.~L.}\ \bibnamefont {Chuang}}, \ and\ \bibinfo {author} {\bibfnamefont {S.}~\bibnamefont {Sugiura}},\ }\href {\doibase 10.22331/q-2023-11-06-1168} {\bibfield  {journal} {\bibinfo  {journal} {Quantum}\ }\textbf {\bibinfo {volume} {7}},\ \bibinfo {pages} {1168} (\bibinfo {year} {2023})}\BibitemShut {NoStop}%
\bibitem [{\citenamefont {Watkins}\ \emph {et~al.}(2024)\citenamefont {Watkins}, \citenamefont {Wiebe}, \citenamefont {Roggero},\ and\ \citenamefont {Lee}}]{PRXQuantum.5.040316}%
  \BibitemOpen
  \bibfield  {author} {\bibinfo {author} {\bibfnamefont {J.}~\bibnamefont {Watkins}}, \bibinfo {author} {\bibfnamefont {N.}~\bibnamefont {Wiebe}}, \bibinfo {author} {\bibfnamefont {A.}~\bibnamefont {Roggero}}, \ and\ \bibinfo {author} {\bibfnamefont {D.}~\bibnamefont {Lee}},\ }\href {\doibase 10.1103/PRXQuantum.5.040316} {\bibfield  {journal} {\bibinfo  {journal} {PRX Quantum}\ }\textbf {\bibinfo {volume} {5}},\ \bibinfo {pages} {040316} (\bibinfo {year} {2024})}\BibitemShut {NoStop}%
\bibitem [{\citenamefont {Cao}\ \emph {et~al.}(2023)\citenamefont {Cao}, \citenamefont {Jin},\ and\ \citenamefont {Liu}}]{cao2023quantumsimulationtimedependenthamiltonians}%
  \BibitemOpen
  \bibfield  {author} {\bibinfo {author} {\bibfnamefont {Y.}~\bibnamefont {Cao}}, \bibinfo {author} {\bibfnamefont {S.}~\bibnamefont {Jin}}, \ and\ \bibinfo {author} {\bibfnamefont {N.}~\bibnamefont {Liu}},\ }\href {https://arxiv.org/abs/2312.02817} {\enquote {\bibinfo {title} {Quantum simulation for time-dependent hamiltonians -- with applications to non-autonomous ordinary and partial differential equations},}\ } (\bibinfo {year} {2023}),\ \Eprint {http://arxiv.org/abs/2312.02817} {arXiv:2312.02817 [quant-ph]} \BibitemShut {NoStop}%
\bibitem [{\citenamefont {An}\ \emph {et~al.}(2022)\citenamefont {An}, \citenamefont {Fang},\ and\ \citenamefont {Lin}}]{An_2022}%
  \BibitemOpen
  \bibfield  {author} {\bibinfo {author} {\bibfnamefont {D.}~\bibnamefont {An}}, \bibinfo {author} {\bibfnamefont {D.}~\bibnamefont {Fang}}, \ and\ \bibinfo {author} {\bibfnamefont {L.}~\bibnamefont {Lin}},\ }\href {\doibase 10.22331/q-2022-04-15-690} {\bibfield  {journal} {\bibinfo  {journal} {Quantum}\ }\textbf {\bibinfo {volume} {6}},\ \bibinfo {pages} {690} (\bibinfo {year} {2022})}\BibitemShut {NoStop}%
\bibitem [{\citenamefont {Berry}\ \emph {et~al.}(2020)\citenamefont {Berry}, \citenamefont {Childs}, \citenamefont {Su}, \citenamefont {Wang},\ and\ \citenamefont {Wiebe}}]{Berry_2020_tdH}%
  \BibitemOpen
  \bibfield  {author} {\bibinfo {author} {\bibfnamefont {D.~W.}\ \bibnamefont {Berry}}, \bibinfo {author} {\bibfnamefont {A.~M.}\ \bibnamefont {Childs}}, \bibinfo {author} {\bibfnamefont {Y.}~\bibnamefont {Su}}, \bibinfo {author} {\bibfnamefont {X.}~\bibnamefont {Wang}}, \ and\ \bibinfo {author} {\bibfnamefont {N.}~\bibnamefont {Wiebe}},\ }\href {\doibase 10.22331/q-2020-04-20-254} {\bibfield  {journal} {\bibinfo  {journal} {Quantum}\ }\textbf {\bibinfo {volume} {4}},\ \bibinfo {pages} {254} (\bibinfo {year} {2020})}\BibitemShut {NoStop}%
\bibitem [{\citenamefont {Trotter}(1959)}]{Trotter1959}%
  \BibitemOpen
  \bibfield  {author} {\bibinfo {author} {\bibfnamefont {H.~F.}\ \bibnamefont {Trotter}},\ }\href {https://doi.org/10.2307/2033649} {\bibfield  {journal} {\bibinfo  {journal} {Proceedings of the American Mathematical Society}\ }\textbf {\bibinfo {volume} {10}},\ \bibinfo {pages} {545} (\bibinfo {year} {1959})}\BibitemShut {NoStop}%
\bibitem [{\citenamefont {Suzuki}(1976)}]{Suzuki1976}%
  \BibitemOpen
  \bibfield  {author} {\bibinfo {author} {\bibfnamefont {M.}~\bibnamefont {Suzuki}},\ }\href {https://academic.oup.com/ptp/article/56/5/1454/1860476} {\bibfield  {journal} {\bibinfo  {journal} {Progress of Theoretical.Physics}\ }\textbf {\bibinfo {volume} {56}} (\bibinfo {year} {1976})}\BibitemShut {NoStop}%
\bibitem [{\citenamefont {Suzuki}(1991)}]{Suzuki1990}%
  \BibitemOpen
  \bibfield  {author} {\bibinfo {author} {\bibfnamefont {M.}~\bibnamefont {Suzuki}},\ }\href {\doibase 10.1063/1.529425} {\bibfield  {journal} {\bibinfo  {journal} {Journal of Mathematical Physics}\ }\textbf {\bibinfo {volume} {32}},\ \bibinfo {pages} {400} (\bibinfo {year} {1991})}\BibitemShut {NoStop}%
\bibitem [{\citenamefont {Granet}\ and\ \citenamefont {Dreyer}(2024{\natexlab{a}})}]{Granet2024}%
  \BibitemOpen
  \bibfield  {author} {\bibinfo {author} {\bibfnamefont {E.}~\bibnamefont {Granet}}\ and\ \bibinfo {author} {\bibfnamefont {H.}~\bibnamefont {Dreyer}},\ }\href {\doibase 10.1038/s41534-024-00877-y} {\bibfield  {journal} {\bibinfo  {journal} {npj Quantum Information}\ }\textbf {\bibinfo {volume} {10}} (\bibinfo {year} {2024}{\natexlab{a}}),\ 10.1038/s41534-024-00877-y}\BibitemShut {NoStop}%
\bibitem [{\citenamefont {Childs}\ and\ \citenamefont {Wiebe}(2012)}]{ChildsLCU}%
  \BibitemOpen
  \bibfield  {author} {\bibinfo {author} {\bibfnamefont {A.~M.}\ \bibnamefont {Childs}}\ and\ \bibinfo {author} {\bibfnamefont {N.}~\bibnamefont {Wiebe}},\ }\href@noop {} {\bibfield  {journal} {\bibinfo  {journal} {Quantum Info. Comput.}\ }\textbf {\bibinfo {volume} {12}},\ \bibinfo {pages} {901–924} (\bibinfo {year} {2012})}\BibitemShut {NoStop}%
\bibitem [{\citenamefont {Gily\'{e}n}\ \emph {et~al.}(2019)\citenamefont {Gily\'{e}n}, \citenamefont {Su}, \citenamefont {Low},\ and\ \citenamefont {Wiebe}}]{qsvtguilyen}%
  \BibitemOpen
  \bibfield  {author} {\bibinfo {author} {\bibfnamefont {A.}~\bibnamefont {Gily\'{e}n}}, \bibinfo {author} {\bibfnamefont {Y.}~\bibnamefont {Su}}, \bibinfo {author} {\bibfnamefont {G.~H.}\ \bibnamefont {Low}}, \ and\ \bibinfo {author} {\bibfnamefont {N.}~\bibnamefont {Wiebe}},\ }in\ \href {\doibase 10.1145/3313276.3316366} {\emph {\bibinfo {booktitle} {Proceedings of the 51st Annual ACM SIGACT Symposium on Theory of Computing}}},\ \bibinfo {series and number} {STOC 2019}\ (\bibinfo  {publisher} {Association for Computing Machinery},\ \bibinfo {address} {New York, NY, USA},\ \bibinfo {year} {2019})\ p.\ \bibinfo {pages} {193–204}\BibitemShut {NoStop}%
\bibitem [{\citenamefont {Martyn}\ \emph {et~al.}(2021)\citenamefont {Martyn}, \citenamefont {Rossi}, \citenamefont {Tan},\ and\ \citenamefont {Chuang}}]{grandunification}%
  \BibitemOpen
  \bibfield  {author} {\bibinfo {author} {\bibfnamefont {J.~M.}\ \bibnamefont {Martyn}}, \bibinfo {author} {\bibfnamefont {Z.~M.}\ \bibnamefont {Rossi}}, \bibinfo {author} {\bibfnamefont {A.~K.}\ \bibnamefont {Tan}}, \ and\ \bibinfo {author} {\bibfnamefont {I.~L.}\ \bibnamefont {Chuang}},\ }\href {\doibase 10.1103/PRXQuantum.2.040203} {\bibfield  {journal} {\bibinfo  {journal} {PRX Quantum}\ }\textbf {\bibinfo {volume} {2}},\ \bibinfo {pages} {040203} (\bibinfo {year} {2021})}\BibitemShut {NoStop}%
\bibitem [{\citenamefont {Low}\ \emph {et~al.}(2019)\citenamefont {Low}, \citenamefont {Kliuchnikov},\ and\ \citenamefont {Wiebe}}]{Low2019}%
  \BibitemOpen
  \bibfield  {author} {\bibinfo {author} {\bibfnamefont {G.~H.}\ \bibnamefont {Low}}, \bibinfo {author} {\bibfnamefont {V.}~\bibnamefont {Kliuchnikov}}, \ and\ \bibinfo {author} {\bibfnamefont {N.}~\bibnamefont {Wiebe}},\ }\href {http://arxiv.org/abs/1907.11679} {\  (\bibinfo {year} {2019})}\BibitemShut {NoStop}%
\bibitem [{\citenamefont {Zeng}\ \emph {et~al.}(2022)\citenamefont {Zeng}, \citenamefont {Sun}, \citenamefont {Jiang},\ and\ \citenamefont {Zhao}}]{Zeng2022}%
  \BibitemOpen
  \bibfield  {author} {\bibinfo {author} {\bibfnamefont {P.}~\bibnamefont {Zeng}}, \bibinfo {author} {\bibfnamefont {J.}~\bibnamefont {Sun}}, \bibinfo {author} {\bibfnamefont {L.}~\bibnamefont {Jiang}}, \ and\ \bibinfo {author} {\bibfnamefont {Q.}~\bibnamefont {Zhao}},\ }\href {http://arxiv.org/abs/2212.04566} {\  (\bibinfo {year} {2022})}\BibitemShut {NoStop}%
\bibitem [{\citenamefont {Low}\ \emph {et~al.}(2023)\citenamefont {Low}, \citenamefont {Su}, \citenamefont {Tong},\ and\ \citenamefont {Tran}}]{YuanSu}%
  \BibitemOpen
  \bibfield  {author} {\bibinfo {author} {\bibfnamefont {G.~H.}\ \bibnamefont {Low}}, \bibinfo {author} {\bibfnamefont {Y.}~\bibnamefont {Su}}, \bibinfo {author} {\bibfnamefont {Y.}~\bibnamefont {Tong}}, \ and\ \bibinfo {author} {\bibfnamefont {M.~C.}\ \bibnamefont {Tran}},\ }\href {\doibase 10.1103/PRXQuantum.4.020323} {\bibfield  {journal} {\bibinfo  {journal} {PRX Quantum}\ }\textbf {\bibinfo {volume} {4}},\ \bibinfo {pages} {020323} (\bibinfo {year} {2023})}\BibitemShut {NoStop}%
\bibitem [{\citenamefont {Zhuk}\ \emph {et~al.}(2023)\citenamefont {Zhuk}, \citenamefont {Robertson},\ and\ \citenamefont {Bravyi}}]{Zhuk2023}%
  \BibitemOpen
  \bibfield  {author} {\bibinfo {author} {\bibfnamefont {S.}~\bibnamefont {Zhuk}}, \bibinfo {author} {\bibfnamefont {N.}~\bibnamefont {Robertson}}, \ and\ \bibinfo {author} {\bibfnamefont {S.}~\bibnamefont {Bravyi}},\ }\href {http://arxiv.org/abs/2306.12569} {\  (\bibinfo {year} {2023})}\BibitemShut {NoStop}%
\bibitem [{\citenamefont {Vazquez}\ \emph {et~al.}(2023)\citenamefont {Vazquez}, \citenamefont {Egger}, \citenamefont {Ochsner},\ and\ \citenamefont {Woerner}}]{Vazquez2023}%
  \BibitemOpen
  \bibfield  {author} {\bibinfo {author} {\bibfnamefont {A.~C.}\ \bibnamefont {Vazquez}}, \bibinfo {author} {\bibfnamefont {D.~J.}\ \bibnamefont {Egger}}, \bibinfo {author} {\bibfnamefont {D.}~\bibnamefont {Ochsner}}, \ and\ \bibinfo {author} {\bibfnamefont {S.}~\bibnamefont {Woerner}},\ }\href {\doibase 10.22331/q-2023-07-25-1067} {\bibfield  {journal} {\bibinfo  {journal} {Quantum}\ }\textbf {\bibinfo {volume} {7}} (\bibinfo {year} {2023}),\ 10.22331/q-2023-07-25-1067}\BibitemShut {NoStop}%
\bibitem [{\citenamefont {Haah}\ \emph {et~al.}(2018)\citenamefont {Haah}, \citenamefont {Hastings}, \citenamefont {Kothari},\ and\ \citenamefont {Low}}]{Haah2018}%
  \BibitemOpen
  \bibfield  {author} {\bibinfo {author} {\bibfnamefont {J.}~\bibnamefont {Haah}}, \bibinfo {author} {\bibfnamefont {M.~B.}\ \bibnamefont {Hastings}}, \bibinfo {author} {\bibfnamefont {R.}~\bibnamefont {Kothari}}, \ and\ \bibinfo {author} {\bibfnamefont {G.~H.}\ \bibnamefont {Low}},\ }\href {\doibase 10.1137/18M1231511} {\  (\bibinfo {year} {2018}),\ 10.1137/18M1231511}\BibitemShut {NoStop}%
\bibitem [{\citenamefont {Childs}\ and\ \citenamefont {Su}(2019)}]{Childs2019}%
  \BibitemOpen
  \bibfield  {author} {\bibinfo {author} {\bibfnamefont {A.~M.}\ \bibnamefont {Childs}}\ and\ \bibinfo {author} {\bibfnamefont {Y.}~\bibnamefont {Su}},\ }\href {\doibase 10.1103/PhysRevLett.123.050503} {\  (\bibinfo {year} {2019}),\ 10.1103/PhysRevLett.123.050503}\BibitemShut {NoStop}%
\bibitem [{\citenamefont {Wecker}\ \emph {et~al.}(2015)\citenamefont {Wecker}, \citenamefont {Hastings}, \citenamefont {Wiebe}, \citenamefont {Clark}, \citenamefont {Nayak},\ and\ \citenamefont {Troyer}}]{Wecker2015}%
  \BibitemOpen
  \bibfield  {author} {\bibinfo {author} {\bibfnamefont {D.}~\bibnamefont {Wecker}}, \bibinfo {author} {\bibfnamefont {M.~B.}\ \bibnamefont {Hastings}}, \bibinfo {author} {\bibfnamefont {N.}~\bibnamefont {Wiebe}}, \bibinfo {author} {\bibfnamefont {B.~K.}\ \bibnamefont {Clark}}, \bibinfo {author} {\bibfnamefont {C.}~\bibnamefont {Nayak}}, \ and\ \bibinfo {author} {\bibfnamefont {M.}~\bibnamefont {Troyer}},\ }\href {\doibase 10.1103/PhysRevA.92.062318} {\  (\bibinfo {year} {2015}),\ 10.1103/PhysRevA.92.062318}\BibitemShut {NoStop}%
\bibitem [{\citenamefont {Childs}\ \emph {et~al.}(2021{\natexlab{a}})\citenamefont {Childs}, \citenamefont {Su}, \citenamefont {Tran}, \citenamefont {Wiebe},\ and\ \citenamefont {Zhu}}]{Childs_2021}%
  \BibitemOpen
  \bibfield  {author} {\bibinfo {author} {\bibfnamefont {A.~M.}\ \bibnamefont {Childs}}, \bibinfo {author} {\bibfnamefont {Y.}~\bibnamefont {Su}}, \bibinfo {author} {\bibfnamefont {M.~C.}\ \bibnamefont {Tran}}, \bibinfo {author} {\bibfnamefont {N.}~\bibnamefont {Wiebe}}, \ and\ \bibinfo {author} {\bibfnamefont {S.}~\bibnamefont {Zhu}},\ }\href {\doibase 10.1103/physrevx.11.011020} {\bibfield  {journal} {\bibinfo  {journal} {Physical Review X}\ }\textbf {\bibinfo {volume} {11}} (\bibinfo {year} {2021}{\natexlab{a}}),\ 10.1103/physrevx.11.011020}\BibitemShut {NoStop}%
\bibitem [{\citenamefont {Tran}\ \emph {et~al.}(2020)\citenamefont {Tran}, \citenamefont {Chu}, \citenamefont {Su}, \citenamefont {Childs},\ and\ \citenamefont {Gorshkov}}]{Tran_2020}%
  \BibitemOpen
  \bibfield  {author} {\bibinfo {author} {\bibfnamefont {M.~C.}\ \bibnamefont {Tran}}, \bibinfo {author} {\bibfnamefont {S.-K.}\ \bibnamefont {Chu}}, \bibinfo {author} {\bibfnamefont {Y.}~\bibnamefont {Su}}, \bibinfo {author} {\bibfnamefont {A.~M.}\ \bibnamefont {Childs}}, \ and\ \bibinfo {author} {\bibfnamefont {A.~V.}\ \bibnamefont {Gorshkov}},\ }\href {\doibase 10.1103/physrevlett.124.220502} {\bibfield  {journal} {\bibinfo  {journal} {Physical Review Letters}\ }\textbf {\bibinfo {volume} {124}} (\bibinfo {year} {2020}),\ 10.1103/physrevlett.124.220502}\BibitemShut {NoStop}%
\bibitem [{\citenamefont {Cho}\ \emph {et~al.}(2022)\citenamefont {Cho}, \citenamefont {Berry},\ and\ \citenamefont {Hsieh}}]{Cho2022}%
  \BibitemOpen
  \bibfield  {author} {\bibinfo {author} {\bibfnamefont {C.~H.}\ \bibnamefont {Cho}}, \bibinfo {author} {\bibfnamefont {D.~W.}\ \bibnamefont {Berry}}, \ and\ \bibinfo {author} {\bibfnamefont {M.-H.}\ \bibnamefont {Hsieh}},\ }\href {\doibase 10.1103/PhysRevA.109.062431} {\  (\bibinfo {year} {2022}),\ 10.1103/PhysRevA.109.062431}\BibitemShut {NoStop}%
\bibitem [{\citenamefont {Childs}\ \emph {et~al.}(2018{\natexlab{b}})\citenamefont {Childs}, \citenamefont {Ostrander},\ and\ \citenamefont {Su}}]{Childs_Ostrander_Su2018}%
  \BibitemOpen
  \bibfield  {author} {\bibinfo {author} {\bibfnamefont {A.~M.}\ \bibnamefont {Childs}}, \bibinfo {author} {\bibfnamefont {A.}~\bibnamefont {Ostrander}}, \ and\ \bibinfo {author} {\bibfnamefont {Y.}~\bibnamefont {Su}},\ }\href {\doibase 10.22331/q-2019-09-02-182} {\  (\bibinfo {year} {2018}{\natexlab{b}}),\ 10.22331/q-2019-09-02-182}\BibitemShut {NoStop}%
\bibitem [{\citenamefont {Campbell}(2018)}]{Campbell2018}%
  \BibitemOpen
  \bibfield  {author} {\bibinfo {author} {\bibfnamefont {E.}~\bibnamefont {Campbell}},\ }\href {\doibase 10.1103/PhysRevLett.123.070503} {\  (\bibinfo {year} {2018}),\ 10.1103/PhysRevLett.123.070503}\BibitemShut {NoStop}%
\bibitem [{\citenamefont {Chen}\ \emph {et~al.}(2021{\natexlab{b}})\citenamefont {Chen}, \citenamefont {Huang}, \citenamefont {Kueng},\ and\ \citenamefont {Tropp}}]{Chen2021}%
  \BibitemOpen
  \bibfield  {author} {\bibinfo {author} {\bibfnamefont {C.~F.}\ \bibnamefont {Chen}}, \bibinfo {author} {\bibfnamefont {H.~Y.}\ \bibnamefont {Huang}}, \bibinfo {author} {\bibfnamefont {R.}~\bibnamefont {Kueng}}, \ and\ \bibinfo {author} {\bibfnamefont {J.~A.}\ \bibnamefont {Tropp}},\ }\href {\doibase 10.1103/PRXQuantum.2.040305} {\bibfield  {journal} {\bibinfo  {journal} {PRX Quantum}\ }\textbf {\bibinfo {volume} {2}} (\bibinfo {year} {2021}{\natexlab{b}}),\ 10.1103/PRXQuantum.2.040305}\BibitemShut {NoStop}%
\bibitem [{\citenamefont {Chertkov}\ \emph {et~al.}(2024)\citenamefont {Chertkov}, \citenamefont {Chen}, \citenamefont {Lubasch}, \citenamefont {Hayes},\ and\ \citenamefont {Foss-Feig}}]{Chertkov2024}%
  \BibitemOpen
  \bibfield  {author} {\bibinfo {author} {\bibfnamefont {E.}~\bibnamefont {Chertkov}}, \bibinfo {author} {\bibfnamefont {Y.-H.}\ \bibnamefont {Chen}}, \bibinfo {author} {\bibfnamefont {M.}~\bibnamefont {Lubasch}}, \bibinfo {author} {\bibfnamefont {D.}~\bibnamefont {Hayes}}, \ and\ \bibinfo {author} {\bibfnamefont {M.}~\bibnamefont {Foss-Feig}},\ }\href {http://arxiv.org/abs/2410.10794} {\  (\bibinfo {year} {2024})}\BibitemShut {NoStop}%
\bibitem [{\citenamefont {Granet}\ and\ \citenamefont {Dreyer}(2024{\natexlab{b}})}]{Granet2024b}%
  \BibitemOpen
  \bibfield  {author} {\bibinfo {author} {\bibfnamefont {E.}~\bibnamefont {Granet}}\ and\ \bibinfo {author} {\bibfnamefont {H.}~\bibnamefont {Dreyer}},\ }\href {http://arxiv.org/abs/2409.04254} {\  (\bibinfo {year} {2024}{\natexlab{b}})}\BibitemShut {NoStop}%
\bibitem [{\citenamefont {Chen}(2024)}]{chen2024trottererrortimescaling}%
  \BibitemOpen
  \bibfield  {author} {\bibinfo {author} {\bibfnamefont {Y.-H.}\ \bibnamefont {Chen}},\ }\href {https://arxiv.org/abs/2409.16634} {\enquote {\bibinfo {title} {Trotter error time scaling separation via commutant decomposition},}\ } (\bibinfo {year} {2024}),\ \Eprint {http://arxiv.org/abs/2409.16634} {arXiv:2409.16634 [quant-ph]} \BibitemShut {NoStop}%
\bibitem [{\citenamefont {Meister}\ \emph {et~al.}(2022)\citenamefont {Meister}, \citenamefont {Benjamin},\ and\ \citenamefont {Campbell}}]{Meister2022tailoringterm}%
  \BibitemOpen
  \bibfield  {author} {\bibinfo {author} {\bibfnamefont {R.}~\bibnamefont {Meister}}, \bibinfo {author} {\bibfnamefont {S.~C.}\ \bibnamefont {Benjamin}}, \ and\ \bibinfo {author} {\bibfnamefont {E.~T.}\ \bibnamefont {Campbell}},\ }\href {\doibase 10.22331/q-2022-02-02-637} {\bibfield  {journal} {\bibinfo  {journal} {{Quantum}}\ }\textbf {\bibinfo {volume} {6}},\ \bibinfo {pages} {637} (\bibinfo {year} {2022})}\BibitemShut {NoStop}%
\bibitem [{\citenamefont {Berry}\ \emph {et~al.}(2015{\natexlab{a}})\citenamefont {Berry}, \citenamefont {Childs}, \citenamefont {Cleve}, \citenamefont {Kothari},\ and\ \citenamefont {Somma}}]{Berry_etal_2015}%
  \BibitemOpen
  \bibfield  {author} {\bibinfo {author} {\bibfnamefont {D.~W.}\ \bibnamefont {Berry}}, \bibinfo {author} {\bibfnamefont {A.~M.}\ \bibnamefont {Childs}}, \bibinfo {author} {\bibfnamefont {R.}~\bibnamefont {Cleve}}, \bibinfo {author} {\bibfnamefont {R.}~\bibnamefont {Kothari}}, \ and\ \bibinfo {author} {\bibfnamefont {R.~D.}\ \bibnamefont {Somma}},\ }\href {\doibase 10.1103/PhysRevLett.114.090502} {\bibfield  {journal} {\bibinfo  {journal} {Phys. Rev. Lett.}\ }\textbf {\bibinfo {volume} {114}},\ \bibinfo {pages} {090502} (\bibinfo {year} {2015}{\natexlab{a}})}\BibitemShut {NoStop}%
\bibitem [{\citenamefont {Berry}\ \emph {et~al.}(2014)\citenamefont {Berry}, \citenamefont {Childs}, \citenamefont {Cleve}, \citenamefont {Kothari},\ and\ \citenamefont {Somma}}]{Berry1}%
  \BibitemOpen
  \bibfield  {author} {\bibinfo {author} {\bibfnamefont {D.~W.}\ \bibnamefont {Berry}}, \bibinfo {author} {\bibfnamefont {A.~M.}\ \bibnamefont {Childs}}, \bibinfo {author} {\bibfnamefont {R.}~\bibnamefont {Cleve}}, \bibinfo {author} {\bibfnamefont {R.}~\bibnamefont {Kothari}}, \ and\ \bibinfo {author} {\bibfnamefont {R.~D.}\ \bibnamefont {Somma}},\ }in\ \href {\doibase 10.1145/2591796.2591854} {\emph {\bibinfo {booktitle} {Proceedings of the Forty-Sixth Annual ACM Symposium on Theory of Computing}}},\ \bibinfo {series and number} {STOC '14}\ (\bibinfo  {publisher} {Association for Computing Machinery},\ \bibinfo {address} {New York, NY, USA},\ \bibinfo {year} {2014})\ p.\ \bibinfo {pages} {283–292}\BibitemShut {NoStop}%
\bibitem [{\citenamefont {Loaiza}\ \emph {et~al.}(2023)\citenamefont {Loaiza}, \citenamefont {Khah}, \citenamefont {Wiebe},\ and\ \citenamefont {Izmaylov}}]{Loaiza_2023}%
  \BibitemOpen
  \bibfield  {author} {\bibinfo {author} {\bibfnamefont {I.}~\bibnamefont {Loaiza}}, \bibinfo {author} {\bibfnamefont {A.~M.}\ \bibnamefont {Khah}}, \bibinfo {author} {\bibfnamefont {N.}~\bibnamefont {Wiebe}}, \ and\ \bibinfo {author} {\bibfnamefont {A.~F.}\ \bibnamefont {Izmaylov}},\ }\href {\doibase 10.1088/2058-9565/acd577} {\bibfield  {journal} {\bibinfo  {journal} {Quantum Science and Technology}\ }\textbf {\bibinfo {volume} {8}},\ \bibinfo {pages} {035019} (\bibinfo {year} {2023})}\BibitemShut {NoStop}%
\bibitem [{\citenamefont {Babbush}\ \emph {et~al.}(2018{\natexlab{a}})\citenamefont {Babbush}, \citenamefont {Gidney}, \citenamefont {Berry}, \citenamefont {Wiebe}, \citenamefont {McClean}, \citenamefont {Paler}, \citenamefont {Fowler},\ and\ \citenamefont {Neven}}]{BabbushPhase}%
  \BibitemOpen
  \bibfield  {author} {\bibinfo {author} {\bibfnamefont {R.}~\bibnamefont {Babbush}}, \bibinfo {author} {\bibfnamefont {C.}~\bibnamefont {Gidney}}, \bibinfo {author} {\bibfnamefont {D.~W.}\ \bibnamefont {Berry}}, \bibinfo {author} {\bibfnamefont {N.}~\bibnamefont {Wiebe}}, \bibinfo {author} {\bibfnamefont {J.}~\bibnamefont {McClean}}, \bibinfo {author} {\bibfnamefont {A.}~\bibnamefont {Paler}}, \bibinfo {author} {\bibfnamefont {A.}~\bibnamefont {Fowler}}, \ and\ \bibinfo {author} {\bibfnamefont {H.}~\bibnamefont {Neven}},\ }\href {\doibase 10.1103/PhysRevX.8.041015} {\bibfield  {journal} {\bibinfo  {journal} {Phys. Rev. X}\ }\textbf {\bibinfo {volume} {8}},\ \bibinfo {pages} {041015} (\bibinfo {year} {2018}{\natexlab{a}})}\BibitemShut {NoStop}%
\bibitem [{\citenamefont {Lee}\ \emph {et~al.}(2021)\citenamefont {Lee}, \citenamefont {Berry}, \citenamefont {Gidney}, \citenamefont {Huggins}, \citenamefont {McClean}, \citenamefont {Wiebe},\ and\ \citenamefont {Babbush}}]{LeeTensorHyper}%
  \BibitemOpen
  \bibfield  {author} {\bibinfo {author} {\bibfnamefont {J.}~\bibnamefont {Lee}}, \bibinfo {author} {\bibfnamefont {D.~W.}\ \bibnamefont {Berry}}, \bibinfo {author} {\bibfnamefont {C.}~\bibnamefont {Gidney}}, \bibinfo {author} {\bibfnamefont {W.~J.}\ \bibnamefont {Huggins}}, \bibinfo {author} {\bibfnamefont {J.~R.}\ \bibnamefont {McClean}}, \bibinfo {author} {\bibfnamefont {N.}~\bibnamefont {Wiebe}}, \ and\ \bibinfo {author} {\bibfnamefont {R.}~\bibnamefont {Babbush}},\ }\href {\doibase 10.1103/PRXQuantum.2.030305} {\bibfield  {journal} {\bibinfo  {journal} {PRX Quantum}\ }\textbf {\bibinfo {volume} {2}},\ \bibinfo {pages} {030305} (\bibinfo {year} {2021})}\BibitemShut {NoStop}%
\bibitem [{\citenamefont {Su}\ \emph {et~al.}(2021)\citenamefont {Su}, \citenamefont {Berry}, \citenamefont {Wiebe}, \citenamefont {Rubin},\ and\ \citenamefont {Babbush}}]{Babbush1Q}%
  \BibitemOpen
  \bibfield  {author} {\bibinfo {author} {\bibfnamefont {Y.}~\bibnamefont {Su}}, \bibinfo {author} {\bibfnamefont {D.~W.}\ \bibnamefont {Berry}}, \bibinfo {author} {\bibfnamefont {N.}~\bibnamefont {Wiebe}}, \bibinfo {author} {\bibfnamefont {N.}~\bibnamefont {Rubin}}, \ and\ \bibinfo {author} {\bibfnamefont {R.}~\bibnamefont {Babbush}},\ }\href {\doibase 10.1103/PRXQuantum.2.040332} {\bibfield  {journal} {\bibinfo  {journal} {PRX Quantum}\ }\textbf {\bibinfo {volume} {2}},\ \bibinfo {pages} {040332} (\bibinfo {year} {2021})}\BibitemShut {NoStop}%
\bibitem [{\citenamefont {Shokrian~Zini}\ \emph {et~al.}(2023)\citenamefont {Shokrian~Zini}, \citenamefont {Delgado}, \citenamefont {dos Reis}, \citenamefont {Moreno~Casares}, \citenamefont {Mueller}, \citenamefont {Voigt},\ and\ \citenamefont {Arrazola}}]{ShokrianZini2023quantumsimulationof}%
  \BibitemOpen
  \bibfield  {author} {\bibinfo {author} {\bibfnamefont {M.}~\bibnamefont {Shokrian~Zini}}, \bibinfo {author} {\bibfnamefont {A.}~\bibnamefont {Delgado}}, \bibinfo {author} {\bibfnamefont {R.}~\bibnamefont {dos Reis}}, \bibinfo {author} {\bibfnamefont {P.~A.}\ \bibnamefont {Moreno~Casares}}, \bibinfo {author} {\bibfnamefont {J.~E.}\ \bibnamefont {Mueller}}, \bibinfo {author} {\bibfnamefont {A.-C.}\ \bibnamefont {Voigt}}, \ and\ \bibinfo {author} {\bibfnamefont {J.~M.}\ \bibnamefont {Arrazola}},\ }\href {\doibase 10.22331/q-2023-07-10-1049} {\bibfield  {journal} {\bibinfo  {journal} {{Quantum}}\ }\textbf {\bibinfo {volume} {7}},\ \bibinfo {pages} {1049} (\bibinfo {year} {2023})}\BibitemShut {NoStop}%
\bibitem [{\citenamefont {Kalev}\ and\ \citenamefont {Hen}(2021)}]{Kalev2021quantumalgorithm}%
  \BibitemOpen
  \bibfield  {author} {\bibinfo {author} {\bibfnamefont {A.}~\bibnamefont {Kalev}}\ and\ \bibinfo {author} {\bibfnamefont {I.}~\bibnamefont {Hen}},\ }\href {\doibase 10.22331/q-2021-04-08-426} {\bibfield  {journal} {\bibinfo  {journal} {{Quantum}}\ }\textbf {\bibinfo {volume} {5}},\ \bibinfo {pages} {426} (\bibinfo {year} {2021})}\BibitemShut {NoStop}%
\bibitem [{\citenamefont {Berry}\ \emph {et~al.}(2015{\natexlab{b}})\citenamefont {Berry}, \citenamefont {Childs},\ and\ \citenamefont {Kothari}}]{Berry2}%
  \BibitemOpen
  \bibfield  {author} {\bibinfo {author} {\bibfnamefont {D.~W.}\ \bibnamefont {Berry}}, \bibinfo {author} {\bibfnamefont {A.~M.}\ \bibnamefont {Childs}}, \ and\ \bibinfo {author} {\bibfnamefont {R.}~\bibnamefont {Kothari}},\ }in\ \href {\doibase 10.1109/FOCS.2015.54} {\emph {\bibinfo {booktitle} {Proceedings of the 2015 IEEE 56th Annual Symposium on Foundations of Computer Science (FOCS)}}},\ \bibinfo {series and number} {FOCS '15}\ (\bibinfo  {publisher} {IEEE Computer Society},\ \bibinfo {address} {USA},\ \bibinfo {year} {2015})\ p.\ \bibinfo {pages} {792–809}\BibitemShut {NoStop}%
\bibitem [{\citenamefont {Yan}\ \emph {et~al.}(2022)\citenamefont {Yan}, \citenamefont {Wei}, \citenamefont {Jiang}, \citenamefont {Wang}, \citenamefont {Duan}, \citenamefont {Ma},\ and\ \citenamefont {Long}}]{Amp1}%
  \BibitemOpen
  \bibfield  {author} {\bibinfo {author} {\bibfnamefont {B.}~\bibnamefont {Yan}}, \bibinfo {author} {\bibfnamefont {S.}~\bibnamefont {Wei}}, \bibinfo {author} {\bibfnamefont {H.}~\bibnamefont {Jiang}}, \bibinfo {author} {\bibfnamefont {H.}~\bibnamefont {Wang}}, \bibinfo {author} {\bibfnamefont {Q.}~\bibnamefont {Duan}}, \bibinfo {author} {\bibfnamefont {Z.}~\bibnamefont {Ma}}, \ and\ \bibinfo {author} {\bibfnamefont {G.-L.}\ \bibnamefont {Long}},\ }\href {\doibase 10.1038/s41598-022-15093-x} {\bibfield  {journal} {\bibinfo  {journal} {Scientific Reports}\ }\textbf {\bibinfo {volume} {12}},\ \bibinfo {pages} {14339} (\bibinfo {year} {2022})}\BibitemShut {NoStop}%
\bibitem [{\citenamefont {Berry}\ \emph {et~al.}(2017)\citenamefont {Berry}, \citenamefont {Childs}, \citenamefont {Ostrander},\ and\ \citenamefont {Wang}}]{dif1}%
  \BibitemOpen
  \bibfield  {author} {\bibinfo {author} {\bibfnamefont {D.~W.}\ \bibnamefont {Berry}}, \bibinfo {author} {\bibfnamefont {A.~M.}\ \bibnamefont {Childs}}, \bibinfo {author} {\bibfnamefont {A.}~\bibnamefont {Ostrander}}, \ and\ \bibinfo {author} {\bibfnamefont {G.}~\bibnamefont {Wang}},\ }\href {\doibase 10.1007/s00220-017-3002-y} {\bibfield  {journal} {\bibinfo  {journal} {Communications in Mathematical Physics}\ }\textbf {\bibinfo {volume} {356}},\ \bibinfo {pages} {1057} (\bibinfo {year} {2017})}\BibitemShut {NoStop}%
\bibitem [{\citenamefont {Liu}\ \emph {et~al.}(2021)\citenamefont {Liu}, \citenamefont {Kolden}, \citenamefont {Krovi}, \citenamefont {Loureiro}, \citenamefont {Trivisa},\ and\ \citenamefont {Childs}}]{dif2}%
  \BibitemOpen
  \bibfield  {author} {\bibinfo {author} {\bibfnamefont {J.-P.}\ \bibnamefont {Liu}}, \bibinfo {author} {\bibfnamefont {H.~{\O}.}\ \bibnamefont {Kolden}}, \bibinfo {author} {\bibfnamefont {H.~K.}\ \bibnamefont {Krovi}}, \bibinfo {author} {\bibfnamefont {N.~F.}\ \bibnamefont {Loureiro}}, \bibinfo {author} {\bibfnamefont {K.}~\bibnamefont {Trivisa}}, \ and\ \bibinfo {author} {\bibfnamefont {A.~M.}\ \bibnamefont {Childs}},\ }\href {\doibase 10.1073/pnas.2026805118} {\bibfield  {journal} {\bibinfo  {journal} {Proceedings of the National Academy of Sciences}\ }\textbf {\bibinfo {volume} {118}},\ \bibinfo {pages} {e2026805118} (\bibinfo {year} {2021})},\ \Eprint {http://arxiv.org/abs/https://www.pnas.org/doi/pdf/10.1073/pnas.2026805118} {https://www.pnas.org/doi/pdf/10.1073/pnas.2026805118} \BibitemShut {NoStop}%
\bibitem [{\citenamefont {Childs}\ \emph {et~al.}(2021{\natexlab{b}})\citenamefont {Childs}, \citenamefont {Liu},\ and\ \citenamefont {Ostrander}}]{Childs2021highprecision}%
  \BibitemOpen
  \bibfield  {author} {\bibinfo {author} {\bibfnamefont {A.~M.}\ \bibnamefont {Childs}}, \bibinfo {author} {\bibfnamefont {J.-P.}\ \bibnamefont {Liu}}, \ and\ \bibinfo {author} {\bibfnamefont {A.}~\bibnamefont {Ostrander}},\ }\href {\doibase 10.22331/q-2021-11-10-574} {\bibfield  {journal} {\bibinfo  {journal} {{Quantum}}\ }\textbf {\bibinfo {volume} {5}},\ \bibinfo {pages} {574} (\bibinfo {year} {2021}{\natexlab{b}})}\BibitemShut {NoStop}%
\bibitem [{\citenamefont {Ge}\ \emph {et~al.}(2019)\citenamefont {Ge}, \citenamefont {Tura},\ and\ \citenamefont {Cirac}}]{CiracGS}%
  \BibitemOpen
  \bibfield  {author} {\bibinfo {author} {\bibfnamefont {Y.}~\bibnamefont {Ge}}, \bibinfo {author} {\bibfnamefont {J.}~\bibnamefont {Tura}}, \ and\ \bibinfo {author} {\bibfnamefont {J.~I.}\ \bibnamefont {Cirac}},\ }\href {\doibase 10.1063/1.5027484} {\bibfield  {journal} {\bibinfo  {journal} {Journal of Mathematical Physics}\ }\textbf {\bibinfo {volume} {60}},\ \bibinfo {pages} {022202} (\bibinfo {year} {2019})},\ \Eprint {http://arxiv.org/abs/https://pubs.aip.org/aip/jmp/article-pdf/doi/10.1063/1.5027484/13434463/022202\_1\_online.pdf} {https://pubs.aip.org/aip/jmp/article-pdf/doi/10.1063/1.5027484/13434463/022202\_1\_online.pdf} \BibitemShut {NoStop}%
\bibitem [{\citenamefont {Keen}\ \emph {et~al.}(2021)\citenamefont {Keen}, \citenamefont {Dumitrescu},\ and\ \citenamefont {Wang}}]{keen2021quantum}%
  \BibitemOpen
  \bibfield  {author} {\bibinfo {author} {\bibfnamefont {T.}~\bibnamefont {Keen}}, \bibinfo {author} {\bibfnamefont {E.}~\bibnamefont {Dumitrescu}}, \ and\ \bibinfo {author} {\bibfnamefont {Y.}~\bibnamefont {Wang}},\ }\href@noop {} {\enquote {\bibinfo {title} {Quantum algorithms for ground-state preparation and green's function calculation},}\ } (\bibinfo {year} {2021}),\ \Eprint {http://arxiv.org/abs/2112.05731} {arXiv:2112.05731 [quant-ph]} \BibitemShut {NoStop}%
\bibitem [{\citenamefont {He}\ \emph {et~al.}(2022)\citenamefont {He}, \citenamefont {Zhang},\ and\ \citenamefont {Wang}}]{PhysRevA.106.032420}%
  \BibitemOpen
  \bibfield  {author} {\bibinfo {author} {\bibfnamefont {M.-Q.}\ \bibnamefont {He}}, \bibinfo {author} {\bibfnamefont {D.-B.}\ \bibnamefont {Zhang}}, \ and\ \bibinfo {author} {\bibfnamefont {Z.~D.}\ \bibnamefont {Wang}},\ }\href {\doibase 10.1103/PhysRevA.106.032420} {\bibfield  {journal} {\bibinfo  {journal} {Phys. Rev. A}\ }\textbf {\bibinfo {volume} {106}},\ \bibinfo {pages} {032420} (\bibinfo {year} {2022})}\BibitemShut {NoStop}%
\bibitem [{\citenamefont {Ralli}\ \emph {et~al.}(2021)\citenamefont {Ralli}, \citenamefont {Love}, \citenamefont {Tranter},\ and\ \citenamefont {Coveney}}]{AlexisLCU}%
  \BibitemOpen
  \bibfield  {author} {\bibinfo {author} {\bibfnamefont {A.}~\bibnamefont {Ralli}}, \bibinfo {author} {\bibfnamefont {P.~J.}\ \bibnamefont {Love}}, \bibinfo {author} {\bibfnamefont {A.}~\bibnamefont {Tranter}}, \ and\ \bibinfo {author} {\bibfnamefont {P.~V.}\ \bibnamefont {Coveney}},\ }\href {\doibase 10.1103/PhysRevResearch.3.033195} {\bibfield  {journal} {\bibinfo  {journal} {Phys. Rev. Res.}\ }\textbf {\bibinfo {volume} {3}},\ \bibinfo {pages} {033195} (\bibinfo {year} {2021})}\BibitemShut {NoStop}%
\bibitem [{\citenamefont {Rall}(2020)}]{RallPhysical}%
  \BibitemOpen
  \bibfield  {author} {\bibinfo {author} {\bibfnamefont {P.}~\bibnamefont {Rall}},\ }\href {\doibase 10.1103/PhysRevA.102.022408} {\bibfield  {journal} {\bibinfo  {journal} {Phys. Rev. A}\ }\textbf {\bibinfo {volume} {102}},\ \bibinfo {pages} {022408} (\bibinfo {year} {2020})}\BibitemShut {NoStop}%
\bibitem [{\citenamefont {Tong}\ \emph {et~al.}(2021)\citenamefont {Tong}, \citenamefont {An}, \citenamefont {Wiebe},\ and\ \citenamefont {Lin}}]{LinLinGFLCU}%
  \BibitemOpen
  \bibfield  {author} {\bibinfo {author} {\bibfnamefont {Y.}~\bibnamefont {Tong}}, \bibinfo {author} {\bibfnamefont {D.}~\bibnamefont {An}}, \bibinfo {author} {\bibfnamefont {N.}~\bibnamefont {Wiebe}}, \ and\ \bibinfo {author} {\bibfnamefont {L.}~\bibnamefont {Lin}},\ }\href {\doibase 10.1103/PhysRevA.104.032422} {\bibfield  {journal} {\bibinfo  {journal} {Phys. Rev. A}\ }\textbf {\bibinfo {volume} {104}},\ \bibinfo {pages} {032422} (\bibinfo {year} {2021})}\BibitemShut {NoStop}%
\bibitem [{\citenamefont {Chowdhury}\ and\ \citenamefont {Somma}(2017)}]{SommaGibbsSampling}%
  \BibitemOpen
  \bibfield  {author} {\bibinfo {author} {\bibfnamefont {A.~N.}\ \bibnamefont {Chowdhury}}\ and\ \bibinfo {author} {\bibfnamefont {R.~D.}\ \bibnamefont {Somma}},\ }\href@noop {} {\bibfield  {journal} {\bibinfo  {journal} {Quantum Info. Comput.}\ }\textbf {\bibinfo {volume} {17}},\ \bibinfo {pages} {41–64} (\bibinfo {year} {2017})}\BibitemShut {NoStop}%
\bibitem [{\citenamefont {van Apeldoorn}\ \emph {et~al.}(2020)\citenamefont {van Apeldoorn}, \citenamefont {Gily{\'{e}}n}, \citenamefont {Gribling},\ and\ \citenamefont {de~Wolf}}]{vanApeldoorn2020quantumsdpsolvers}%
  \BibitemOpen
  \bibfield  {author} {\bibinfo {author} {\bibfnamefont {J.}~\bibnamefont {van Apeldoorn}}, \bibinfo {author} {\bibfnamefont {A.}~\bibnamefont {Gily{\'{e}}n}}, \bibinfo {author} {\bibfnamefont {S.}~\bibnamefont {Gribling}}, \ and\ \bibinfo {author} {\bibfnamefont {R.}~\bibnamefont {de~Wolf}},\ }\href {\doibase 10.22331/q-2020-02-14-230} {\bibfield  {journal} {\bibinfo  {journal} {{Quantum}}\ }\textbf {\bibinfo {volume} {4}},\ \bibinfo {pages} {230} (\bibinfo {year} {2020})}\BibitemShut {NoStop}%
\bibitem [{\citenamefont {Toyoizumi}\ \emph {et~al.}(2023)\citenamefont {Toyoizumi}, \citenamefont {Yamamoto},\ and\ \citenamefont {Hoshino}}]{toyoizumi2023hamiltoniansimulationusingquantum}%
  \BibitemOpen
  \bibfield  {author} {\bibinfo {author} {\bibfnamefont {K.}~\bibnamefont {Toyoizumi}}, \bibinfo {author} {\bibfnamefont {N.}~\bibnamefont {Yamamoto}}, \ and\ \bibinfo {author} {\bibfnamefont {K.}~\bibnamefont {Hoshino}},\ }\href {https://arxiv.org/abs/2304.08937} {\enquote {\bibinfo {title} {Hamiltonian simulation using quantum singular value transformation: complexity analysis and application to the linearized vlasov-poisson equation},}\ } (\bibinfo {year} {2023}),\ \Eprint {http://arxiv.org/abs/2304.08937} {arXiv:2304.08937 [quant-ph]} \BibitemShut {NoStop}%
\bibitem [{\citenamefont {Berry}\ \emph {et~al.}(2024)\citenamefont {Berry}, \citenamefont {Motlagh}, \citenamefont {Pantaleoni},\ and\ \citenamefont {Wiebe}}]{Berry2024}%
  \BibitemOpen
  \bibfield  {author} {\bibinfo {author} {\bibfnamefont {D.~W.}\ \bibnamefont {Berry}}, \bibinfo {author} {\bibfnamefont {D.}~\bibnamefont {Motlagh}}, \bibinfo {author} {\bibfnamefont {G.}~\bibnamefont {Pantaleoni}}, \ and\ \bibinfo {author} {\bibfnamefont {N.}~\bibnamefont {Wiebe}},\ }\href {\doibase 10.1103/physreva.110.012612} {\bibfield  {journal} {\bibinfo  {journal} {Physical Review A}\ }\textbf {\bibinfo {volume} {110}} (\bibinfo {year} {2024}),\ 10.1103/physreva.110.012612}\BibitemShut {NoStop}%
\bibitem [{\citenamefont {Brown}\ \emph {et~al.}(2006)\citenamefont {Brown}, \citenamefont {Clark},\ and\ \citenamefont {Chuang}}]{Brown2006}%
  \BibitemOpen
  \bibfield  {author} {\bibinfo {author} {\bibfnamefont {K.~R.}\ \bibnamefont {Brown}}, \bibinfo {author} {\bibfnamefont {R.~J.}\ \bibnamefont {Clark}}, \ and\ \bibinfo {author} {\bibfnamefont {I.~L.}\ \bibnamefont {Chuang}},\ }\href {\doibase 10.1103/physrevlett.97.050504} {\bibfield  {journal} {\bibinfo  {journal} {Physical Review Letters}\ }\textbf {\bibinfo {volume} {97}} (\bibinfo {year} {2006}),\ 10.1103/physrevlett.97.050504}\BibitemShut {NoStop}%
\bibitem [{\citenamefont {Barends}\ \emph {et~al.}(2015)\citenamefont {Barends}, \citenamefont {Lamata}, \citenamefont {Kelly}, \citenamefont {García-Álvarez}, \citenamefont {Fowler}, \citenamefont {Megrant}, \citenamefont {Jeffrey}, \citenamefont {White}, \citenamefont {Sank}, \citenamefont {Mutus}, \citenamefont {Campbell}, \citenamefont {Chen}, \citenamefont {Chen}, \citenamefont {Chiaro}, \citenamefont {Dunsworth}, \citenamefont {Hoi}, \citenamefont {Neill}, \citenamefont {O'Malley}, \citenamefont {Quintana}, \citenamefont {Roushan}, \citenamefont {Vainsencher}, \citenamefont {Wenner}, \citenamefont {Solano},\ and\ \citenamefont {Martinis}}]{Barends2015}%
  \BibitemOpen
  \bibfield  {author} {\bibinfo {author} {\bibfnamefont {R.}~\bibnamefont {Barends}}, \bibinfo {author} {\bibfnamefont {L.}~\bibnamefont {Lamata}}, \bibinfo {author} {\bibfnamefont {J.}~\bibnamefont {Kelly}}, \bibinfo {author} {\bibfnamefont {L.}~\bibnamefont {García-Álvarez}}, \bibinfo {author} {\bibfnamefont {A.~G.}\ \bibnamefont {Fowler}}, \bibinfo {author} {\bibfnamefont {A.}~\bibnamefont {Megrant}}, \bibinfo {author} {\bibfnamefont {E.}~\bibnamefont {Jeffrey}}, \bibinfo {author} {\bibfnamefont {T.~C.}\ \bibnamefont {White}}, \bibinfo {author} {\bibfnamefont {D.}~\bibnamefont {Sank}}, \bibinfo {author} {\bibfnamefont {J.~Y.}\ \bibnamefont {Mutus}}, \bibinfo {author} {\bibfnamefont {B.}~\bibnamefont {Campbell}}, \bibinfo {author} {\bibfnamefont {Y.}~\bibnamefont {Chen}}, \bibinfo {author} {\bibfnamefont {Z.}~\bibnamefont {Chen}}, \bibinfo {author} {\bibfnamefont {B.}~\bibnamefont {Chiaro}}, \bibinfo {author} {\bibfnamefont {A.}~\bibnamefont {Dunsworth}}, \bibinfo {author} {\bibfnamefont {I.~C.}\
  \bibnamefont {Hoi}}, \bibinfo {author} {\bibfnamefont {C.}~\bibnamefont {Neill}}, \bibinfo {author} {\bibfnamefont {P.~J.}\ \bibnamefont {O'Malley}}, \bibinfo {author} {\bibfnamefont {C.}~\bibnamefont {Quintana}}, \bibinfo {author} {\bibfnamefont {P.}~\bibnamefont {Roushan}}, \bibinfo {author} {\bibfnamefont {A.}~\bibnamefont {Vainsencher}}, \bibinfo {author} {\bibfnamefont {J.}~\bibnamefont {Wenner}}, \bibinfo {author} {\bibfnamefont {E.}~\bibnamefont {Solano}}, \ and\ \bibinfo {author} {\bibfnamefont {J.~M.}\ \bibnamefont {Martinis}},\ }\href {\doibase 10.1038/ncomms8654} {\bibfield  {journal} {\bibinfo  {journal} {Nature Communications}\ }\textbf {\bibinfo {volume} {6}} (\bibinfo {year} {2015}),\ 10.1038/ncomms8654}\BibitemShut {NoStop}%
\bibitem [{\citenamefont {Lanyon}\ \emph {et~al.}(2011)\citenamefont {Lanyon}, \citenamefont {Hempel}, \citenamefont {Nigg}, \citenamefont {Müller}, \citenamefont {Gerritsma}, \citenamefont {Zähringer}, \citenamefont {Schindler}, \citenamefont {Barreiro}, \citenamefont {Rambach}, \citenamefont {Kirchmair}, \citenamefont {Hennrich}, \citenamefont {Zoller}, \citenamefont {Blatt},\ and\ \citenamefont {Roos}}]{Lanyon2011}%
  \BibitemOpen
  \bibfield  {author} {\bibinfo {author} {\bibfnamefont {B.~P.}\ \bibnamefont {Lanyon}}, \bibinfo {author} {\bibfnamefont {C.}~\bibnamefont {Hempel}}, \bibinfo {author} {\bibfnamefont {D.}~\bibnamefont {Nigg}}, \bibinfo {author} {\bibfnamefont {M.}~\bibnamefont {Müller}}, \bibinfo {author} {\bibfnamefont {R.}~\bibnamefont {Gerritsma}}, \bibinfo {author} {\bibfnamefont {F.}~\bibnamefont {Zähringer}}, \bibinfo {author} {\bibfnamefont {P.}~\bibnamefont {Schindler}}, \bibinfo {author} {\bibfnamefont {J.~T.}\ \bibnamefont {Barreiro}}, \bibinfo {author} {\bibfnamefont {M.}~\bibnamefont {Rambach}}, \bibinfo {author} {\bibfnamefont {G.}~\bibnamefont {Kirchmair}}, \bibinfo {author} {\bibfnamefont {M.}~\bibnamefont {Hennrich}}, \bibinfo {author} {\bibfnamefont {P.}~\bibnamefont {Zoller}}, \bibinfo {author} {\bibfnamefont {R.}~\bibnamefont {Blatt}}, \ and\ \bibinfo {author} {\bibfnamefont {C.~F.}\ \bibnamefont {Roos}},\ }\href {\doibase 10.1126/science.1208001} {\bibfield  {journal} {\bibinfo  {journal} {Science}\
  }\textbf {\bibinfo {volume} {334}},\ \bibinfo {pages} {57–61} (\bibinfo {year} {2011})}\BibitemShut {NoStop}%
\bibitem [{\citenamefont {Raeisi}\ \emph {et~al.}(2012)\citenamefont {Raeisi}, \citenamefont {Wiebe},\ and\ \citenamefont {Sanders}}]{Raeisi2012}%
  \BibitemOpen
  \bibfield  {author} {\bibinfo {author} {\bibfnamefont {S.}~\bibnamefont {Raeisi}}, \bibinfo {author} {\bibfnamefont {N.}~\bibnamefont {Wiebe}}, \ and\ \bibinfo {author} {\bibfnamefont {B.~C.}\ \bibnamefont {Sanders}},\ }\href {\doibase 10.1088/1367-2630/14/10/103017} {\bibfield  {journal} {\bibinfo  {journal} {New Journal of Physics}\ }\textbf {\bibinfo {volume} {14}} (\bibinfo {year} {2012}),\ 10.1088/1367-2630/14/10/103017}\BibitemShut {NoStop}%
\bibitem [{\citenamefont {Yu}\ \emph {et~al.}(2022)\citenamefont {Yu}, \citenamefont {Chi}, \citenamefont {Zhai}, \citenamefont {Huang}, \citenamefont {Gong},\ and\ \citenamefont {Wang}}]{yu2022simulating}%
  \BibitemOpen
  \bibfield  {author} {\bibinfo {author} {\bibfnamefont {Y.}~\bibnamefont {Yu}}, \bibinfo {author} {\bibfnamefont {Y.}~\bibnamefont {Chi}}, \bibinfo {author} {\bibfnamefont {C.}~\bibnamefont {Zhai}}, \bibinfo {author} {\bibfnamefont {J.}~\bibnamefont {Huang}}, \bibinfo {author} {\bibfnamefont {Q.}~\bibnamefont {Gong}}, \ and\ \bibinfo {author} {\bibfnamefont {J.}~\bibnamefont {Wang}},\ }\href {http://arxiv.org/abs/2211.06723} {\  (\bibinfo {year} {2022})}\BibitemShut {NoStop}%
\bibitem [{\citenamefont {Boyd}(2024)}]{Boyd2023}%
  \BibitemOpen
  \bibfield  {author} {\bibinfo {author} {\bibfnamefont {G.}~\bibnamefont {Boyd}},\ }\href {https://arxiv.org/abs/2312.00696} {\enquote {\bibinfo {title} {Low-overhead parallelisation of lcu via commuting operators},}\ } (\bibinfo {year} {2024}),\ \Eprint {http://arxiv.org/abs/2312.00696} {arXiv:2312.00696 [quant-ph]} \BibitemShut {NoStop}%
\bibitem [{\citenamefont {Chakraborty}(2024)}]{Chakraborty2023}%
  \BibitemOpen
  \bibfield  {author} {\bibinfo {author} {\bibfnamefont {S.}~\bibnamefont {Chakraborty}},\ }\href {\doibase 10.22331/q-2024-10-10-1496} {\bibfield  {journal} {\bibinfo  {journal} {Quantum}\ }\textbf {\bibinfo {volume} {8}},\ \bibinfo {pages} {1496} (\bibinfo {year} {2024})}\BibitemShut {NoStop}%
\bibitem [{\citenamefont {Loaiza}\ and\ \citenamefont {Izmaylov}(2023)}]{loaiza2023blockinvariantsymmetryshiftpreprocessing}%
  \BibitemOpen
  \bibfield  {author} {\bibinfo {author} {\bibfnamefont {I.}~\bibnamefont {Loaiza}}\ and\ \bibinfo {author} {\bibfnamefont {A.~F.}\ \bibnamefont {Izmaylov}},\ }\href {https://arxiv.org/abs/2304.13772} {\enquote {\bibinfo {title} {Block-invariant symmetry shift: Preprocessing technique for second-quantized hamiltonians to improve their decompositions to linear combination of unitaries},}\ } (\bibinfo {year} {2023}),\ \Eprint {http://arxiv.org/abs/2304.13772} {arXiv:2304.13772 [quant-ph]} \BibitemShut {NoStop}%
\bibitem [{qua(2024)}]{quantinuum_h11}%
  \BibitemOpen
  \href {https://www.quantinuum.com/products/h1} {\enquote {\bibinfo {title} {Quantinuum system model h1 product data sheet},}\ } (\bibinfo {year} {Version 6.1. April 15, 2024})\BibitemShut {NoStop}%
\bibitem [{\citenamefont {Babbush}\ \emph {et~al.}(2018{\natexlab{b}})\citenamefont {Babbush}, \citenamefont {Gidney}, \citenamefont {Berry}, \citenamefont {Wiebe}, \citenamefont {McClean}, \citenamefont {Paler}, \citenamefont {Fowler},\ and\ \citenamefont {Neven}}]{Babbush_2018}%
  \BibitemOpen
  \bibfield  {author} {\bibinfo {author} {\bibfnamefont {R.}~\bibnamefont {Babbush}}, \bibinfo {author} {\bibfnamefont {C.}~\bibnamefont {Gidney}}, \bibinfo {author} {\bibfnamefont {D.~W.}\ \bibnamefont {Berry}}, \bibinfo {author} {\bibfnamefont {N.}~\bibnamefont {Wiebe}}, \bibinfo {author} {\bibfnamefont {J.}~\bibnamefont {McClean}}, \bibinfo {author} {\bibfnamefont {A.}~\bibnamefont {Paler}}, \bibinfo {author} {\bibfnamefont {A.}~\bibnamefont {Fowler}}, \ and\ \bibinfo {author} {\bibfnamefont {H.}~\bibnamefont {Neven}},\ }\href {\doibase 10.1103/physrevx.8.041015} {\bibfield  {journal} {\bibinfo  {journal} {Physical Review X}\ }\textbf {\bibinfo {volume} {8}} (\bibinfo {year} {2018}{\natexlab{b}}),\ 10.1103/physrevx.8.041015}\BibitemShut {NoStop}%
\bibitem [{\citenamefont {Kikuchi}\ \emph {et~al.}(2023)\citenamefont {Kikuchi}, \citenamefont {Mc~Keever}, \citenamefont {Coopmans}, \citenamefont {Lubasch},\ and\ \citenamefont {Benedetti}}]{Kikuchi_23}%
  \BibitemOpen
  \bibfield  {author} {\bibinfo {author} {\bibfnamefont {Y.}~\bibnamefont {Kikuchi}}, \bibinfo {author} {\bibfnamefont {C.}~\bibnamefont {Mc~Keever}}, \bibinfo {author} {\bibfnamefont {L.}~\bibnamefont {Coopmans}}, \bibinfo {author} {\bibfnamefont {M.}~\bibnamefont {Lubasch}}, \ and\ \bibinfo {author} {\bibfnamefont {M.}~\bibnamefont {Benedetti}},\ }\href {\doibase 10.1088/2058-9565/ab8e92} {\bibfield  {journal} {\bibinfo  {journal} {npj Quantum Information}\ }\textbf {\bibinfo {volume} {9}},\ \bibinfo {pages} {93} (\bibinfo {year} {2023})}\BibitemShut {NoStop}%
\bibitem [{\citenamefont {Vallury}\ \emph {et~al.}(2020)\citenamefont {Vallury}, \citenamefont {Jones}, \citenamefont {Hill},\ and\ \citenamefont {Hollenberg}}]{Vallury2020}%
  \BibitemOpen
  \bibfield  {author} {\bibinfo {author} {\bibfnamefont {H.~J.}\ \bibnamefont {Vallury}}, \bibinfo {author} {\bibfnamefont {M.~A.}\ \bibnamefont {Jones}}, \bibinfo {author} {\bibfnamefont {C.~D.}\ \bibnamefont {Hill}}, \ and\ \bibinfo {author} {\bibfnamefont {L.~C.}\ \bibnamefont {Hollenberg}},\ }\href {\doibase 10.22331/Q-2020-12-15-373} {\bibfield  {journal} {\bibinfo  {journal} {Quantum}\ }\textbf {\bibinfo {volume} {4}} (\bibinfo {year} {2020}),\ 10.22331/Q-2020-12-15-373}\BibitemShut {NoStop}%
\bibitem [{\citenamefont {Bergholm}\ \emph {et~al.}(2005)\citenamefont {Bergholm}, \citenamefont {Vartiainen}, \citenamefont {M\"ott\"onen},\ and\ \citenamefont {Salomaa}}]{Bergholm_2005}%
  \BibitemOpen
  \bibfield  {author} {\bibinfo {author} {\bibfnamefont {V.}~\bibnamefont {Bergholm}}, \bibinfo {author} {\bibfnamefont {J.~J.}\ \bibnamefont {Vartiainen}}, \bibinfo {author} {\bibfnamefont {M.}~\bibnamefont {M\"ott\"onen}}, \ and\ \bibinfo {author} {\bibfnamefont {M.~M.}\ \bibnamefont {Salomaa}},\ }\href {\doibase 10.1103/PhysRevA.71.052330} {\bibfield  {journal} {\bibinfo  {journal} {Phys. Rev. A}\ }\textbf {\bibinfo {volume} {71}},\ \bibinfo {pages} {052330} (\bibinfo {year} {2005})}\BibitemShut {NoStop}%
\bibitem [{\citenamefont {Shende}\ \emph {et~al.}(2006)\citenamefont {Shende}, \citenamefont {Bullock},\ and\ \citenamefont {Markov}}]{Shende_2006}%
  \BibitemOpen
  \bibfield  {author} {\bibinfo {author} {\bibfnamefont {V.}~\bibnamefont {Shende}}, \bibinfo {author} {\bibfnamefont {S.}~\bibnamefont {Bullock}}, \ and\ \bibinfo {author} {\bibfnamefont {I.}~\bibnamefont {Markov}},\ }\href {\doibase 10.1109/tcad.2005.855930} {\bibfield  {journal} {\bibinfo  {journal} {{IEEE} Transactions on Computer-Aided Design of Integrated Circuits and Systems}\ }\textbf {\bibinfo {volume} {25}},\ \bibinfo {pages} {1000} (\bibinfo {year} {2006})}\BibitemShut {NoStop}%
\bibitem [{\citenamefont {Tudorovskaya}\ and\ \citenamefont {Muñoz~Ramo}(2024)}]{Tudorovskaya_2024}%
  \BibitemOpen
  \bibfield  {author} {\bibinfo {author} {\bibfnamefont {M.}~\bibnamefont {Tudorovskaya}}\ and\ \bibinfo {author} {\bibfnamefont {D.}~\bibnamefont {Muñoz~Ramo}},\ }\href {\doibase 10.1103/physreva.109.032612} {\bibfield  {journal} {\bibinfo  {journal} {Physical Review A}\ }\textbf {\bibinfo {volume} {109}} (\bibinfo {year} {2024}),\ 10.1103/physreva.109.032612}\BibitemShut {NoStop}%
\bibitem [{\citenamefont {Sawaya}\ \emph {et~al.}(2020)\citenamefont {Sawaya}, \citenamefont {Menke}, \citenamefont {Kyaw}, \citenamefont {Johri}, \citenamefont {Aspuru-Guzik},\ and\ \citenamefont {Guerreschi}}]{Sawaya_2020}%
  \BibitemOpen
  \bibfield  {author} {\bibinfo {author} {\bibfnamefont {N.~P.}\ \bibnamefont {Sawaya}}, \bibinfo {author} {\bibfnamefont {T.}~\bibnamefont {Menke}}, \bibinfo {author} {\bibfnamefont {T.~H.}\ \bibnamefont {Kyaw}}, \bibinfo {author} {\bibfnamefont {S.}~\bibnamefont {Johri}}, \bibinfo {author} {\bibfnamefont {A.}~\bibnamefont {Aspuru-Guzik}}, \ and\ \bibinfo {author} {\bibfnamefont {G.~G.}\ \bibnamefont {Guerreschi}},\ }\href {\doibase 10.1038/s41534-020-0278-0} {\bibfield  {journal} {\bibinfo  {journal} {npj Quantum Information}\ }\textbf {\bibinfo {volume} {6}} (\bibinfo {year} {2020}),\ 10.1038/s41534-020-0278-0}\BibitemShut {NoStop}%
\bibitem [{\citenamefont {Bravyi}\ \emph {et~al.}(2017)\citenamefont {Bravyi}, \citenamefont {Gambetta}, \citenamefont {Mezzacapo},\ and\ \citenamefont {Temme}}]{Bravyi:2017eoo}%
  \BibitemOpen
  \bibfield  {author} {\bibinfo {author} {\bibfnamefont {S.}~\bibnamefont {Bravyi}}, \bibinfo {author} {\bibfnamefont {J.~M.}\ \bibnamefont {Gambetta}}, \bibinfo {author} {\bibfnamefont {A.}~\bibnamefont {Mezzacapo}}, \ and\ \bibinfo {author} {\bibfnamefont {K.}~\bibnamefont {Temme}},\ }\href@noop {} {\  (\bibinfo {year} {2017})},\ \Eprint {http://arxiv.org/abs/1701.08213} {arXiv:1701.08213 [quant-ph]} \BibitemShut {NoStop}%
\bibitem [{\citenamefont {Yamamoto}\ \emph {et~al.}(2024)\citenamefont {Yamamoto}, \citenamefont {Duffield}, \citenamefont {Kikuchi},\ and\ \citenamefont {Mu\~noz Ramo}}]{PhysRevResearch.6.013221}%
  \BibitemOpen
  \bibfield  {author} {\bibinfo {author} {\bibfnamefont {K.}~\bibnamefont {Yamamoto}}, \bibinfo {author} {\bibfnamefont {S.}~\bibnamefont {Duffield}}, \bibinfo {author} {\bibfnamefont {Y.}~\bibnamefont {Kikuchi}}, \ and\ \bibinfo {author} {\bibfnamefont {D.}~\bibnamefont {Mu\~noz Ramo}},\ }\href {\doibase 10.1103/PhysRevResearch.6.013221} {\bibfield  {journal} {\bibinfo  {journal} {Phys. Rev. Res.}\ }\textbf {\bibinfo {volume} {6}},\ \bibinfo {pages} {013221} (\bibinfo {year} {2024})}\BibitemShut {NoStop}%
\bibitem [{git(2025)}]{github}%
  \BibitemOpen
  \href {https://github.com/CQCL/qtnm-tts} {\enquote {\bibinfo {title} {Truncated taylor series lcu},}\ } (\bibinfo {year} {2025})\BibitemShut {NoStop}%
\bibitem [{\citenamefont {Shaw}\ \emph {et~al.}(2020)\citenamefont {Shaw}, \citenamefont {Lougovski}, \citenamefont {Stryker},\ and\ \citenamefont {Wiebe}}]{Shaw2020quantumalgorithms}%
  \BibitemOpen
  \bibfield  {author} {\bibinfo {author} {\bibfnamefont {A.~F.}\ \bibnamefont {Shaw}}, \bibinfo {author} {\bibfnamefont {P.}~\bibnamefont {Lougovski}}, \bibinfo {author} {\bibfnamefont {J.~R.}\ \bibnamefont {Stryker}}, \ and\ \bibinfo {author} {\bibfnamefont {N.}~\bibnamefont {Wiebe}},\ }\href {\doibase 10.22331/q-2020-08-10-306} {\bibfield  {journal} {\bibinfo  {journal} {{Quantum}}\ }\textbf {\bibinfo {volume} {4}},\ \bibinfo {pages} {306} (\bibinfo {year} {2020})}\BibitemShut {NoStop}%
\bibitem [{\citenamefont {Norambuena}\ \emph {et~al.}(2020)\citenamefont {Norambuena}, \citenamefont {Tancara},\ and\ \citenamefont {Coto}}]{Norambuena_2020}%
  \BibitemOpen
  \bibfield  {author} {\bibinfo {author} {\bibfnamefont {A.}~\bibnamefont {Norambuena}}, \bibinfo {author} {\bibfnamefont {D.}~\bibnamefont {Tancara}}, \ and\ \bibinfo {author} {\bibfnamefont {R.}~\bibnamefont {Coto}},\ }\href {\doibase 10.1088/1361-6404/ab8360} {\bibfield  {journal} {\bibinfo  {journal} {European Journal of Physics}\ }\textbf {\bibinfo {volume} {41}},\ \bibinfo {pages} {045404} (\bibinfo {year} {2020})}\BibitemShut {NoStop}%
\bibitem [{\citenamefont {Tranter}\ \emph {et~al.}(2022)\citenamefont {Tranter}, \citenamefont {Di~Paola}, \citenamefont {Mu\~{n}oz Ramo}, \citenamefont {Manrique}, \citenamefont {Gowland}, \citenamefont {Plekhanov}, \citenamefont {Greene-Diniz}, \citenamefont {Christopoulou}, \citenamefont {Prokopiou}, \citenamefont {Keen}, \citenamefont {Polyak}, \citenamefont {Khan}, \citenamefont {Pilipczuk}, \citenamefont {Kirsopp}, \citenamefont {Yamamoto}, \citenamefont {Tudorovskaya}, \citenamefont {Krompiec}, \citenamefont {Sze}, \citenamefont {Fitzpatrick}, \citenamefont {Anderson},\ and\ \citenamefont {Bhasker}}]{inquanto}%
  \BibitemOpen
  \bibfield  {author} {\bibinfo {author} {\bibfnamefont {A.}~\bibnamefont {Tranter}}, \bibinfo {author} {\bibfnamefont {C.}~\bibnamefont {Di~Paola}}, \bibinfo {author} {\bibfnamefont {D.}~\bibnamefont {Mu\~{n}oz Ramo}}, \bibinfo {author} {\bibfnamefont {D.~Z.}\ \bibnamefont {Manrique}}, \bibinfo {author} {\bibfnamefont {D.}~\bibnamefont {Gowland}}, \bibinfo {author} {\bibfnamefont {E.}~\bibnamefont {Plekhanov}}, \bibinfo {author} {\bibfnamefont {G.}~\bibnamefont {Greene-Diniz}}, \bibinfo {author} {\bibfnamefont {G.}~\bibnamefont {Christopoulou}}, \bibinfo {author} {\bibfnamefont {G.}~\bibnamefont {Prokopiou}}, \bibinfo {author} {\bibfnamefont {H.~D.~J.}\ \bibnamefont {Keen}}, \bibinfo {author} {\bibfnamefont {I.}~\bibnamefont {Polyak}}, \bibinfo {author} {\bibfnamefont {I.~T.}\ \bibnamefont {Khan}}, \bibinfo {author} {\bibfnamefont {J.}~\bibnamefont {Pilipczuk}}, \bibinfo {author} {\bibfnamefont {J.~J.~M.}\ \bibnamefont {Kirsopp}}, \bibinfo {author} {\bibfnamefont {K.}~\bibnamefont {Yamamoto}}, \bibinfo
  {author} {\bibfnamefont {M.}~\bibnamefont {Tudorovskaya}}, \bibinfo {author} {\bibfnamefont {M.}~\bibnamefont {Krompiec}}, \bibinfo {author} {\bibfnamefont {M.}~\bibnamefont {Sze}}, \bibinfo {author} {\bibfnamefont {N.}~\bibnamefont {Fitzpatrick}}, \bibinfo {author} {\bibfnamefont {R.~J.}\ \bibnamefont {Anderson}}, \ and\ \bibinfo {author} {\bibfnamefont {V.}~\bibnamefont {Bhasker}},\ }\href {https://www.quantinuum.com/products-solutions/inquanto} {\enquote {\bibinfo {title} {{InQuanto: Quantum Computational Chemistry}},}\ } (\bibinfo {year} {2022})\BibitemShut {NoStop}%
\bibitem [{\citenamefont {Mottonen}\ \emph {et~al.}(2004)\citenamefont {Mottonen}, \citenamefont {Vartiainen}, \citenamefont {Bergholm},\ and\ \citenamefont {Salomaa}}]{Mottonen_04}%
  \BibitemOpen
  \bibfield  {author} {\bibinfo {author} {\bibfnamefont {M.}~\bibnamefont {Mottonen}}, \bibinfo {author} {\bibfnamefont {J.~J.}\ \bibnamefont {Vartiainen}}, \bibinfo {author} {\bibfnamefont {V.}~\bibnamefont {Bergholm}}, \ and\ \bibinfo {author} {\bibfnamefont {M.~M.}\ \bibnamefont {Salomaa}},\ }\href {\doibase 10.1103/PhysRevLett.93.130502} {\enquote {\bibinfo {title} {Quantum circuits for general multiqubit gates},}\ } (\bibinfo {year} {2004})\BibitemShut {NoStop}%
\end{thebibliography}%

 \onecolumngrid
 \appendix
 \section{Amplified LCU success probability}\label{appendix:oaa}

The success probability of LCU is given by
\begin{equation}
        p = \frac{1}{|\alpha|
_1^2}\langle \psi| \Upsilon^\dagger \Upsilon | \psi \rangle,  
\end{equation}
where $\Upsilon$ is the truncated $\exp(-iHt)$, expressed as a linear combination of Pauli strings, and $|\alpha|_1$ is the sum of absolute coefficients of the Pauli terms. This decreases with increasing $|\alpha|_1$ or number of LCU terms. To improve $p$, an amplifier $A = -WRW^\dagger R $  can be employed by attaching it on the LCU circuit before postselection, where $W$ is the LCU operator acting on both the ancilla and system registers, and $R = (I - 2|\bar{0}\rangle\langle\bar{0}|) \otimes I$ is the reflection operator. We derive the outcome of a single application of $A$ as follows:
\begin{align}
    W |\Psi\rangle &= W |0\rangle |\psi\rangle = \sqrt{p} |\Phi\rangle + \sqrt{1-p} |\Phi^\perp\rangle = \sqrt{p} |0\rangle \Upsilon \frac{|\psi\rangle}{|\alpha|_1} + \sqrt{1-p} |\Phi^\perp\rangle,\label{eq:LCU_op}\\
    W^\dagger |\Phi\rangle &= \sqrt{p} |\Psi\rangle + \sqrt{1-p} |\Psi^\perp\rangle, \label{eq:U_dag_Phi}\\
    W^\dagger |\Phi^\perp\rangle &= \sqrt{1-p} |\Psi\rangle - \sqrt{p} |\Psi^\perp\rangle, \label{eq:U_dag_PhiPerp}\\
    RW |\Psi\rangle &= \sqrt{p} |\Phi\rangle - \sqrt{1-p} |\Phi^\perp\rangle \\
    W^\dagger RW |\Psi\rangle &= (2p-1) |\Psi\rangle + 2\sqrt{p(1-p)}|\Psi^\perp\rangle \\
    RW^\dagger RW |\Psi\rangle &= (2p-1) |\Psi\rangle - 2\sqrt{p(1-p)}|\Psi^\perp\rangle\\
    WRW^\dagger RW |\Psi\rangle &= \sqrt{p}(4p-3) |\Phi\rangle + \sqrt{1-p}(4p-1) |\Phi^\perp\rangle = \sqrt{p}(4p-3)|0\rangle \Upsilon \frac{|\psi\rangle}{ |\alpha|_1}  + \sqrt{1-p}(4p-1) |\Phi^\perp\rangle.
\end{align}
Since the amplitude amplification operator is defined as $A = -WRW^\dagger R$, postselection of the $|0\rangle$ ancilla register leads to a state $\frac{\sqrt{p}(4p-3)}{|\alpha|_1 }\Upsilon |\psi\rangle$. Thus, the amplified amplitude is given by
\begin{align}\label{eq:p_amp}
        p_{amp} = p (4p-3)^2.
\end{align}
This implies that if $p\ll 0.5$, then after one round of amplification, the success probability should be improved by about a factor of $9$. 
Equation~(\ref{eq:p_amp}) is exact where we assumed $\Upsilon^\dagger \Upsilon \approx I$. Otherwise, the more general expression if $\langle\psi|\Upsilon^\dagger \Upsilon|\psi\rangle \not\approx 1$ is given by
\begin{align}
\label{eq:p_amp_gen}
    p_{amp} =\frac{1}{|\alpha|_1^2} \left\langle \psi \left| \left( 3\Upsilon - \frac{4}{|\alpha|_1^2}\Upsilon \Upsilon^\dagger \Upsilon \right)^\dagger \left( 3\Upsilon - \frac{4}{|\alpha|_1^2}\Upsilon \Upsilon^\dagger \Upsilon \right) \right|\psi \right\rangle.
\end{align}
As $A$ utilizes two LCUs, the circuit depth of an amplified LCU triples for every application of $A$. 


 \section{\textit{Multiplexor} Compilation} \label{sec:app_m}
\subsection{\textit{Multiplexed-U(2)}}\label{sec:app mul-u2}
A $k$ control quantum multiplexor gate targeting a single qubit can be decomposed into a quantum circuit with $2^k - 1$ \CX gates, $2^k$ single qubit gates and a $(k+1)$-qubit diagonal gate with a recursive algorithm that applies a series of \textit{demultiplexing} steps, as shown earlier in Fig. \ref{fig:vdecomp}. The construction we explain here was first introduced by Bergholm et al in \cite{Bergholm_2005}. We will modify the notation by defining a multiplexor with $k$ control qubits and $n$ target qubits as $M^k_n(V)$.

\subsubsection{2-qubit multiplexed-U(2)}
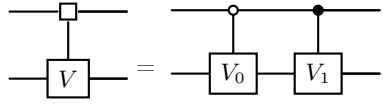
\begin{figure*}[!ht]
    \begin{quantikz}
        & \mctrl{1}& \qw \\
        & \gate{V}& \qw \\
    \end{quantikz} =
    \begin{quantikz}
        & \octrl{1}& \ctrl{1}& \qw \\
        & \gate{V_0}& \gate{V_1}& \qw \\
    \end{quantikz}
    \caption{$M^1_1(V)$ gate}
    \label{fig:m1v1}
\end{figure*}

We will first look at how to decompose a quantum multiplexor gate $M^1_1(V)$ with one control qubit and one target qubit, before seeing how to extend this to $M^k_1(V)$ for $k>1$. The unitary matrix of $M^1_1(V) = V_0\bigoplus V_1$ is block-diagonal, with $U(2)$ unitary $V_0$ in the top left corner and $U(2)$ unitary $V_1$ in the bottom right corner. We may also consider it as a pair of controlled gates as in Fig. \ref{fig:m1v1}. For decomposing $M^1_1(V)$ to single and two-qubit gates, the aim is to rewrite the matrix representation  $M^1_1(V)$ such that it is a sequence of matrices with known gate implementations.
This can be achieved by solving the matrix equation for the parameterisable $U(2)$ matrices $W, L$ and $U(2)$ diagonal matrix $r$,
\begin{equation}
    \begin{pmatrix}
    V_0 & \\
    & V_1 \\
    \end{pmatrix} =
    \begin{pmatrix}
    r^\dagger & \\
    & r \\
    \end{pmatrix}
    \begin{pmatrix}
    W & \\
    &W  \\
    \end{pmatrix}
    \begin{pmatrix}
    d & \\
    & d^\dagger\\
    \end{pmatrix}
    \begin{pmatrix}
    L & \\
    & L \\
    \end{pmatrix}
    \label{eq:constant demul}
\end{equation}
where $d = \begin{pmatrix}e^{i\pi/4} & 0\\0 & e^{-i\pi/4}\end{pmatrix}$.

Equation~(\ref{eq:constant demul}) can be written as two equations:
\begin{equation}
    V_0 = r^\dagger WdL,
    \label{eq:v0}
\end{equation}
\begin{equation}
    V_1 = r Wd^\dagger L.
    \label{eq:v1}
\end{equation}

To solve for $W$, $L$ and $r$, we rewrite these equations as 
\begin{equation}
    rV_0V_1^\dagger r  =rr^\dagger WdL L^\dagger d W^\dagger r^\dagger r =  Wd^2W^\dagger.
    \label{eq:matrixrewrite}
\end{equation}
$Wd^2W^\dagger$ can be obtained via eigen-decomposition on $rV_0V_1^\dagger r$. Since we know $d^2 = \begin{pmatrix}e^{i\pi/2} & 0\\0 & e^{-i\pi/2}\end{pmatrix}$, we find $r$ such that the eigenvalues of $rV_0V_1^\dagger r$ are $e^{\pm i\pi/2}$.
As $V_0V_1^\dagger$ is a $2\times2$ unitary matrix, it can be expressed as 
\begin{equation}
    V_0V_1^\dagger  = 
     \begin{pmatrix} 
     x_0 & x_1 \\
     -\overline{x_1} & \overline{x_0}
     \end{pmatrix}e^{i\theta/2},
\end{equation}
where $|x_0|^2 + |x_1|^2 = 1$. The characteristic polynomial of $rV_0V_1^\dagger r$ is therefore $det(rV_0V_1^\dagger r - \lambda I) = \lambda^2 - \lambda(\gamma_0^2x_0 + \gamma_1^2\overline{x_0})e^{i\theta/2} + \gamma_0^2\gamma_1^2e^{i\theta}$, where $r=\begin{pmatrix}\gamma_0 & \\ & \gamma_1 \end{pmatrix}$ (ommitting the simplification steps). The roots have the expression \begin{equation}\lambda = \frac{(\gamma_0^2x_0 + \gamma_1^2\overline{x_0})e^{i\theta/2} \pm \sqrt{(\gamma_0^2x_0 + \gamma_1^2\overline{x_0})^2 e^{i\theta} - 4\gamma_0^2\gamma_1^2e^{i\theta}}}{2}.
\label{eq:root expression}\end{equation}
Since the roots are the eigenvalues of $rV_0V_1^\dagger r$, and we know they are $\pm i$, summing the expressions for these two roots, we obtain
\begin{equation}
    \gamma_0^2x_0 + \gamma_1^2\overline{x_0} = 0.
    \label{eq:gamma condition 0.5}
\end{equation}
Equation~(\ref{eq:gamma condition 0.5}) implies
\begin{equation}
    \gamma_0^2e^{i\arg{x_0}} + \gamma_1^2e^{-i\arg{x_0}} = 0,
    \label{eq:gamma condition 1}
\end{equation}
where $\arg{x_0}$ is the argument of $x_0$. In addition, Eq.~(\ref{eq:root expression}) becomes
\begin{equation}
\lambda = \frac{\pm \sqrt{ - 4\gamma_0^2\gamma_1^2e^{i\theta}}}{2} = \pm  i\gamma_0\gamma_1e^{i\theta/2}.
\label{eq:gamma condition 2}\end{equation}
In order to satisfy Eq.~(\ref{eq:gamma condition 1}), we set 
$\gamma_0 = e^{i(\frac{-\arg{x_0}}{2}+\eta)}$ and $\gamma_1 = e^{i(\frac{\arg{x_0}}{2}+\eta + \pi/2)}$, where $\eta$ is a real number which we solve next. We substitute $\gamma_0$ and $\gamma_1$ in Eq.~(\ref{eq:gamma condition 2}) to obtain
\begin{equation}
\lambda = \pm  ie^{i(2\eta+\pi/2+\theta/2)}
\end{equation}
hence $\eta = \frac{-\theta-\pi}{4}$, and putting everything together,
\begin{equation}
    r=\begin{pmatrix}e^{i(-\arg{x_0} -\frac{\theta}{2} -\frac{\pi}{2})/2} & 0\\0 & e^{i(\arg{x_0} -\frac{\theta}{2} +\frac{\pi}{2})/2} \end{pmatrix}.
\end{equation}

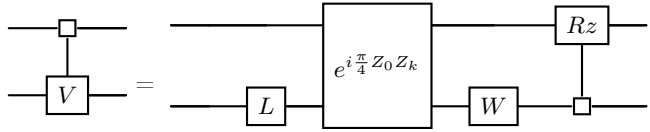
\begin{figure*}[!htbp]
    \begin{quantikz}
        & \mctrl{1}& \qw \\
        & \gate{V}& \qw \\
    \end{quantikz} 
    =
    \begin{quantikz}
       &\qw& \qw& \gate[wires=2]{e^{i\frac{\pi}{4}Z_0Z_k}} & \qw& \gate{Rz} & \qw\\
        &\qw& \gate{L}& \qw & \gate{W} & \mctrl{-1} & \qw\\
    \end{quantikz}
    \caption{Demultiplex $M^1_1(V)$}
    \label{fig:2-q multiplexor}
\end{figure*}

With $r$ solved, $W$ can be obtained immediately by performing eigen-decomposition on $rV_0V_1^\dagger r$ and subsequently $L$ can be solved using Eq.~(\ref{eq:v0}). To implement the matrix decomposition in Eq.~(\ref{eq:constant demul}), recognise that $\begin{pmatrix}d & \\& d^\dagger \end{pmatrix}$ is the 2-qubit gate $e^{i\frac{\pi}{4}\Z\Z}$. $\begin{pmatrix}W & \\&W \end{pmatrix}$ and $\begin{pmatrix}L & \\&L \end{pmatrix}$ are single qubit unitary gates on the second qubit (i.e. $\begin{pmatrix}W & \\&W \end{pmatrix}$ = $\identity\otimes W$). 
Lastly, note that 
$$\begin{pmatrix}r^\dagger & \\& r\end{pmatrix}=\begin{pmatrix}\overline{\gamma_0}&&&\\&\overline{\gamma_1}&&\\&&\gamma_0&\\&&&\gamma_1\end{pmatrix}$$
and the submatrices $\begin{pmatrix}\overline{\gamma_0} & \\& \gamma_0\end{pmatrix}, \begin{pmatrix}\overline{\gamma_1} & \\& \gamma_1\end{pmatrix}$ correspond to two \Rz gates. Therefore, $\begin{pmatrix}r^\dagger & \\& r\end{pmatrix}$ is a multiplexed-\Rz gate controlled by the second qubit. To conclude, Eq.~(\ref{eq:constant demul}) in circuit form is shown in Fig.~\ref{fig:2-q multiplexor}. 
The multiplexed-\Rz gate has an alternative construction explained in the following section.


\subsubsection{Multi-qubit multiplexed-$U(2)$}
\begin{figure*}[!htbp]
    \begin{quantikz}
        &\qwbundle{k}& \mctrl{1}& \qw \\
        &\qw& \gate{V}& \qw \\
    \end{quantikz} 
    =
    \begin{quantikz}
       &\qw& \qw& \gate[wires=3, nwires=2]{e^{i\frac{\pi}{4}\Z_0\Z_k}} & \qw& \gate{\Rz} & \qw\\
        &\qwbundle{k-1}& \mctrl{1}& \qw & \mctrl{1}& \mctrl{-1} & \qw\\
        &\qw& \gate{L}& \qw & \gate{W} & \mctrl{-1} & \qw\\
    \end{quantikz}
    \caption{Demultiplex $M^k_1(V)$}
    \label{fig:multiq demul}
\end{figure*}
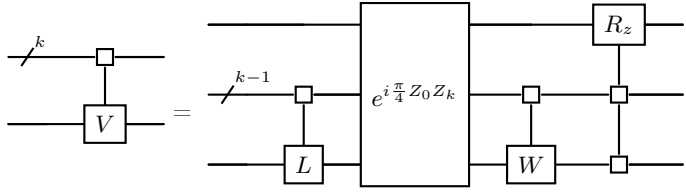

The problem is now to decompose an arbitrary multiplexed-$U(2)$ gate with $k$ control qubits. This gate has a block-diagonal matrix representation where each block is a $U(2)$ unitary. To decompose it, we can  pair every $U(2)$ matrix $V_i$ from the top half of the matrix with the corresponding $V_{i+2^{k-1}}$ in the bottom half at the same position, and apply Eq.~(\ref{eq:constant demul}) for each $V_i$ and $V_{i+2^{k-1}}$ pair. This process is illustrated both in Eq.~(\ref{eq:multiq demul}) and in circuit form in Fig.~\ref{fig:multiq demul}.
\newcommand{\bigmtxt}[1]{
  \scalebox{1.5}{$#1$}
}
\newcommand{\smmtxt}[1]{
  \scalebox{0.5}{$#1$}
}
\begin{equation}
    \resizebox{\textwidth}{!}{$
    \left(\begin{array}{ccc|ccc}
    V_0& & & & &\\
     &\smmtxt{\ddots} & & & &\\ 
     & &V_{j-1}& & & \\ \hline 
     & & &V_j& &\\
     & & & &\smmtxt{\ddots} &\\ 
     & & & & &V_{2^k-1}
    \end{array}\right)
    =
    \underbrace{\left(\begin{array}{ccc|ccc}
    r_0^\dagger& & & & &\\
     &\smmtxt{\ddots} & & & &\\ 
     & &r_{j-1}^\dagger& & & \\ \hline 
     & & &r_0& &\\
     & & & &\smmtxt{\ddots} &\\ 
     & & & & &r_{j-1}
    \end{array}\right)}_{Multiplexed-\Rz}
    \underbrace{\left(\begin{array}{ccc|ccc}
    W_0& & & & &\\
     &\smmtxt{\ddots} & & & &\\ 
     & &W_{j-1}& & & \\ \hline 
     & & &W_0& &\\
     & & & &\smmtxt{\ddots} &\\ 
     & & & & &W_{j-1}
    \end{array}\right)}_{I \otimes M^{k-1}_1(W)}
    \underbrace{\left(\begin{array}{ccc|ccc}
    d& & & & &\\
     &\smmtxt{\ddots} & & & &\\ 
     & &d& & & \\ \hline 
     & & &d^\dagger& &\\
     & & & &\smmtxt{\ddots} &\\ 
     & & & & &d^\dagger
    \end{array}\right)}_{e^{i\frac{\pi}{4}Z_0Z_k}}
    \underbrace{\left(\begin{array}{ccc|ccc}
    L_0& & & & &\\
     &\smmtxt{\ddots} & & & &\\ 
     & &L_{j-1}& & & \\ \hline 
     & & &L_0& &\\
     & & & &\smmtxt{\ddots} &\\ 
     & & & & &L_{j-1}
    \end{array}\right)}_{\identity \otimes M^{k-1}_1(L)}
    $}
    \label{eq:multiq demul}
\end{equation}

First note that the $e^{i\frac{\pi}{4}\Z_0\Z_k}$ gate only acts on the top and bottom qubits. Then, to see why the first matrix is a multiplexed-\Rz gate, we can write each $r_i:= \text{diag}(e^{\frac{1}{2}i\pi a_i}, e^{\frac{1}{2}i\pi b_i})$, so the entire matrix is of the form $\text{diag}(e^{-\frac{1}{2}i\pi a_0}, e^{-\frac{1}{2}i\pi b_0}, \cdots, e^{-\frac{1}{2}i\pi a_{j-1}}, e^{-\frac{1}{2}i\pi b_{j-1}}, e^{\frac{1}{2}i\pi a_0}, e^{\frac{1}{2}i\pi b_0}, \cdots, e^{\frac{1}{2}i\pi a_{j-1}}, e^{\frac{1}{2}i\pi b_{j-1}})$. Note that every entry from the first half together with its corresponding entry from the second half form an \Rz gate (i.e. $\text{diag}(e^{-\frac{1}{2}i\pi a_0}, e^{\frac{1}{2}i\pi a_0}) = \Rz(a_0$)). Therefore, the matrix is a multiplexed rotation $M^\mathcal{C}_\mathcal{T}(R)$ where $R = [\Rz(a_0),\cdots,\Rz(a_{j-1}), \Rz(b_0),\cdots,\Rz(b_{j-1})]$, $\mathcal{T}=[0]$, and $\mathcal{C} = [k, 1, 2, \cdots, k-1]$.

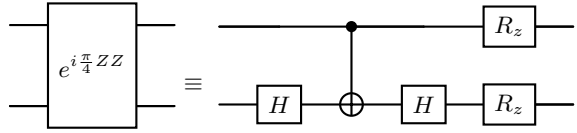
\begin{figure*}[!htbp]
    \begin{quantikz}
        \qw& \gate[2]{e^{i\frac{\pi}{4}\Z\Z}}& \qw \\
        \qw& \qw& \qw \\
    \end{quantikz} 
    $\equiv$
    \begin{quantikz}
       \qw& \qw& \ctrl{1} & \qw& \gate{\Rz}&\qw\\
        \qw& \gate{H}& \targ{}& \gate{H}& \gate{\Rz}&\qw\\
    \end{quantikz} 
    \caption{Rewrite $e^{i\frac{\pi}{4}\Z\Z}$}
    \label{fig:zzphase}
\end{figure*}

\begin{figure}[!htbp]
    \begin{quantikz}
        \qw& \mctrl{1}& \qw \\
        \qw& \mctrl{1}& \qw \\
        \qw& \gate{}& \qw \\
    \end{quantikz} 
    =
    \begin{quantikz}
       \qw& \qw& \gate[wires=3, nwires={2}]{e^{i\frac{\pi}{4}\Z_0\Z_2}} & \qw& \gate{\Rz} & \qw\\
        \qw& \mctrl{1}& \qw & \mctrl{1}& \mctrl{-1} & \qw\\
        \qw& \gate{}& \qw & \gate{} & \mctrl{-1} & \qw\\
    \end{quantikz} 
    = 
    \begin{quantikz}
       \qw&\qw    &\qw                            &\qw     &\qw        &\gate[wires=3, nwires={2}]{e^{i\frac{\pi}{4}\Z_0\Z_2}} &\qw      & \gate{\Rz} & \qw\\
       \qw&\qw    &\gate[2]{e^{i\frac{\pi}{4}\Z\Z}} &\qw     &\gate{\Rz}  &\qw                                                  &\mctrl{1}& \mctrl{-1} & \qw\\
       \qw&\gate{}&\qw                            &\gate{} &\mctrl{-1} &\qw                                                  &\gate{}  & \mctrl{-1} & \qw\\
    \end{quantikz} 
    = 
    \begin{quantikz}
       \qw&\qw    &\qw                            &\qw     &\gate[wires=3, nwires={2}]{e^{i\frac{\pi}{4}\Z_0\Z_2}}&\qw       &\qw      & \gate{\Rz} & \qw\\
       \qw&\qw    &\gate[2]{e^{i\frac{\pi}{4}\Z\Z}} &\qw     &\qw                                                 &\gate{\Rz} &\mctrl{1}& \mctrl{-1} & \qw\\
       \qw&\gate{}&\qw                            &\gate{} &\qw                                                 &\mctrl{-1}&\gate{}  & \mctrl{-1} & \qw\\
    \end{quantikz}
    = 
    \begin{quantikz}
       \qw&\qw    &\qw                            &\qw     &\gate[wires=3, nwires={2}]{e^{i\frac{\pi}{4}\Z_0\Z_2}}&\qw                              &\qw      & \gate{\Rz} & \qw\\
       \qw&\qw    &\gate[2]{e^{i\frac{\pi}{4}\Z\Z}} &\qw     &\qw                                                 &\gate{\Rz}\gategroup[2,steps=2,style={dotted, cap=round, inner sep=2pt}]{merge} &\mctrl{1}& \mctrl{-1} & \qw\\
       \qw&\gate{}&\qw                            &\gate{} &\qw                                                 &\mctrl{-1}                       &\gate{}  & \mctrl{-1} & \qw
    \end{quantikz} 
    = 
    \begin{quantikz}
       \qw&\qw    &\qw                            &\qw     &\gate[wires=3, nwires={2}]{e^{i\frac{\pi}{4}\Z_0\Z_2}}&\qw      & \gate{\Rz} & \qw\\
       \qw&\qw    &\gate[2]{e^{i\frac{\pi}{4}\Z\Z}} &\qw     &\qw                                                 &\mctrl{1}& \mctrl{-1} & \qw\\
       \qw&\gate{}&\qw                            &\gate{} &\qw                                                 &\gate{}  & \mctrl{-1} & \qw\\
    \end{quantikz}
    = 
    \begin{quantikz}
       \qw&\qw    &\qw                            &\qw     &\gate[wires=3, nwires={2}]{e^{i\frac{\pi}{4}\Z_0\Z_2}}&\qw    &\qw                           &\qw    &\qw           &\gate{\Rz} & \qw\\
       \qw&\qw    &\gate[2]{e^{i\frac{\pi}{4}\Z\Z}} &\qw     &\qw                                                 &\qw    &\gate[2]{e^{i\frac{\pi}{4}\Z\Z}}&\qw    &\gate{\Rz}     &\mctrl{-1} & \qw\\
       \qw&\gate{}&\qw                            &\gate{} &\qw                                                 &\gate{}&\qw                           &\gate{}&\mctrl{-1}    &\mctrl{-1} & \qw\\
    \end{quantikz}
    = 
    \begin{quantikz}
       \qw&\qw    &\qw       &\qw     &\ctrl{2} &\qw    &\qw     &\qw    &\qw           &\gate{\Rz} & \qw\\
       \qw&\qw    &\ctrl{1}  &\qw     &\qw      &\qw    &\ctrl{1}&\qw    &\gate{\Rz}     &\mctrl{-1} & \qw\\
       \qw&\gate{}&\targ{}   &\gate{} &\targ{}  &\gate{}&\targ{} &\gate{}&\mctrl{-1}    &\mctrl{-1} & \qw\\
    \end{quantikz}
    \caption{Decomposition of a 2-control multiplexor}
    \label{fig:3multiplexdec}
\end{figure}
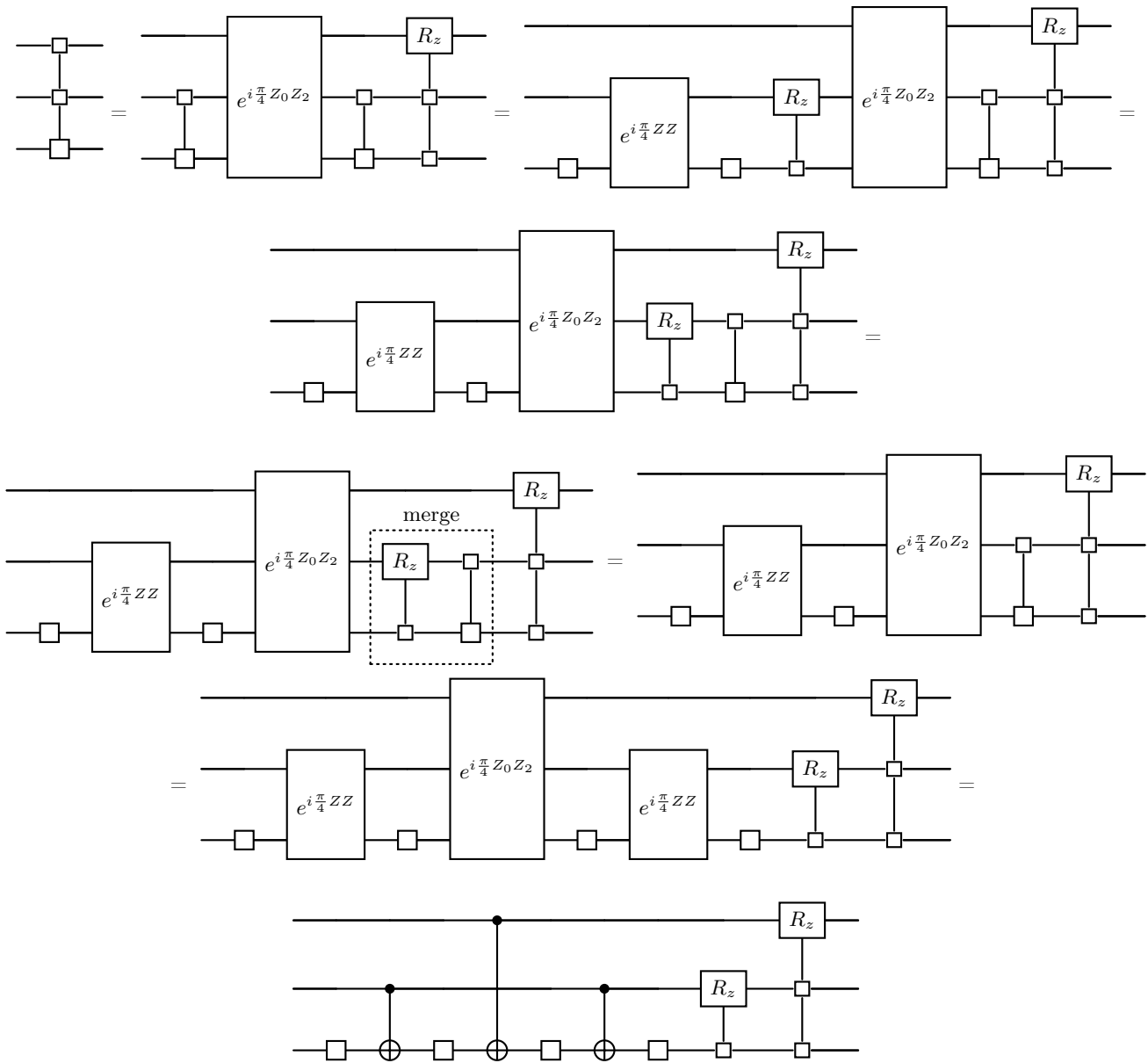

We proceed the recursion by decomposing the left multiplexor $M^{k-1}_1(L)$ first. The resulting multiplexed-\Rz gate from decomposing $M^{k-1}_1(L)$ commutes with $e^{i\frac{\pi}{4}\Z_0\Z_k}$ and can be merged with the $M^{k-1}_1(W)$ multiplexor on the right. The recursion finishes when the circuit is left with only single qubit gates, $e^{i\frac{\pi}{4}\Z_0\Z_k}$ gates and multiplexed-\Rz gates. Furthermore, a $e^{i\frac{\pi}{4}\Z\Z}$ gate can be rewritten (up to a global phase $e^{-i\pi/4}$) into a \cx gate and 4 single qubit gates, as shown in Fig.~\ref{fig:zzphase}. The \Rz on the top qubit can be merged with a multiplexed-\Rz gate on its right, and the bottom single-qubit gates can be absorbed into their surrounding single qubit unitaries. Finally, the multiplexed-\Rz gates can be synthesised into elementary gates \cite{Mottonen_04}. See Appendix~\ref{sec:app mul-rotation} for explanation. Figure~\ref{fig:3multiplexdec} shows how this decomposition works for a 2-control single-target multiplexor.


\subsection{\textit{Multiplexed-Rotation}}\label{sec:app mul-rotation}

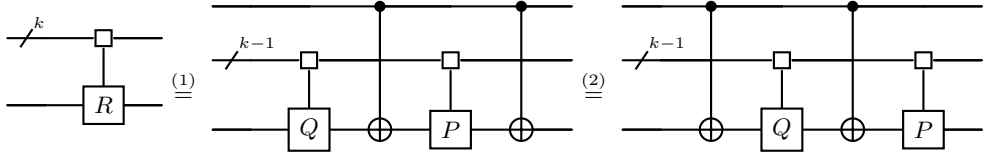
\begin{figure}[!htbp]
    \begin{quantikz}
        &\qwbundle{k}& \mctrl{1}& \qw \\
        &\qw& \gate{R}& \qw \\
    \end{quantikz} 
    $\stackrel{\text{(1)}}{=}$
    \begin{quantikz}
       &\qw& \qw& \ctrl{2} & \qw& \ctrl{2} & \qw\\
        &\qwbundle{k-1}& \mctrl{1}& \qw & \mctrl{1}& \qw & \qw\\
        &\qw& \gate{Q}& \targ{} & \gate{P} & \targ{} & \qw\\
    \end{quantikz}
    $\stackrel{\text{(2)}}{=}$
    \begin{quantikz}
       &\qw&\ctrl{2}& \qw& \ctrl{2} & \qw& \qw\\
        &\qwbundle{k-1}&\qw& \mctrl{1}& \qw & \mctrl{1}& \qw\\
        &\qw& \targ{}&\gate{Q}& \targ{} & \gate{P}& \qw\\
    \end{quantikz}
    \caption{Two ways to demultiplex $M^k_1(R)$ where $R, P, Q$ consist of rotations around a fixed axis that is perpendicular to the $x$-axis.}
    \label{fig:multiq demul rotation}
\end{figure}

Multiplexed rotations can be decomposed recursively in a way that is similar to multiplexed-U(2) decomposition, as shown in Fig.~\ref{fig:multiq demul rotation}.

The first recursive relation in Fig.~\ref{fig:multiq demul rotation}(1) can be written in the matrix form

\begin{equation}
    \resizebox{\textwidth}{!}{$
    \left(\begin{array}{ccc|ccc}
    R(\alpha_0)& & & & &\\
     &\smmtxt{\ddots} & & & &\\ 
     & &R(\alpha_{j-1})& & & \\ \hline 
     & & &R(\alpha_j)& &\\
     & & & &\smmtxt{\ddots} &\\ 
     & & & & &R(\alpha_{2^k-1})
    \end{array}\right)
    =
    \underbrace{\left(\begin{array}{ccc|ccc}
    I& & & & &\\
     &\smmtxt{\ddots} & & & &\\ 
     & &I& & & \\ \hline 
     & & &X& &\\
     & & & &\smmtxt{\ddots} &\\ 
     & & & & &X
    \end{array}\right)}_{CX}
    \underbrace{\left(\begin{array}{ccc|ccc}
    R(\beta_0)& & & & &\\
     &\smmtxt{\ddots} & & & &\\ 
     & &R(\beta_{j-1})& & & \\ \hline 
     & & &R(\beta_0)& &\\
     & & & &\smmtxt{\ddots} &\\ 
     & & & & &R(\beta_{j-1})
    \end{array}\right)}_{I \otimes M^{k-1}_1(P)}
    \underbrace{\left(\begin{array}{ccc|ccc}
    I& & & & &\\
     &\smmtxt{\ddots} & & & &\\ 
     & &I& & & \\ \hline 
     & & &X& &\\
     & & & &\smmtxt{\ddots} &\\ 
     & & & & &X
    \end{array}\right)}_{CX}
    \underbrace{\left(\begin{array}{ccc|ccc}
    R(\gamma_0)& & & & &\\
     &\smmtxt{\ddots} & & & &\\ 
     & &R(\gamma_{j-1})& & & \\ \hline 
     & & &R(\gamma_0)& &\\
     & & & &\smmtxt{\ddots} &\\ 
     & & & & &R(\gamma_{j-1})
    \end{array}\right)}_{I \otimes M^{k-1}_1(Q)}
    $}
    \label{eq:multiq demul rotation}
\end{equation}
where $\gamma_i = \frac{\alpha_i+\alpha_{2^{k-1}+i}}{2}$ and $\beta_i = \frac{\alpha_i-\alpha_{2^{k-1}+i}}{2}$. To see why this equation holds, note that the following identities hold:
$$I R(\beta_i) I R(\gamma_i) = R(\alpha_i)$$ $$XR(\beta_i)XR(\gamma_i) = R(-\beta_i)R(\gamma_i) = R(\alpha_{2^{k-1}+i})$$
The recursive relation in Fig.~\ref{fig:multiq demul rotation}(2) is verified in the same way.

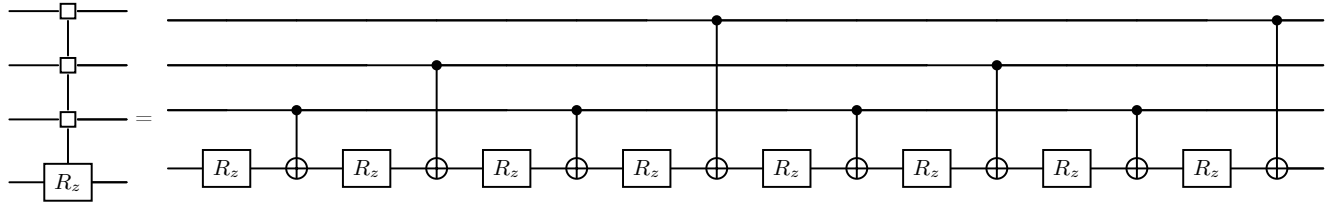
\begin{figure*}[!htbp]
  \begin{adjustbox}{width=\textwidth}
    \begin{quantikz}
      \qw    &\mctrl{1}       &\qw\\
      \qw    &\mctrl{1}       &\qw\\
      \qw    &\mctrl{1}       &\qw\\
      \qw    &\gate{\Rz}              &\qw\\ 
    \end{quantikz}
    =
    \begin{quantikz}
      \qw&\qw&\qw&\qw&\qw&\qw&\qw&\qw&\ctrl{3}&\qw&\qw&\qw&\qw&\qw&\qw&\qw&\ctrl{3}&\qw\\
      \qw&\qw&\qw&\qw&\ctrl{2}&\qw&\qw&\qw&\qw&\qw&\qw&\qw&\ctrl{2}&\qw&\qw&\qw&\qw&\qw\\
      \qw&\qw&\ctrl{1}&\qw&\qw&\qw&\ctrl{1}&\qw&\qw&\qw&\ctrl{1}&\qw&\qw&\qw&\ctrl{1}&\qw&\qw&\qw\\
      \qw&\gate{\Rz}&\targ{}&\gate{\Rz}&\targ{}&\gate{\Rz}&\targ{}&\gate{\Rz}&\targ{}&\gate{\Rz}&\targ{}&\gate{\Rz}&\targ{}&\gate{\Rz}&\targ{}&\gate{\Rz}&\targ{}&\qw\\
    \end{quantikz}
  \end{adjustbox}
  \caption{
    Recursive decomposition for $M^\CC_n(\Rz)$
    where $n=1$ and $\CC=3$, producing an
    alternating sequence of $2^k$ \CX gates and $2^k$ \Rz gates.} 
  \label{fig:rzdecomp}
\end{figure*}

The whole decomposition works by first demultiplexing $M^{k-1}_1(R)$ using Fig.~\ref{fig:multiq demul rotation}(1). In the recursive step, we always decompose the left and right multiplexors using Figs.~\ref{fig:multiq demul rotation}(1) and~\ref{fig:multiq demul rotation}(2), respectively. This gives redundant \CX that can be removed, resulting in a final circuit with $2^k$ \CX gates and $2^k$ single qubit rotations. This is shown for a 3-control-multiplexed-\Rz in Fig.~\ref{fig:rzdecomp}.

\subsection{Scaling}
\label{appedix:scaling}

The \select oracle construction using the proposed multiplexor construction requires no additional ancilla qubits, requiring at most $2^k(2n+1)-n-2$ \CX gates, where $k$ is the size of the
control register and $n$ the size of the target register. Depth is reduced by re-ordering the sequence of \CX gates between different multiplexors. A promising alternative construction for \select is  \textit{unary iteration} \cite{Babbush_2018}. The \textit{LCU direct} method proposed in \cite{AlexisLCU} describes unary iteration and gives a \CX count of $O(2^{k-1}(4n+17)-31)$, with $k-1$ ancilla qubits. Assuming $k\ge2$ and $n\ge1$, we find that the multiplexor construction only produces circuits with fewer \CX gates when $n\le12$. Additionally, unary iteration focuses on minimising the number of non-Clifford resources in \select, while the multiplexor construction does not. Therefore, we consider the multiplexor construction to be a useful near-term construction that is expected to be outperformed by unary iteration for larger problems.

 \clearpage
\section{Jaynes-Cummings Hamiltonian}\label{appendix:jc}

A simple light-matter interaction system is described by a Jaynes-Cummings Hamiltonian\footnote{For neatness, we set $\hbar=1$ so it is not explicit in the expression.}:
\begin{align}
    H_{JC} = \omega_c \hat{b}^\dagger \hat{b} + \omega_a \frac{\hat{\sigma}_z}{2} + g \left( \hat{b}\hat{\sigma}_+ + \hat{b}^\dagger\hat{\sigma}_- \right),
\end{align}
where $\omega_c$ and $\omega_a$ are the cavity or light field and atomic transition frequencies, and $g$ is the light-atom coupling constant. 
The light field mode is expressed in terms of the bosonic creation and annihilation operators $\hat{b}^\dagger$, $\hat{b}$, while the two-level atom is treated as a spin-half system, where $\hat{\sigma}_\pm = \frac{1}{2}\left( \hat{\sigma}_x \pm i \hat{\sigma}_y \right)$ is the raising or lowering operator of the atom\footnote{$\hat{\sigma}_+ = |e\rangle\langle g| = |0\rangle\langle1|$ and $\hat{\sigma}_- = |g\rangle\langle e| = |1\rangle\langle 0| $}, and $\hat{\sigma}_z = |e\rangle\langle e| - |g\rangle\langle g| $ is the atomic inversion operator.  Binary encoding \cite{Sawaya_2020} is employed to map the field operators into qubit Pauli operators while the atomic mode operators are already expressed in qubit Paulis. 

\begin{figure}[p]
     \centering
         \includegraphics[width=1\textwidth]{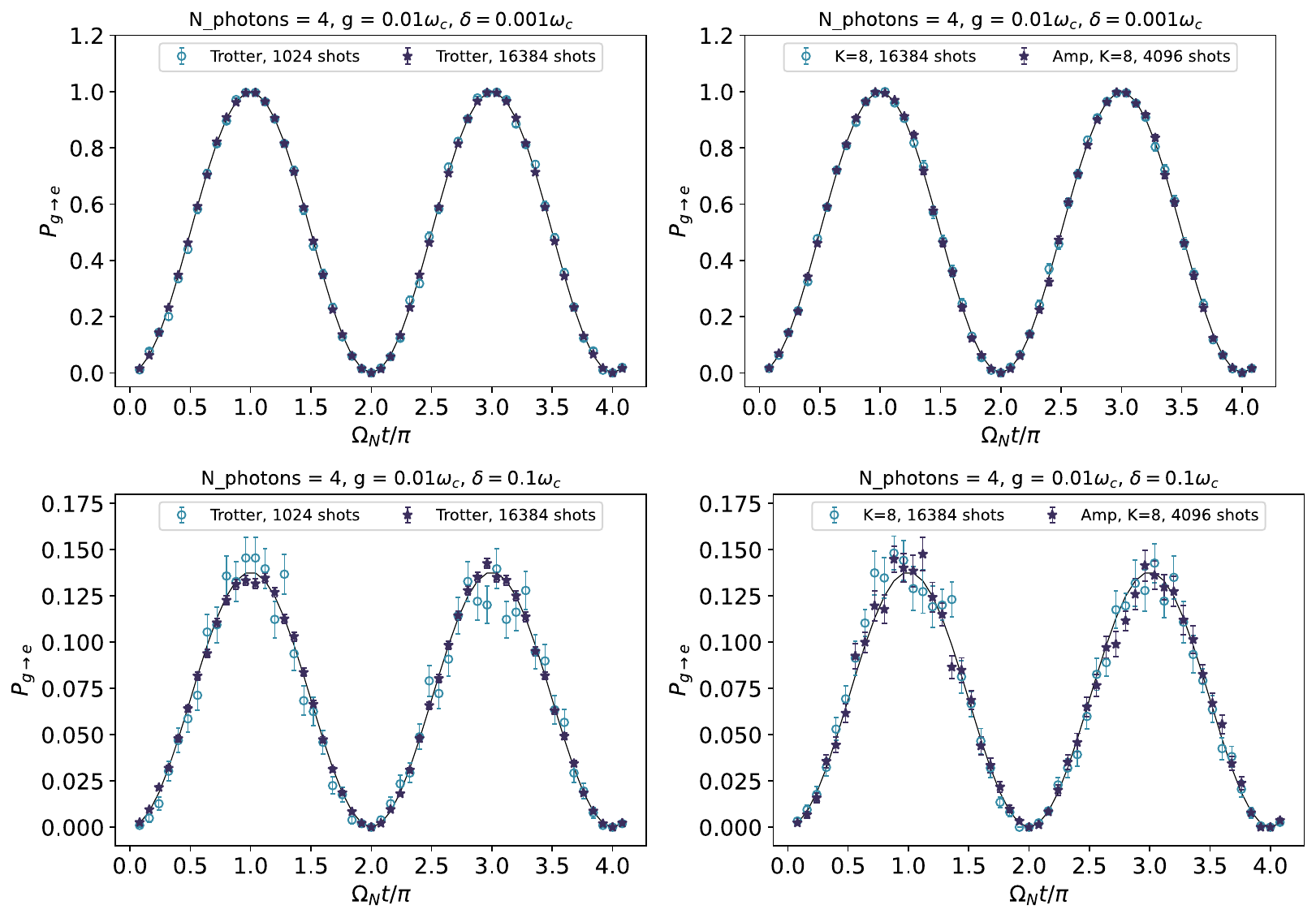}
     \caption{Transition probability in Jaynes-Cummings system with $N=4$, $g=0.01\omega_c$, $\delta=0.001\omega_c$ (top) and $\delta=0.1\omega_c$ (bottom), calculated using the Trotter method (left) with 4 qubits and LCU method (right) using 10 qubits.  Solid curve is the analytical transition probability Eq.~(\ref{eq:JC_P_g2e}). }
        \label{fig:JC_transition}
\end{figure}

\begin{figure}[p]
     \centering
         \includegraphics[width=1\textwidth]{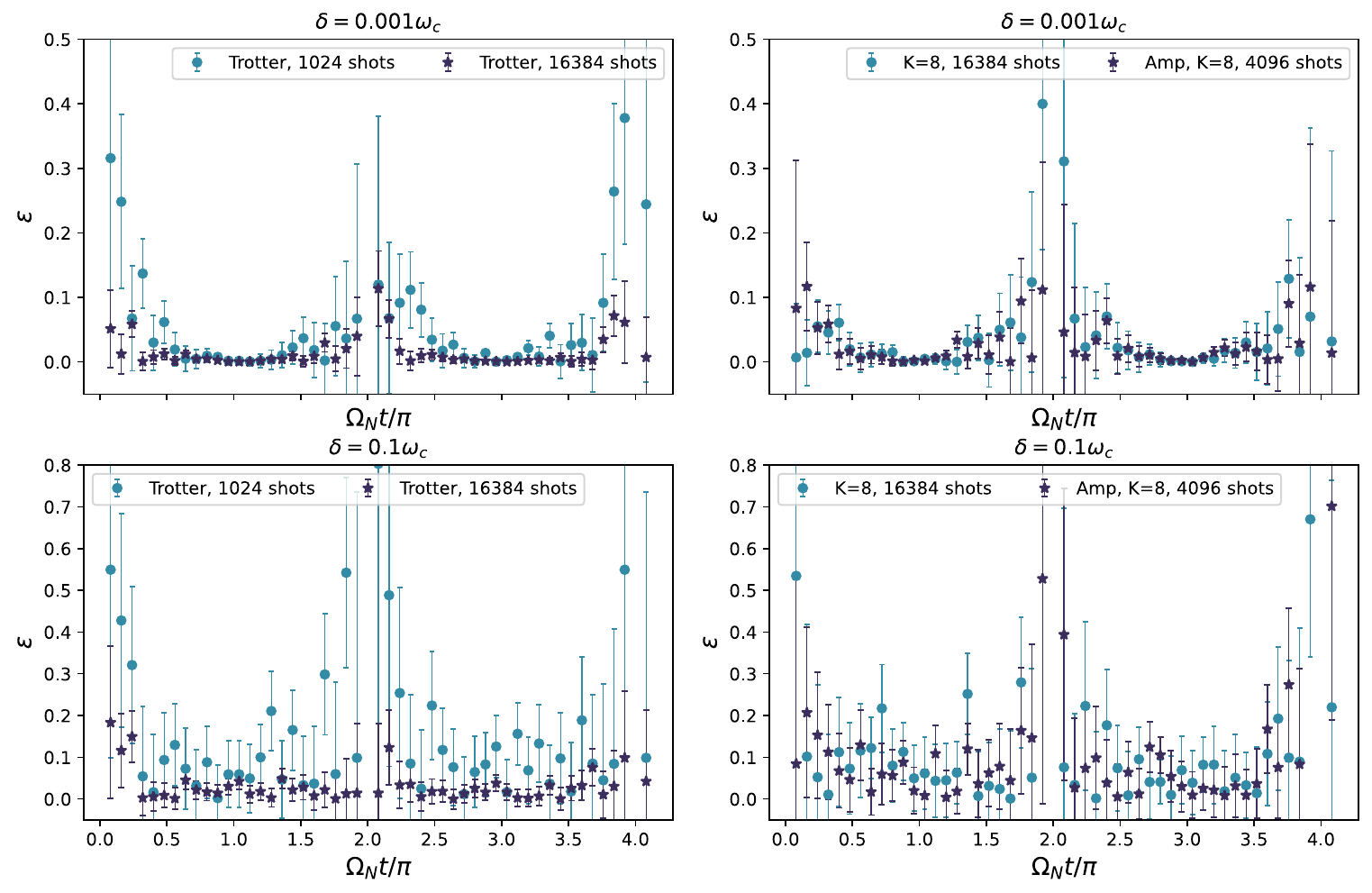}
     \caption{Simulations error with respect to the analytical solution of Jaynes-Cummings system transition probability using the Trotter (left) and LCU (right) methods.}
       \label{fig:Error_JC}
\end{figure}

If the system starts with $N$ photons and an atom in the ground state, $|N, g\rangle$, then the probability of finding it in excited state  $|N-1, e\rangle$, at time $t$, oscillates according to
\begin{align}
\label{eq:JC_P_g2e}
    P_{g\rightarrow e} = \left| \left\langle N-1, e \left| e^{-iH_{JC}t/\hbar} \right| N, g \right\rangle \right|^2 = \frac{4g^2 N}{\Omega_N^2} \sin^2{(\frac{\Omega_N t}{2})},
\end{align}
where $\Omega_N = \sqrt{4g^2N + \delta^2}$ is the Rabi frequency, and $\delta = \omega_a - \omega_c$ is the detuning frequency. When the light and atomic transition frequencies are close to resonant ($\delta \rightarrow 0$, single-mode cavity is resonant with the energy difference between the two energy levels), the maximum probability of the atom flipping to excited state approaches one. In Fig.~\ref{fig:JC_transition}, we demonstrate this dynamics for $N=4$, and $\delta/\omega_c=0.001, 0.1$. Results from a naive implementation of Trotterization are included for comparison. Four qubits are used to define the state; $6$ additional ancillary qubits (coming from $40$ Pauli strings of the expanded $e^{-iH_{JC}t}$ ) are required if applying the LCU methods, without using the reduced method described in Sec.~\ref{sec:sq_overlap}. For the LCU simulations, the cutoff $K$ is chosen such that the precision is $|1 - \sqrt{\langle \psi_0 | \Upsilon^\dagger \Upsilon| \psi_0\rangle}| \leq 10^{-3}$. On the emulator H1-1E, we run the Trotter circuit with $2^{10}, 2^{14}$ shots, the LCU circuit with $2^{14}$, and LCU with amplification with $2^{12}$ shots. In Fig.~\ref{fig:Error_JC}, we plot the simulation error $\epsilon$ with respect to the analytical solution in Eq.~(\ref{eq:JC_P_g2e}), i.e., $\epsilon = |P_{g\rightarrow e} - y_{sim} |/P_{g\rightarrow e}$. Near resonance (e.g. $\delta/\omega_c = 0.001$), both circuit methods provide accurate and precise simulations. We observe enlarged $\epsilon$ when the target quantity is close to $0$. Sampling errors from both become apparent as $\delta$ increases. More shots may be needed for Trotter, and amplification for LCU to reduce the error. 

\begin{figure}[p]
     \centering
         \includegraphics[width=1\textwidth]{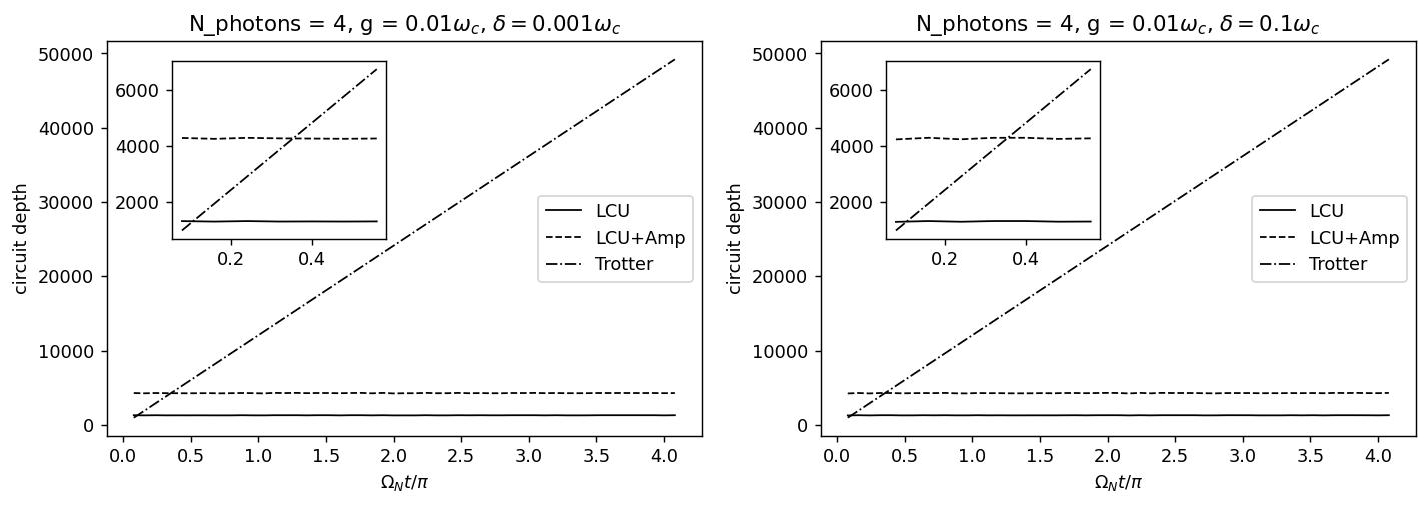}
     \caption{Circuit depth. Trotter vs LCU. The parameters $K=8$ and $\tau = 4\pi/1250/\Omega_N$ are constant in time.}
       \label{fig:JC_depth}
\end{figure}

The LCU method is expected to have an overhead cost coming from the ancillary qubits. However, when long time dynamics is the subject of interest, the method seems to perform better than Trotter considering the accuracy, precision and circuit depth. We plot the circuit depths of Trotterization, LCU, LCU+OAA in Fig.~\ref{fig:JC_depth}. Trotter starts with low-depth circuit at $t\approx 0$ but LCU (+OAA) soon outperforms it by having almost constant depth for the time interval considered while Trotter depth increases linearly.

Here, we consider both LCU and LCU with OAA methods, without the LCU reduction technique described in Sec.~\ref{sec:sq_overlap}, and compare these with the naïve first order Trotter approach. These simulations are performed on Quantinuum's emulator H1-1E without the noise model. The LCU, without OAA, circuits use $10$ qubits and have an average depth of $1280$ and $714$ \CX gates. Upon employing the preselection technique in Sec.~\ref{sec:sq_overlap}, the number of LCU terms is reduced to $2$, requiring only $1$ ancillary qubit. This is a good illustration of the effectivity of Sec.~\ref{sec:sq_overlap}, but may be inconsequential for testing the LCU algorithm on a hardware. 
 \clearpage
\section{$\Upsilon_\parallel$ and $\Upsilon_\perp$ grouping} \label{appendix:grouping}

In Sec.~\ref{sec:sq_overlap}, we suggest preselecting Pauli terms from the $\Upsilon_\parallel$, where $\langle \psi_0 | \Upsilon_\parallel |\psi_0 \rangle \neq 0$ and implementing the LCU on this group. The Paulis in $\Upsilon_\perp$ group are then irrelevant. Here, we outline the procedure that divides the Pauli terms in the qubit representation of $\exp(-iH_{RH}t)$ into the two groups.

Given an initial state that spans the basis $|q_0 q_1 q_2 q_3 q_4\rangle \in \lbrace |00000\rangle,  |01000\rangle,  |00011\rangle,  |01011\rangle \rbrace$, we make the following observations: (1) $q_0=0$ and $q_2=0$, and (2) $q_3=q_4$. Therefore, a Pauli string $\mathcal{P}_j = P_0 P_1 P_2 P_3 P_4$, where $P_i \in \lbrace X,Y,Z,I \rbrace$,  that goes to $\Upsilon_\parallel$ has the following properties: (i) $P_0, P_2 \in \lbrace I, Z \rbrace$; (ii) $P_3 \in \lbrace I, Z \rbrace$ iff $P_4 \in \lbrace I, Z \rbrace$; or,  $P_3 \in \lbrace X, Y \rbrace$ iff $P_4 \in \lbrace X, Y \rbrace$. Pauli operators that do not obey these properties are automatically placed in $\Upsilon_\perp$. Given these conditions, as $K\rightarrow \infty$ when all the possible Pauli strings may appear in the expanded $\exp({-iH_{RH}t})$, $87.5\%$ reduction is achieved. For the Pauli operators in $\Upsilon_\parallel$, we do the following simplifications: 
\begin{enumerate}
    \item $Z_0 \rightarrow I_0$, $Z_2 \rightarrow I_2$; $Z_3 Z_4 \rightarrow I_3 I_4$,
    \item $P_0 P_1 P_2 (X_3 X_4 + Y_3 Y_4) \rightarrow 0$;  $P_0 P_1 P_2 (X_3 X_4 - Y_3 Y_4) \rightarrow 2 P_0 P_1 P_2 X_3 X_4 $, 
    \item $P_0 P_1 P_2 (X_3 Y_4 - Y_3 X_4) \rightarrow 0$; $P_0 P_1 P_2 (X_3 Y_4 + Y_3 X_4) \rightarrow 2 P_0 P_1 P_2 X_3 Y_4$, 
    \item $P_0 P_1 P_2 (I_3 Z_4 - Z_3 I_4) \rightarrow 0$;  $P_0 P_1 P_2 (I_3 Z_4 + Z_3 I_4) \rightarrow 2 P_0 P_1 P_2 I_3 Z_4 $.
\end{enumerate}
 \clearpage
\section{Supporting figures} \label{appendix:more_figs}

\begin{figure}[!htbp]
         \centering
         \includegraphics[width=0.6\textwidth]{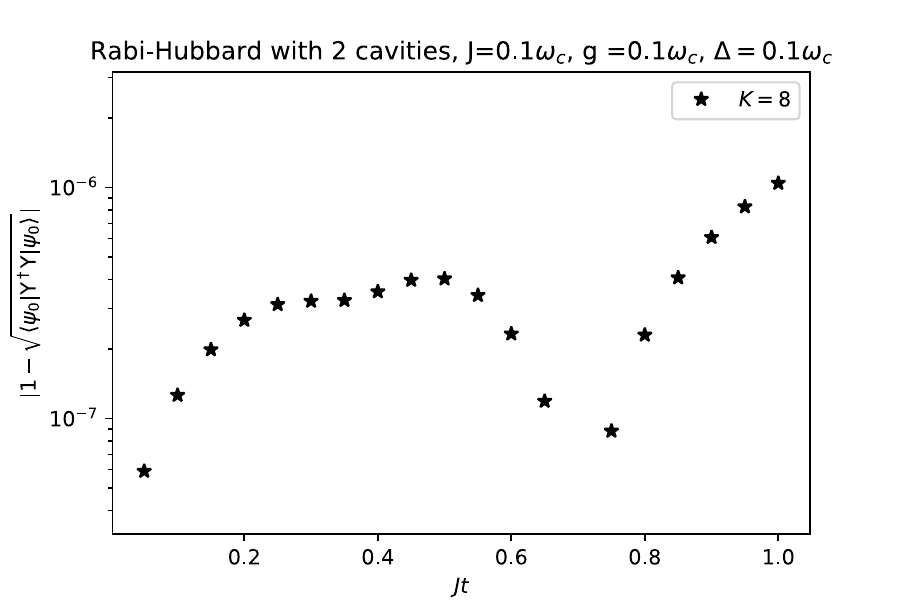}
         \caption{Precision of the Taylor expanded and truncated time evolution operator with the Rabi-Hubbard Hamiltonian, cutoff $K=8$, and discarding terms with coefficients $|\alpha_\ell|<10^{-8}$. }
         \label{fig:RH_lcu_precision}
\end{figure}

\begin{figure}[!htbp]
         \centering
         \includegraphics[width=0.6\textwidth]{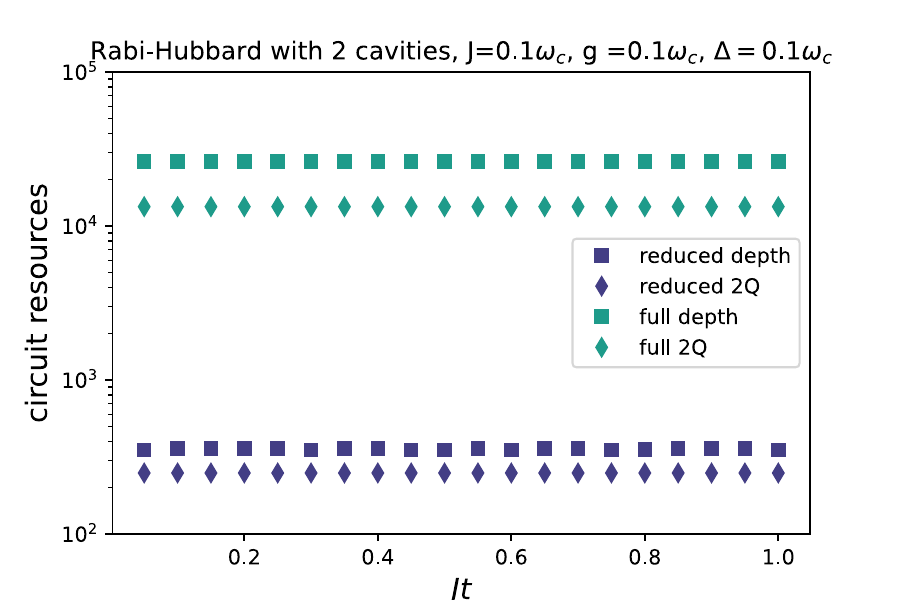}
         \caption{LCU circuit resources for the Rabi-Hubbard propagator with and without using the reduction technique in Sec.~\ref{sec:sq_overlap}. A full propagator circuit, minimally optimized using pytket optimization level 1, has about $26300$ depth with $13400$ 2Q gates and $14$ qubits. A reduced propagator circuit, minimally optimized using pytket optimization level 1, has about $360$ depth with $250$ 2Q gates and $9$ qubits.}
         \label{fig:RH_lcu_resources_opt1}
\end{figure}

\begin{figure}[!htbp]
         \centering
         \includegraphics[width=\textwidth]{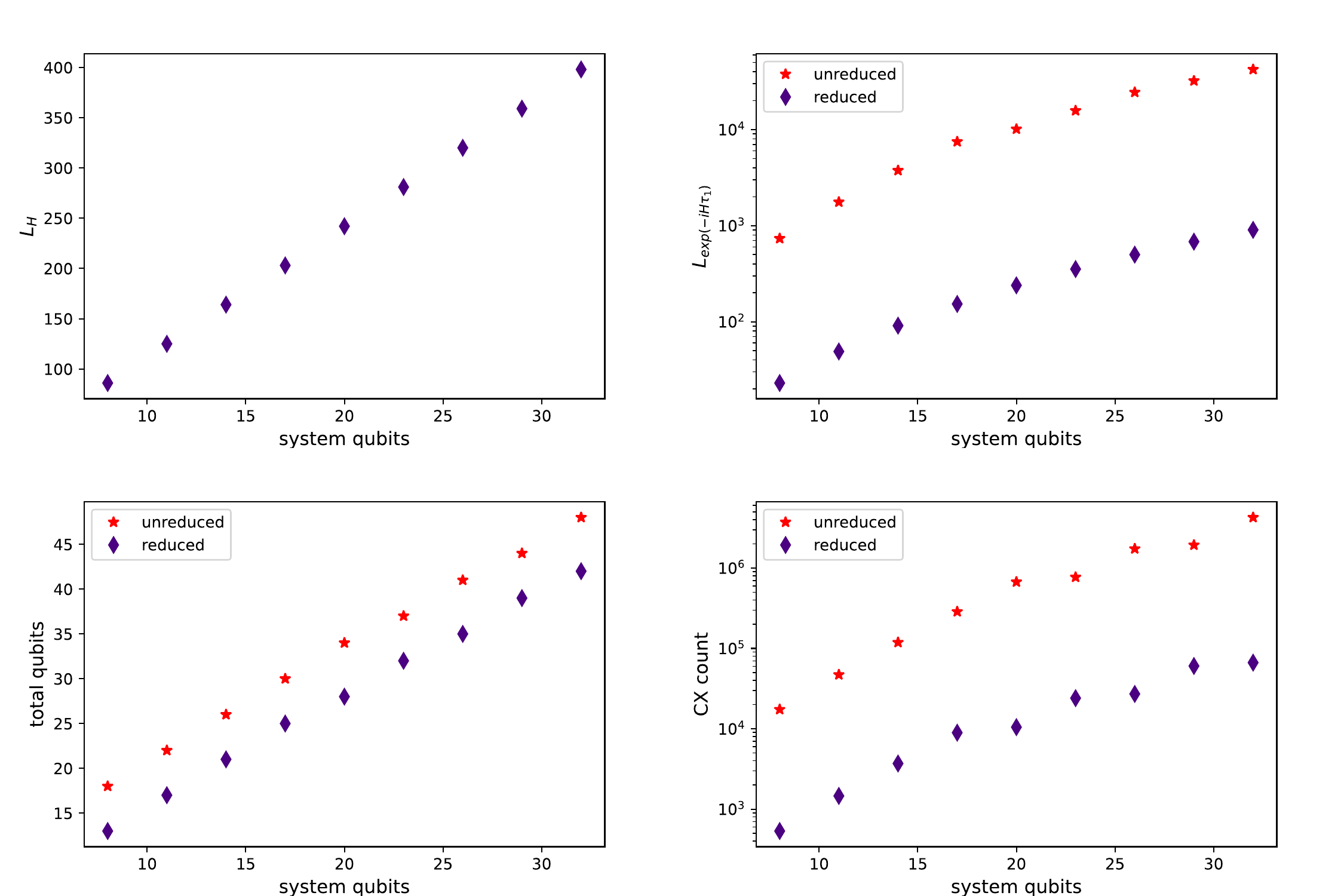}
         \caption{Larger Rabi-Hubbard systems. Number of cavities is varied from $3$ to $11$. The maximum photon number each cavity can have is set to $3$. (Top left) Number of Pauli strings in the resulting tapered qubit Hamiltonian. (Top right) Number of Pauli strings in the Taylor expanded and truncated ($K=6$) propagator at $J\tau_1=0.02$. Data labeled by `reduced' are obtained by doing the filtering technique discussed in Sec.~\ref{sec:sq_overlap}. (Bottom left) Total qubits (system and prepare registers combined) used for LCU implementation. (Bottom right) Estimated \CX gates with a multiplexor compilation strategy.}
         \label{fig:LargeRH}
\end{figure}

\end{document}